\documentclass[final]{agujournal2019}
\usepackage{url} 

\usepackage{lineno}
\usepackage{soul}
\numlinesfalse

\usepackage{amsmath,amssymb}
\usepackage{xcolor,colortbl}
\usepackage{makecell}

\usepackage{longtable}
\usepackage{booktabs}
\usepackage{multirow}
\usepackage{siunitx}

\usepackage{pifont}
\newcommand{\xmark}{\ding{53}}

\usepackage{comment}
\usepackage{textcomp,mathcomp}
\usepackage{algorithm}
\usepackage{algpseudocode}

\usepackage{xr}

\definecolor{lightgray}{cmyk}{0,0,0,0.17}

\draftfalse

\journalname{Geochemistry, Geophysics, Geosystems}

\begin{document}

\title{Sensitivity analysis of the thermal structure within subduction zones using reduced-order modeling}

\authors{Gabrielle M. Hobson \affil{1}, Dave A. May \affil{1}}

\affiliation{1}{University of California San Diego, San Diego, CA, 92093}

\correspondingauthor{Gabrielle M. Hobson}{ghobson@ucsd.edu}

\

\

\textbf{This manuscript is an ArXiv preprint and has been submitted for possible publication in a peer reviewed journal. Please note that this has not been peer-reviewed before and is currently undergoing peer review for the first time. Subsequent versions of this manuscript may have slightly different content.}

\begin{keypoints}
\item Reduced-order models are used to accurately approximate the thermal structure within 2D thermo-mechanical subduction zone models. 
\item Non-intrusive ROMs are highly efficient and are used for global sensitivity analysis exploring variability of slab interface temperatures.
\item Model input variability within previously published ranges results in significant variability in temperature and potential rupture extent.
\end{keypoints}

\begin{abstract}

Megathrust earthquakes are the largest on Earth, capable of causing strong ground shaking and generating tsunamis.
Physical models used to understand megathrust earthquake hazard are limited by existing uncertainties about material properties and governing processes in subduction zones.
A key quantity in megathrust hazard assessment is the distance between the updip and downdip rupture limits.
The thermal structure of a subduction zone exerts a first-order control on the extent of rupture.
We simulate temperature for profiles of the Cascadia, Nankai and Hikurangi subduction zones using a 2D coupled kinematic-dynamic thermal model.
We then build reduced-order models (ROMs) for temperature using the interpolated Proper Orthogonal Decomposition (iPOD).
The resulting ROMs are data-driven, model agnostic, and computationally cheap to evaluate.
Using the ROMs, we can efficiently investigate the sensitivity of temperature to input parameters, physical processes, and modeling choices.
We find that temperature, and by extension the potential rupture extent, is most sensitive to variability in parameters that describe shear heating on the slab interface,
followed by parameters controlling the thermal structure of the incoming lithosphere and coupling between the slab and the mantle.
We quantify the effect of using steady-state vs. time-dependent models, and of uncertainty in the choice of isotherm representing the downdip rupture limit.
We show that variability in input parameters translates to significant differences in estimated moment magnitude.
Our analysis highlights the strong effect of variability in the apparent coefficient of friction, with previously published ranges resulting in pronounced variability in estimated rupture limit depths.

\end{abstract}

\section*{Plain Language Summary}

Megathrust earthquakes are the largest on Earth and can cause destructive ground shaking and generate tsunamis.
Some subduction zones that host megathrust earthquakes every few hundred years are located close to population centers, so understanding where the earthquake might rupture is key to understanding the potential hazard.
The primary factor that controls the extent of earthquake rupture is the temperature near the interface between plates.
When modelers use simulations to estimate temperatures, they must make choices about (a) how to represent physics in the system, and (b) values for model inputs that describe properties of the system.
Because the plate interface is located deep in the Earth, this information cannot be precisely known.
In this study, we present tools to measure variability in the temperature, given a certain amount of variability in model inputs.
We use reduced-order models, which capture the characteristics of a system but are computationally cheaper to evaluate than standard models; they give us the power to calculate robust statistics and learn about the relationships between model input parameters and the estimated extent of rupture.
We show that there is significant variability in the estimated size of earthquakes, given variability in the input parameters.

\section{Introduction} \label{sec:intro}

Megathrust earthquakes are the largest earthquakes on Earth, occurring along thrust faults at convergent plate boundaries with magnitudes of $M_w$ $\geq$ 8.0 or 9.0.
Since they can cause strong ground shaking and generate tsunamis, quantifying the potential for megathrust earthquakes is of primary concern in hazard assessment.
Determining the locations of the updip and downdip rupture limits is a key step in estimating the potential extent of rupture during a megathrust earthquake.
The rupture limits correspond to transitions in the sliding behavior of rocks at the subduction interface, and are primarily controlled by temperature \cite{hyndman_wang_1993, tichelaar_ruff_1993, oleskevich_et_al_1999}.
Thermal models developed over the past few decades have been used to estimate temperatures in subduction zones \cite<e.g.>{hyndman_wang_1993, van_keken_kiefer_peacock_2002, currie_et_al_2004, wada_et_al_2008, community_benchmark_van_keken_2008, wada_wang_2009, syracuse_2010}.
The accuracy of temperatures estimated by these models is subject to uncertainty in (a) the parameters in the governing equations that describe motion and heat flow in the system, and (b) how well the models capture governing processes in subduction zones \cite{peacock_review_2020}.
Our aim is to investigate the effect of these uncertainties on estimates of the potential extent of rupture --- and by extension, megathrust hazard.

Analytical expressions for temperature on the subduction interface have been presented in \citeA{england_may_2021}. 
However, these analytical expressions are designed to be valid only along the plate boundary between the subducting slab and the overlying plate. 
They do not capture the interface between the slab and the mantle wedge, where the thermal regime is dominated by mantle wedge flow. 
Since we aim to investigate quantities of interest that may lie close to or below the mantle wedge corner, we use numerical models to estimate temperature in 2D profiles of subduction zones at a regional scale. 
Models have been developed to investigate temperature in specific subduction zones \cite<e.g.>{hyndman_wang_1993, currie_et_al_2004}, as well as global characteristics of subduction zone thermal structure \cite<e.g.>{syracuse_2010, wada_wang_2009}.
We closely follow the previously established kinematic-dynamic modeling approach to simulate temperature \cite{van_keken_kiefer_peacock_2002, currie_et_al_2004, wada_et_al_2008, community_benchmark_van_keken_2008}, with the aim of investigating uncertainties associated with each part of the model. 
Kinematic-dynamic thermal models estimate temperature in a subduction zone by prescribing the kinematic behavior of the subducting slab and dynamically solving for viscous flow in the mantle wedge. 
Over timescales of thousands to millions of years, the subducting slab is essentially decoupled and sliding beneath the overlying plate. At depth, the subducting slab is mechanically coupled to the overlying mantle, driving viscous flow in the mantle wedge.
The decoupled-to-coupled transition is described by the maximum depth of decoupling and the degree of partial coupling between slab and mantle in the mantle wedge corner.
Sliding along the decoupled interface creates shear heating, the magnitude of which is determined by the apparent coefficient of friction and the plate convergence rate. 
Temperatures in the slab are primarily affected by the thermal structure of the incoming lithosphere, typically modeled using a plate cooling or half-space cooling model that depends on slab age and mantle temperature \cite{community_benchmark_van_keken_2008, stein_stein_1992}.
The thermal structure in the mantle wedge is also strongly affected by the choice of rheology for mantle material when solving for viscous flow \cite{van_keken_kiefer_peacock_2002}.
Temperatures in the overlying plate are typically modeled using a geotherm that depends on the thickness of the overlying plate and the mantle temperature. 
Together, the descriptions of key physical processes in subduction zones result in a model input parameter space with dimension on the order of 10--100 parameters.

The parameters input into thermal models are, by virtue of being parameters describing characteristics of the deep Earth, naturally uncertain.
Shear heating along the subduction interface is assumed to be a function of the apparent shear stress during relative motion across the interface, averaged over many seismic cycles. 
In present-day subduction zones, most of the plate motion is accommodated by seismic events, so the apparent shear stress may be much lower than the static shear stress. 
This motivates the use of the apparent coefficient of friction, $\mu'$, rather than the static coefficient of friction, in the shear heating formulation. 
The apparent coefficient of friction, $\mu'$, has been estimated using 2D models constrained by surface heat flow data \cite{gao_wang_2014}.
Due to scatter in surface heat flow data and uncertainty in crustal radiogenic heat production, $\mu'$ is an ill-constrained parameter with estimates varying between 0.0 and 0.14 \cite{gao_wang_2014, england_2018}.
The maximum depth of decoupling is also uncertain due to limited surface heat flux data; a study of 17 subduction zones showed that for three of the profiles, surface heat flux data are consistent with maximum depths of decoupling in the $\sim$~60--90 km range, and for the remaining 14 profiles a depth of 75 km is not inconsistent with the surface heat flux data \cite{wada_wang_2009}. 
Parameters such as plate convergence rate and age of the subducting lithosphere are generally better constrained, but still carry uncertainty which affects the estimated thermal structure.

In addition to input parameter uncertainties, modeling choices add sources of uncertainty.
Previous work has found significant discrepancies between temperatures recorded by the exhumed rock record and temperatures predicted by models, and argued that these discrepancies may be due to several factors not captured by the thermal models, including shear heating, convection of rock in the shallow mantle and subduction channel,
and hydration reactions due to fluid release from the subducting slab \cite{pennistonDorland_2015}. 
Modeling efforts to better reconcile estimated P-T conditions with the exhumed rock record have involved adding shear heating to models, as well as modeling time-dependent conditions during subduction initiation \cite{van_keken_et_al_2018, holt_condit_2021, van_keken_wilson_2023, peacock_review_2020}. 
In this study, we quantify the effect of both input parameter variability and modeling choices, such as the inclusion of shear heating and the steady-state assumption, on estimated slab interface temperatures and potential rupture extent.

Estimates of the potential extent of coseismic rupture are not only subject to the intrinsic uncertainties associated with the thermal models; they are also subject to uncertainty in the choice of isotherm representing the downdip rupture limit. 
Since temperature is a primary control on rupture limits, the updip and downdip limits of rupture are understood to correspond to specific isotherms \cite{hyndman_wang_1993, tichelaar_ruff_1993}.
The updip limit is considered to be located around the 150°C isotherm, with the downdip limit (the critical temperature for the transition from velocity-weakening/unstable sliding to velocity-strengthening/stable sliding behavior) corresponding to 350{\textcelsius} \cite{hyndman_wang_1993} or 400{\textcelsius} \cite{tichelaar_ruff_1993}. 
The transition zone, which can be partly ruptured coseismically, is considered to be between 350{\textcelsius} and 450{\textcelsius}.
However, a study presenting laboratory experiments on olivine friction indicates that the transition from velocity weakening to velocity strengthening occurs at approximately 600{\textcelsius} \cite{boettcher_hirth_evans_2007}. 
In this study, we quantify how the choice of downdip rupture limit isotherm affects estimates of potential rupture extent.  

Due to the computational cost of evaluating the thermal models, the sensitivity of temperature to variation in model parameters has primarily been explored by varying a subset of model parameters ($\sim$~3--5 parameters) at a few samples in parameter space (typically $<$ 100 as in \citeA{maunder_et_al_2019}, at most 31,000 as in \citeA{morishige_kuwatani}). Similarly, prior studies that have used thermal models to estimate the extent of the seismogenic zone and therefore megathrust earthquake potential, for example in the Cascadia \cite{hyndman_wang_1993} and Makran subduction zones \cite{smith_et_al_2013,khaledzadeh_ghods_2021} among others, have used relatively few samples in a limited parameter space. 
To address the computational cost barrier, we leverage reduced-order modeling techniques that permit accurate approximations of the thermal structure to be evaluated orders of magnitude faster than full-order thermal models.

In this study, we present reduced-order models (ROMs) for the thermal structure of 2D profiles of Cascadia, Nankai and Hikurangi.
The ROMs are accurate and efficient, allowing us to use fully explore parameter space via robust global sensitivity analysis techniques that require many model evaluations.
Using the ROMs, we determine which model input parameters have an effect on the potential extent of rupture.
Further, we quantify how variability in input parameters, modeling choices, and the choice of the downdip limit isotherm translates to variability in estimated rupture extent.
We examine the implications of this variability for estimated moment magnitude for megathrust events.
We also compare the model-estimated rupture limits to observations of the downdip limit of interseismic coupling and slow-slip event depths.
Finally, we investigate whether considering variability in model input parameters results in modeled P-T conditions that span P-T conditions inferred from a global catalog of exhumed rocks.

\section{Subduction Zones Investigated} \label{sec:design_of_experiments}

In this study we examine the thermal structure associated with the Cascadia, Nankai and Hikurangi subduction zones.
These three margins are thought to be mid-to-late in the seismic cycle, and their proximity to population centers makes understanding their potential megathrust hazard a priority.
They exhibit a range of megathrust seismic slip styles, as well as along-strike variations in margin characteristics and aseismic slip processes.
Nankai is a data-rich margin with possibly the most extensive instrumentation and observations of any margin.
It produces Great earthquakes ($M_w > 8$) with an average recurrence interval of 180 years, most recently hosting the 1944 $M_w~8.1$ Tonankai and 1946 $M_w~8.3$ Nankaido earthquakes \cite{ando_1975, sagiya_thatcher_1999}.
Hikurangi displays diverse interseismic slip behavior and has strong along-strike variability in characteristics such as sediment cover, the stress state and seismic properties of the upper plate, and fluid release.
Paleoseismic investigations provide evidence for more frequent rupture of the locked southern Hikurangi margin, with less frequent ruptures of the central margin \cite{clark_2019}, which at present is dominated by slow slip events \cite{wallace_2020}.
Cascadia hosts $M_w > 8.0$ or $M_w > 9.0$ earthquakes every few hundred years \cite{goldfinger_2012}.
It also has along-strike variability in sediment cover, coupling and characteristics of slow-slip events (SSEs).
 It is considered an endmember subduction zone, one of the seismically quietest and hottest on Earth.
 The hot overall thermal structure is due to the young subducting lithosphere (8--10 Myr) and thick insulating sediments \cite{hyndman_wang_1993}.
Nankai is a cooler subduction zone than Cascadia, with an incoming lithosphere age between 15--26 Myr \cite{okino_et_al_1994}, though it has a relatively warm thermal structure compared to global subduction zones \cite{peacock_wang_1999}.
Meanwhile, Hikurangi is a relatively cold end-member with an incoming lithosphere age of $\sim$100 Myr \cite{wada_wang_2009}. 

\begin{figure}
     \centering
     \includegraphics[width=\textwidth]{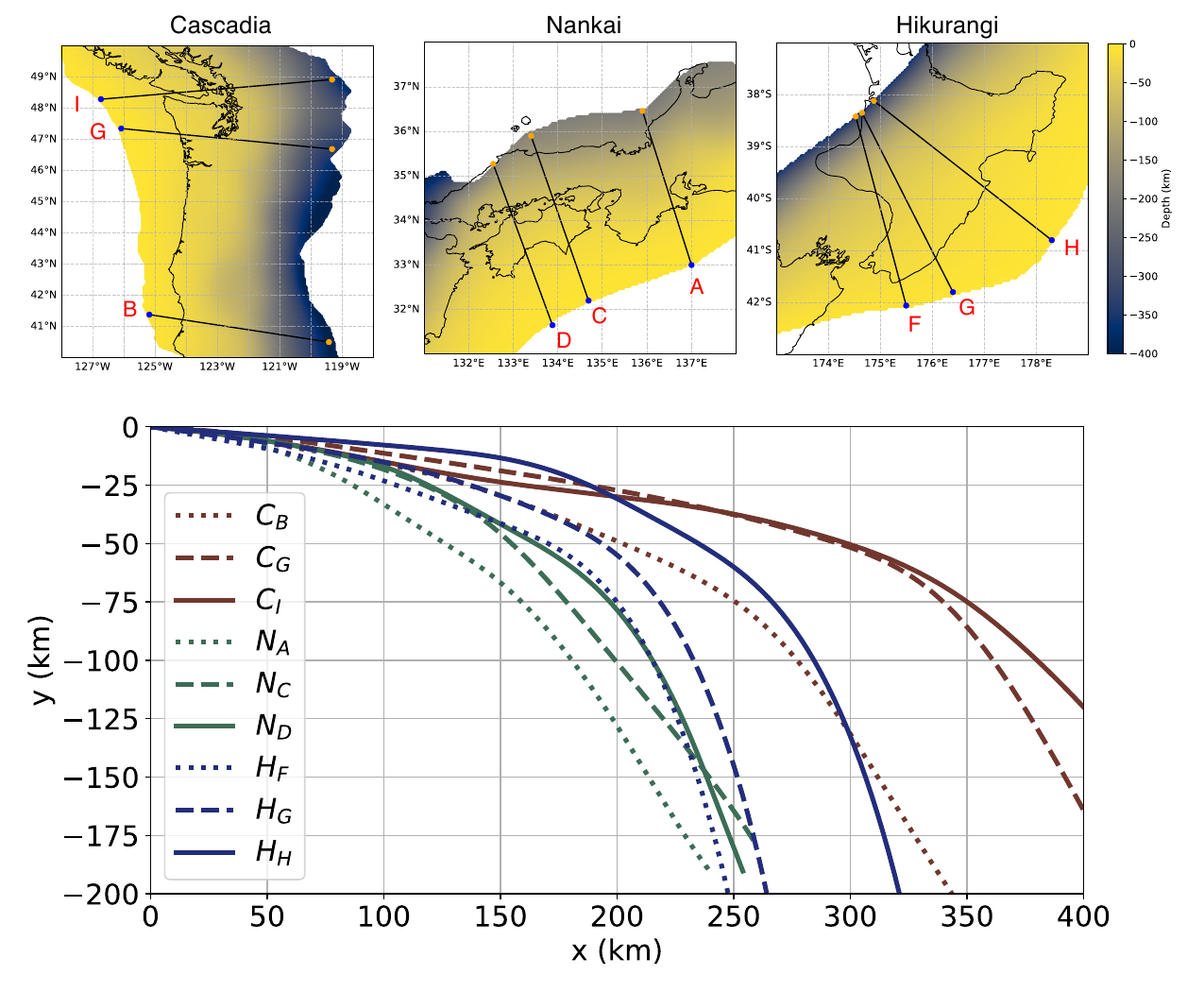}
     \caption{(Upper panels) Map view of the Slab2 data \cite{slab2} for the subduction zones under investigation (Cascadia, Nankai \& Hikurangi), along with the selected profiles used in this study. (Lower panel) Superimposed slab interface profiles using a depth coordinate system in which $y=0$ corresponds to the trench.
     The $x$ coordinate corresponds to the linear (horizontal) distance along each profile with the origin ($x=0$) taken at the blue dot (see upper panels).
     The labels $C_i$, $N_i$, $H_i$ correspond to one of three profiles taken along across the Cascadia, Nankai \& Hikurangi subduction zones respectively.}
     \label{fig:superimposed_profiles_inset}
\end{figure}

For the purposes of this study, we select three profiles per margin (nine total) to keep the analysis to a reasonable scope while still encompassing some along-margin variability in slab interface geometry.
The selected profiles are shown in Figure~\ref{fig:superimposed_profiles_inset} (see also Figure~\ref{fig:superimposed_profiles}).
The profile locations for Hikurangi are chosen to lie in the region of high slip deficit rate at the southern end of the North Island \cite{wallace_kaikoura_2018}.
Similarly, the profiles for Nankai are chosen to lie in the region of historic Nankai megathrust earthquakes \cite{sagiya_thatcher_1999}.
The profiles for Cascadia are chosen to span the margin and to exhibit the range of slab interface shapes across the subduction zone.
Generally, we observe from Figure~\ref{fig:superimposed_profiles_inset} that the Cascadia profiles are shallower dipping than the Hikurangi profiles, which in turn have shallower dip than the Nankai profiles.

\section{Forward Model Framework} \label{sec:forward_model}

\subsection{Governing Equations} \label{sec:gov_eqs}

We consider 2D $(x, y)$ steady-state and time-dependent kinematically-driven regional-scale models of subduction.
In these kinematically-driven models, the subducting slab motion is kinematically prescribed and we solve for the flow in the mantle wedge.
We assume that the mantle wedge flow is solely driven by slab motion, and we neglect the effect of thermal buoyancy, as in  \cite<e.g.>{van_keken_kiefer_peacock_2002,community_benchmark_van_keken_2008,syracuse_2010}.
Viscous coupling between the slab and the adjacent mantle material results in a downward flow in the wedge, which in turn causes hot mantle material to flow into the domain and towards the corner of the wedge.
The forward modeling approach we adopt is a standard approach commonly used to understand the present day thermal structure of subduction zones  \cite<e.g.>{van_keken_kiefer_peacock_2002, wada_et_al_2008, currie_et_al_2004,syracuse_2010,wada_wang_2009}.
In order to describe and identify the specifics of our parameterized subduction model,
below we describe the governing equations, mantle-wedge flow law and parameterized boundary conditions
we consider in our study.

\begin{figure}
     \centering
     \includegraphics[width=\textwidth]{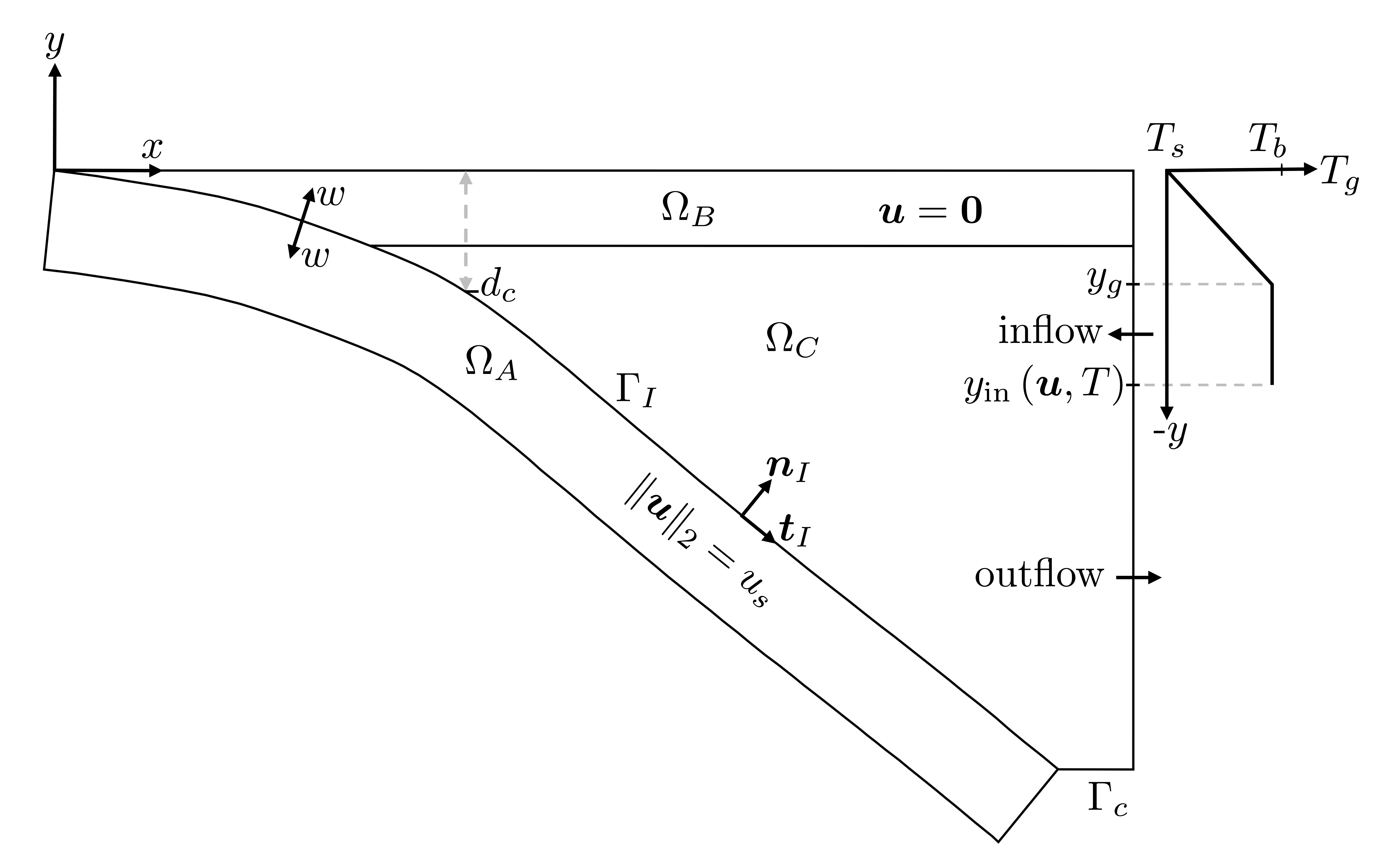}
     \caption{Schematic of the subduction model and boundary conditions.}
     \label{fig:schematic_ge}
\end{figure}

The model domain $\Omega$ is decomposed into three non-overlapping regions defining the slab $\Omega_A$, 
the over-riding plate $\Omega_B$ and the mantle wedge $\Omega_C$ (see Figure~\ref{fig:schematic_ge}).
The domain $\Omega$ has coordinates $(x,y)$ where $y = 0$ corresponds to the trench and $y$ is increasingly negative with depth.
We distinguish between the model depth coordinate $y$ and depth relative to Earth's reference ellipsoid, $z$, used later in this study when comparing to geophysical observations. 
The subduction interface $\Gamma_I$ is given by intersection of the slab boundary
with the boundary of the plate and wedge domains, i.e. $\Gamma_I = (\partial \Omega_A \cap \partial \Omega_B) \cup (\partial \Omega_A \cap \partial \Omega_C)$. We will denote by $\boldsymbol t_I$ the unit vector tangent to $\Gamma_I$ which is oriented down-dip, and $\boldsymbol n_I$ the unit vector normal to subduction interface, constructed such that $\boldsymbol t_I \times \boldsymbol n_I = \hat{\boldsymbol k}$.
We solve the conservation of momentum, mass and energy given by
\begin{linenomath*}
\begin{align}
    \nabla \cdot (2 \eta(\boldsymbol u, T) \,  \underline{\underline{ \dot{\boldsymbol{\epsilon}} }}(\boldsymbol u) ) - \nabla p
    	&= \boldsymbol 0 \quad \text{in } \Omega_C, \label{eqn:momentum} \\
    \nabla \cdot \boldsymbol{u}
    	&= 0 \quad \text{in } \Omega_C, \label{eqn:mass}  \\
    \rho(\boldsymbol x) \ c_p \left( \frac{\partial T}{\partial t} + \boldsymbol u \cdot \nabla T \right)
    	&= \nabla \cdot ( k(\boldsymbol x) \nabla T) + H_{\text{sh}}(\boldsymbol x) \quad \text{in } \Omega,
	\label{eqn:energy}
\end{align}
\end{linenomath*}
where $\boldsymbol u, p, T$ is the velocity, pressure and temperature, $\eta$ the shear viscosity,
$\rho, c_p$ the density and specific heat capacity, $k$ the thermal conductivity and $H_{\text{sh}}$
the volumetric heat production.

The shear viscosity considered is both temperature and strain-rate dependent and is given by
\begin{linenomath*}
\begin{equation}
    \frac{1}{\eta(\boldsymbol u, T)} = \frac{1}{\eta_{\text{diff}}(T) } + \frac{1}{\eta_{\text{disl}}(\boldsymbol u, T) } + \frac{1}{\eta_{\text{max}}},
    \label{eqn:viscosity}
\end{equation}
\end{linenomath*}
where the diffusion creep and dislocation creep mechanisms have shear viscosities given by
\begin{linenomath*}
\begin{align}
    \eta_{\text{diff}}(T) &= A_{\text{diff}} F_2 \exp \left(\frac{E_{\text{diff}}}{R T}\right), \label{eqn:diff_creep} \\
    \eta_{\text{disl}}(\boldsymbol u, T) &= A_{\text{disl}}^{(1/n)} F_2 \ \exp \left( \frac{E_{\text{disl}}}{nRT} \right) \dot{\epsilon}_\text{II}^{(1-n)/n}(\boldsymbol u), \label{eqn:disl_creep}
\end{align}
\end{linenomath*}
where $A$ is a pre-exponential factor, $E$ the activation energy, $n$ the strain-rate exponent,
$R$ the universal gas constant, and $\eta_\text{max} = 10^{26}$ Pa$\,$s is the upper bound on shear viscosity.
The factor $F_2$ accounts for the dependence of the relationship between $\dot{\epsilon}$ and $\dot{\epsilon}_\text{II}$ on the type of experiment performed (see Supporting Information Section~\ref{sec:visco_params}). 
The shear viscosity $\eta(\boldsymbol u, T)$ is also bounded below by $\eta_\text{min} = 10^{18}$ Pa$\,$s. 
In Equation~\eqref{eqn:disl_creep}, $\dot{\epsilon}_\text{II}(\boldsymbol u)$ denotes the second invariant of the strain-rate tensor given by
\begin{linenomath*}
\begin{equation*}
\dot{\epsilon}_\text{II}(\boldsymbol u) = \sqrt{ \frac{1}{2} \, \underline{\underline{ \dot{\boldsymbol{\epsilon}} }}(\boldsymbol u) : \underline{\underline{ \dot{\boldsymbol{\epsilon}} }}(\boldsymbol u) }
\end{equation*}
\end{linenomath*}
in which the deviatoric strain-rate tensor is
\begin{linenomath*}
\begin{align*}
   \underline{\underline{ \dot{\boldsymbol{\epsilon}} }}(\boldsymbol u)
    = \frac{1}{2} \left[ \nabla \boldsymbol u + \nabla \boldsymbol u^\top \right].
\end{align*}
\end{linenomath*}
The input parameters introduced that are related to the rheology are collected in a set $\boldsymbol q^\text{rheo}$ given by
\begin{linenomath}
\begin{equation} \label{eqn:q_rheo}
\boldsymbol q^\text{rheo} =
\left( A_\text{diff}, E_\text{diff}, A_\text{disl}, E_\text{disl}, n \right).
\end{equation}
\end{linenomath}

We assume that the over-riding plate is infinitely rigid and does not deform,
hence for all $\boldsymbol x \in \Omega_B$, we impose that $\boldsymbol u = \boldsymbol 0$.
The over-riding plate is 35 km thick in all experiments (a typical continental crust thickness, \cite<e.g.>{laske_et_al_2013}).
Flow within the slab domain is kinematically prescribed such that the plate behaves as a rigid body
with a speed (convergence velocity) defined by $u_s$.
Such a velocity field is characterized by a vector field in which
the component of velocity tangential to the subduction interface is equal to $u_s$
and the component of velocity normal to the subduction interface is zero, and thus
$\| \boldsymbol u \|_2 = u_s$.
Along $\Gamma_I$, down to the maximum depth of decoupling  $d_c$, we model
partial coupling between the slab and the plate-wedge domain by imposing that
$\boldsymbol u \cdot \boldsymbol t_I = \lambda u_s$, where
$0 \le \lambda \le 1$ and $\lambda$ is the degree of partial coupling.
At depths greater than $d_c$ on $\Gamma_I$, the slab is fully coupled to the mantle
and we impose that
$\boldsymbol u \cdot \boldsymbol t_I = u_s$.
Near the depth of decoupling, the transition from a partially coupled to
fully coupled interface is implemented using a $\tanh$ function (see Equation~\eqref{eqn:tanh})
with a transition length $L = 10$ km (see Figure~\ref{fig:plot_tanh_transition}). 
Along the boundary at the base of the wedge domain $\Gamma_c$, flow of material into the deep mantle is represented by the velocity boundary condition 
$\boldsymbol u = u_\text{base}$, where $u_\text{base}$ is the velocity at the deepest point of $\Gamma_I$ \cite{le_voci_2014}. 
Along the boundary on the right side of the wedge, we impose the Neumann constraint $ \underline{\underline{ \boldsymbol{\sigma}}} \boldsymbol n = 0$. 

The existence of decoupled and partially-coupled parts of the subduction interface implicitly defines the existence of a non-zero slip-rate,
that is $[[\boldsymbol u]]_{\Gamma_I} \ne 0$, where $[[ \cdot ]]_{\Gamma_I}$ denotes the jump operator along the interface $\Gamma_I$.
We assume that this slip-rate across $\Gamma_I$ results in shear heating along the subduction interface.
We consider shear heating to be equal to the product $\mu' [[\boldsymbol u]]_{\Gamma_I} P_{litho}$,
where $\mu'$ is the apparent coefficient of friction and  $P_{litho}=\rho g y$ is the lithostatic pressure.
Hence the source term in Equation~\eqref{eqn:energy} is given by
\begin{linenomath*}
\begin{equation}
    H_{\text{sh}} = 
     \begin{cases} 
    \mu' [[ \boldsymbol u]]_{\Gamma_I} P_{litho} & \text{for} \ \boldsymbol x  \in \Gamma_I\\
    0  & \text{for} \ \boldsymbol x  \notin \Gamma_I\\
    \end{cases}
     \label{eqn:H_sh}
\end{equation}
\end{linenomath*}
The shear heating source is thus only applied along $\Gamma_I$; see Supporting Information Section~\ref{sec:shear_heating} for implementation details. 

The input parameters introduced which relate to imposed (kinematic) subduction velocity are denoted by
\begin{linenomath}
\begin{equation} \label{eqn:q_kin}
\boldsymbol q^\text{kin} =
\left( u_s, d_c, \lambda, \mu' \right).
\end{equation}
\end{linenomath}

The thermal structure of the incoming (subducting) oceanic lithosphere is described using the error-function solution to the half-space cooling model
\begin{linenomath*}
\begin{align}
    T = T_s + (T_b - T_s) \ \text{erf} \left( \frac{ w }{2 \sqrt{\kappa_{\text{slab}} \ a_s}} \right),
\end{align}
\end{linenomath*}
where $T_s = 273$ K is the temperature at the surface, $T_b$ is the mantle inflow temperature,
$\kappa_{\text{slab}} = k_\text{slab}/(\rho_\text{slab} c_p)$ is the thermal diffusivity of the slab
and $a_s$ is the age of the oceanic lithosphere at the trench. To account for the angled boundary, the error-function solution to the half-space cooling model is a function of $w$, 
the distance normal to the slab interface, rather than a function of $y$; this simulates development of the thermal structure in the incoming slab with a ``depth" coordinate relative to the surface of the half-space.  
Dirichlet boundary conditions are set such that $T = T_s$ at the surface.
A linear geotherm $T_g(y)$ between temperatures $T_s$ and $T_b$ is prescribed on the right boundary of the domain, from the surface to a depth $y_g$.
Though the over-riding plate is assumed to be infinitely rigid to 35 km depth, $y_g$ can be varied and need not equal $-35$ km, 
so $T_g(y)$ can extend shallower or deeper than the rigid over-riding plate. 
Any segment of the right boundary where inflow occurs ($\boldsymbol u \cdot \boldsymbol n < 0$) we impose a Dirichlet constraint given by either $T = T_g(\boldsymbol x)$ if $y \ge y_g$, or $T = T_b$ if $y_\text{in} < y < y_g$. 
The inflow-outflow transition depth $y_\text{in}$ depends on $\boldsymbol u, T$ and changes as the nonlinear problem is iteratively solved. 
Outflow occurs ($\boldsymbol u \cdot \boldsymbol n \ge 0$) along all other segments of $\partial \Omega$.
Where outflow occurs we impose the Neumann constraint $(\boldsymbol u T + k \nabla T)\cdot \boldsymbol n = 0$.
The input parameters related to the boundary conditions of the thermal part of the subduction model are collected in the following set:
\begin{linenomath}
\begin{equation} \label{eqn:q_bc}
\boldsymbol q^\text{bc} =
\left(  T_b, a_s, y_g \right).
\end{equation}
\end{linenomath}

In this study we consider both time-dependent models and steady-state models in which we explicitly set
$\partial T / \partial t$ in Equation~\eqref{eqn:energy} to zero.
For the time-dependent calculations, the initial condition is given by
\begin{linenomath*}
\begin{equation}
    T = 
    \begin{cases} 
    T_s + (T_b - T_s) \frac{y}{y_g}, & \text{if} \ y \geq y_g \\
    T_b & \text{if} \ y < y_g
    \end{cases}
     \label{eqn:T_IC}
\end{equation}
\end{linenomath*}

The set of all input parameters to our subduction model that are varied in this study are denoted by the vector $\boldsymbol q$ given by
\begin{linenomath}
\begin{align}
\boldsymbol q &= \left( \boldsymbol q^\text{rheo}, \boldsymbol q^\text{kin}, \boldsymbol q^\text{bc} \right) \nonumber \\
&= \left( u_s, d_c, \lambda, \mu', A_\text{diff}, E_\text{diff}, A_\text{disl}, E_\text{disl}, n, T_b, a_s, y_g \right)
\label{eqn:param}
\end{align}
\end{linenomath}
with $\text{card}(\boldsymbol q) = 12$.
Furthermore, we assume that for a given input $\boldsymbol q$, Equations~\eqref{eqn:momentum}--\eqref{eqn:energy} admit
a unique solution for velocity, pressure and temperature such that the mapping $\mathcal F : \boldsymbol q \rightarrow (\boldsymbol u, p, T)$ is a
one-to-one function. We will denote by $\boldsymbol u(\boldsymbol q)$, $p(\boldsymbol q)$, $T(\boldsymbol q)$ the velocity, pressure and temperature
obtained with the input parameter $\boldsymbol q$.

\subsection{Parameter Intervals and Values} \label{sec:param_intervals}

The forward model takes input parameters, some of which are varied in this study ($q_i \in \boldsymbol q$), 
and the rest of which are held constant in all experiments ($q_k \in \mathbf q_\text{base}$). 
We associate with each parameter $q_i \in \boldsymbol q$ an interval, or range of possible values.
Depending on the parameter $q_i$, the interval may be attributed to measurement uncertainty,
observational uncertainty (for example due to sparse data coverage) or inference uncertainty (for example due to modeling assumptions).
For example, we define the range of possible values of the apparent coefficient of friction by the following closed interval:
$\{ \mu' \, | \, 0 \le \mu' \le 0.14 \}$.
Throughout the remainder of this study we will represent intervals via the notation $[\underline{\alpha}, \bar\alpha]$,
where $\underline{\alpha}, \bar\alpha$ denote the minimum and maximum for some quantity $\alpha$.
In Table~\ref{table:notation_and_ranges} and Table~\ref{table:margin_ranges} we identify the intervals for each parameter varied in this study.
Given the parameter-wise defined intervals, we denote by $\mathcal Q$ the space of all input parameters varied in this study.
$\mathcal Q$ is defined by taking the Cartesian product of the intervals associated with each parameter  $q_i \in \boldsymbol q$.

All other input parameters for the forward model, $q_k \in \mathbf q_\text{base}$, are held constant in all experiments.
The vector $\mathbf q_\text{base}$ is given by
\begin{linenomath}
\begin{align}
\mathbf q_\text{base} = \left(T_s, L, \rho_\text{mantle}, \rho_\text{crust}, \rho_\text{slab}, g, k_\text{mantle}, k_\text{crust}, k_\text{slab}, c_p, s \right). 
\label{eqn:q_base}
\end{align}
\end{linenomath}
Values for these parameters are given in Table~\ref{table:notation_and_ranges}.

\begin{table}[h!]
\begin{center}
\caption{Parameter intervals and the base values used when that parameter is held constant, for the set of all input parameters that are varied in this study, $\boldsymbol q$, and the set of input parameters held constant in all experiments, $\mathbf q_\text{base}$. See Supporting Information for citations and rationales for each parameter range (Table~\ref{table:interval_details}) and base values (Table~\ref{table:base_params_longtable}). $^*$Scaled for stress in Mpa, water fugacity in Mpa, and grain size in $\mu$m (See Supporting Information Section~\ref{sec:visco_params}).}
\label{table:notation_and_ranges}
\begin{tabular}{ c  l l l l l }
Set & Param.  & Definition & Unit &  Value & Range \\ \hline

\multirow{11}{*}{\rotatebox[origin=c]{0}{$\boldsymbol q$}} & $u_s$ & Plate convergence velocity & cm/yr & \multicolumn{2}{l}{Varies by margin, see Table~\ref{table:margin_ranges}. } \\

& $d_c$ & Max depth of decoupling & km  & 80 & [70, 80] \\

& $\lambda$ & Degree of partial coupling & -  & 0.05 & [0, 0.1] \\

& $\mu'$ & Apparent coefficient of friction & -  & 0.03 & [0, 0.14]\\

& $T_b$ & Mantle inflow temperature & K & 1573 & [1550, 1750] \\

& $a_s$ & Age of oceanic lithosphere at trench & Myr & \multicolumn{2}{l}{Varies by margin, see Table~\ref{table:margin_ranges}. } \\

& $y_g$ & Depth of linear geotherm & km  & 35 & [10, 60] \\ 

& $A_\text{diff}$ & Diffusion creep pre-exponential factor & Pa s$^*$ & $ 4.4  \times 10^{-10}$ & [4.4 $ \times 10^{-10}$, 1.3 $ \times 10^{1}$] \\

& $E_\text{diff}$ & Diffusion creep activation energy &  kJ/mol & 335 & [$260$, $450$]  \\

& $A_\text{disl}$ & Dislocation creep pre-exponential factor & Pa s$^*$ & 7.5 $\times 10^{-11}$ & [7.5 $\times 10^{-11}$, 7.7 $ \times 10^{-5}$] \\

& $E_\text{disl}$ & Dislocation creep activation energy & kJ/mol & 540 & [$480$, $560$] \\

& $n$ & Strain-rate exponent & -  & 3.5 & [0, 3.5] \\ \hline

\multirow{11}{*}{\rotatebox[origin=c]{0}{$ \mathbf q_\text{base}$ }} & $T_s$ & Temp at the surface & K & 273 & \\

& $L$ & Partial to full coupling transition length & km & 10 & \\

& $\rho_\text{mantle}$ & Density of the mantle & kg m$^{-3}$ & 3300 & \\

& $\rho_\text{crust}$ & Density of the crust & kg m$^{-3}$ & 2700 & \\

& $\rho_\text{slab}$ & Density of the slab & kg m$^{-3}$ & 3300 &\\

& $g$ & Gravitational acceleration & m s$^{-2}$ & 9.8  & \\

& $k_\text{mantle}$ & Thermal conductivity of mantle & W m$^{-1}$ K$^{-1}$ & 3.1 &  \\

& $k_\text{crust}$ & Thermal conductivity of crust & W m$^{-1}$ K$^{-1}$ & 2.5 & \\

& $k_\text{slab}$ & Thermal conductivity of slab & W m$^{-1}$ K$^{-1}$ & 3.1 &  \\

& $c_p$ & Specific heat capacity & J kg$^{-1}$ K$^{-1}$ & 1250 & \\

& $s$ & Standard deviation of Gaussian & km & 0.2  &\\

\hline
\end{tabular}
\end{center}
\end{table}

\begin{table}
\begin{center}
\caption{Margin-specific input parameter intervals and the base values used when that parameter is held constant. See Table~\ref{table:margin_ranges_longtable} for citations and rationales for each parameter range and base value.}
\label{table:margin_ranges}
\begin{tabular}{  l l l l l  }
Param. & Unit & Margin & Interval & Base value \\ \hline
$u_s$ & cm/yr & Cascadia & [3, 5] & 5 \\

& & Nankai & [3, 5] & 5 \\

& & Hikurangi & [3, 4] & 3 \\

$a_s$ & Myr & Cascadia & [8, 10] & 8 \\

& & Nankai & [15, 26] & 20 \\

& & Hikurangi & [90, 110] & 100 \\

\hline
\end{tabular}
\end{center}
\end{table}

\subsection{Quantities of Interest} \label{sec:qoi}
We are interested in several quantities of interest (QoI), which include:
(i) the solution to the energy equation $T(\boldsymbol q)$;
(ii) the location of the intersections between the slab interface and specific temperature isotherms which are used to infer
the updip rupture limit (150{\textcelsius}) and the downdip rupture limits (350{\textcelsius}, 450{\textcelsius}, 600{\textcelsius});
(iii) the along-slab distances between the updip and downdip limits.
We denote by $I_{150}, I_{350}, I_{450}, I_{600}$ the depth at which the 150{\textcelsius}, 350{\textcelsius}, 450{\textcelsius} and 600{\textcelsius} isotherms intersect the slab interface.
The coordinate of the intersection is denoted by $\boldsymbol I_\alpha = (I_{x, \alpha}, I_\alpha)$, where $\alpha =150, 350, 450, 600$.
If we refer to $I$ without a sub-script, we are implicitly referring to all isotherm depths.
We denote by $D_{350}, D_{450}, D_{600}$ the along-slab interface distance measured between the point $\boldsymbol I_{150}$
and $\boldsymbol I_\beta$, where $\beta =350, 450, 600$.
Again, if we refer to $D$ without a sub-script, we are implicitly referring to all along-slab distances considered.

\subsection{Numerical Implementation} \label{sec:impl}

The slab interface geometry is created by taking 2D vertical cross-sections across 3D surfaces generated from the Slab2 dataset \cite{slab2}.
The mesh is constructed such that the slab inflow boundary is approximately normal to the local slab interface, to facilitate imposing the free-slip boundary conditions.
The model domain is discretized using a locally-refined, unstructured triangular mesh created in GMSH version \verb|4.10.1| \cite{gmsh}.
The mesh has a resolution of 0.5 km near the slab interface and at most 8 km in the far-field.

The coupled equations defining the conservation of mass, conversation of momentum and conservation of energy are solved using the Finite Element Method via FEniCS version \verb|2019.1.0| \cite{fenics}.
Within FEniCS we employ unstructured triangular meshes to efficiently discretize the spatial domain.
The velocity, pressure and temperature are discretized over each triangular finite element using $\boldsymbol P_2$, $P_1$ and $P_2$ function spaces respectively.
The discrete velocity, pressure, and temperature solution vectors are referred to as $\widehat{\mathbf{u}}$, $\widehat{\mathbf{p}}$, and $\widehat{\mathbf{T}}$.
It should be noted that all the discrete solutions are functions of space and the input parameter vector $\mathbf q$, hence
$\widehat{\mathbf{u}} := \widehat{\mathbf{u}}(\mathbf x, \mathbf q)$, $\widehat{\mathbf{p}} := \widehat{\mathbf{p}}(\mathbf x, \mathbf q)$, and $\widehat{\mathbf{T}} := \widehat{\mathbf{T}}(\mathbf x, \mathbf q)$.

Equations~\eqref{eqn:momentum} and \eqref{eqn:mass} are first solved in the slab subdomain $\Omega_A$, with velocity boundary conditions imposed such that, (a) the velocity normal to the inflow and outflow slab boundaries is the slab convergence rate $u_s$, and (b) the curved upper and lower boundaries of the slab subdomain have free-slip boundary conditions. These constraints are weakly imposed using Nitsche's method \cite{venice95, sime_wilson, houston_sime}.

Both steady-state and time-dependent calculations are inherently non-linear due to the coupling of $\boldsymbol u, T$ between the
(i) conservation of momentum (Equation~\eqref{eqn:momentum}) and energy (Equation~\eqref{eqn:energy}) and
(ii) the inflow-outflow boundary condition location, $y_\text{in}$, associated with Equation~\eqref{eqn:energy}.
To solve the non-linear system of discrete PDEs we adopt the method of successive substitution.
For steady-state calculations and the first time step of time-dependent calculations we obtain an initial velocity, pressure and temperature field
by assuming an isoviscous mantle wedge rheology.
Given the initial $\widehat{\mathbf{T}}$, the nonlinear effective viscosity given by Equation~\eqref{eqn:viscosity} is evaluated.
The successive substitution solver proceeds by iterating between solving Equations~\eqref{eqn:momentum}, \eqref{eqn:mass} in the wedge subdomain and solving Equation~\eqref{eqn:energy} (or its steady-state form with $ \partial T / \partial t= 0$) in the entire domain, updating $\eta(\widehat{\mathbf{T}}, \widehat{\mathbf{u}})$.
The solution is considered converged when the difference between successive $\widehat{\mathbf{T}}$ solutions, measured in the $L_{\infty}$ norm, is less than $10^{-1}$ K.

Our code has been verified against analytical expressions for slab interface temperatures presented in \citeA{england_may_2021}; see \ref{sec:appendix_verif_analytical}. 
We also verify against a number of benchmark problems presented in \citeA{community_benchmark_van_keken_2008}; see \ref{sec:appendix_comm_benchmark}.

\section{Reduced-Order Modeling Framework} \label{sec:ROM_framework}

To assess the sensitivity of our parameterized forward model to changes in the input parameters, we require a computationally efficient and accurate way to obtain many instances of our forward model.
Even when using a 2D forward model, direct evaluation of a highly efficient and optimized implementation of the FE method described in Section~\ref{sec:impl} is unlikely computationally fast enough to rigorously explore the 12 dimensional parameter space.
We use model order reduction to build a surrogate model (a reduced-order model, or ROM) that accurately approximates full-order model (FOM) output and is extremely efficient to evaluate. 
Our framework employs a data-driven, non-intrusive approach based on the Proper Orthogonal Decomposition (POD) \cite{lumley}.
This method has the advantage that it does not require information \textit{a priori} about the governing equations of the FOM, nor does it require modification of the FOM software stack in order to construct or evaluate the ROM.
The building block of this ROM is discrete solutions (i.e. vectors) of a physical quantity (in our case temperature).
These characteristics render the ROM framework we present model-and-software agnostic, and therefore applicable to other areas of computational geophysics \cite<e.g.>{rekoske_2023}.

\subsection{POD with the Method of Snapshots} \label{sec:POD-ROM}

The POD is a data-driven model order reduction technique that determines the dominant modes of a dataset.
It finds the optimal basis for the data in the least-squares sense, i.e. with respect to the mean square error.
We construct a dataset from snapshots of the field variable, in this case temperature solutions $\widehat{\mathbf{T}} \in \mathbb R^n$, and seek to find the optimal basis for this dataset using POD.
The snapshots are obtained by solving the governing equations at a given parameter trajectory and storing the solution.
Let the $i$th snapshot be denoted $\widehat{\mathbf{T}}_i = \widehat{\mathbf{T}}(\mathbf q_i)$ for $i = 1,2,\ldots, N$ where $N$ is the total number of snapshots and $N \ll n$.
Furthermore, we denote the set of all parameter vectors via $\mathbb Q = \left( \mathbf q_1, \dots, \mathbf q_N \right)$.
We compose the snapshots into a data matrix $\mathbf{X}$, which has a snapshot in each column.
\begin{linenomath*}
\begin{align}
    \mathbf{X} = \begin{bmatrix}
    \vert & \vert  & & \vert \\
    \widehat{\mathbf{T}}_1, & \widehat{\mathbf{T}}_2, & \ldots, & \widehat{\mathbf{T}}_N \\
    \vert & \vert & & \vert
    \end{bmatrix} \in \mathbb{R}^{n \times N} .
\end{align}
\end{linenomath*}

POD solves the following minimization problem: for snapshots $\widehat{\mathbf{T}}_1, \widehat{\mathbf{T}}_2, \ldots, \widehat{\mathbf{T}}_N$, find a basis consisting of $r$ modes $\left\{ \mathbf{\varphi}_1, \ldots,\mathbf{\varphi}_r \right\}$ such that
\begin{linenomath*}
\begin{align}
    \min_{\mathbf{\varphi}_1, \ldots ,\mathbf{\varphi}_r} \sum_{j=1}^N \left| \left| \ \widehat{\mathbf{T}}_j - \sum_{i=1}^r \langle \widehat{\mathbf{T}}_j, \boldsymbol{\varphi}_i \rangle \boldsymbol \varphi_i \ \right| \right|_2^2
\end{align}
\end{linenomath*}
where $\langle \boldsymbol{\varphi}_i,\boldsymbol{\varphi}_j \rangle = \delta_{ij}$ and $\langle \cdot, \cdot \rangle_{\mathbb{R}^n}$ denotes the inner product in $\mathbb{R}^n$ \cite{volkwein}.
We obtain the POD basis by computing the singular value decomposition (SVD) of the data matrix, following the Method of Snapshots, presented by \citeA{sirovich}.
That is, we exploit the following decomposition
\begin{linenomath*}
\begin{align}
    \mathbf{X} = \mathbf U \mathbf{\Sigma} \mathbf{V}^\top
    	= \left( \mathbf u^{(1)}, \dots, \mathbf u^{(N)} \right) \text{diag}\left( \Sigma_1, \dots, \Sigma_N \right) \left( \mathbf v^{(1)}, \dots, \mathbf v^{(N)} \right)^\top
\end{align}
\end{linenomath*}
where $\mathbf{U} \in \mathbb R^{n \times N}$ is the matrix whose columns contain left singular vectors $\mathbf u^{(i)}$, $\mathbf{\Sigma} \in \mathbb R^{N \times N}$ is a diagonal matrix containing singular values, and $\mathbf V \in \mathbb R^{N \times N}$ is the matrix whose columns are the right singular vectors $\mathbf v^{(i)}$.
The left singular vectors $\mathbf u^{(i)}$ define the optimal basis and are thus taken as the POD basis.

We define the POD coefficients $\mathbf c(\mathbf q_i)$ for $i=1,\ldots,N$ by scaling the right singular vectors $\mathbf{V}$ by the singular values $\mathbf{\Sigma}$,
\begin{linenomath*}
\begin{align}
    \mathbf{C} = \mathbf{\Sigma} \mathbf{V}^\top =  \left( \mathbf c^{(1)}, \dots, \mathbf c^{(N)} \right),
\end{align}
where $\mathbf c^{(i)} = \mathbf c(\mathbf q_i)$.
\end{linenomath*}
These POD coefficients correspond to parameter trajectories $\mathbf q_i$ at which the solutions $\widehat{\mathbf{T}}_1, \widehat{\mathbf{T}}_2,\ldots,\widehat{\mathbf{T}}_N$ in the snapshot matrix were computed,
for example $\widehat{\mathbf{T}}_i = \mathbf U \mathbf c^{(i)}$.
It should be emphasized that the decomposition of $\mathbf X = \mathbf U \mathbf C$ results in a space-parameter modal
 decomposition with all spatial modes being contained in the columns of $\mathbf U$ and all input parameter
 modes contained in $\mathbf C$.

\subsection{Approximating via interpolation}

We wish to approximate the discrete temperature $\widehat{\mathbf{T}}$ at parameter trajectories $\widehat{\mathbf q}$ that are not in the set $\mathbb Q$.
To do this, we interpolate the POD coefficients $\mathbf C$ \cite<e.g.>{ly_tran_2001, bui-thanh_2003, my-ha_2007}.
In this work, we interpolate the coefficients in $\mathbf C$ using radial basis function (RBF) interpolation \cite{walton}.
RBF interpolation represent a class of scattered data interpolants which can be applied to data of any dimension.
The essential feature of RBF interpolants is that they are constructed in terms of
the Euclidian distance between the input scattered data \cite{fasshauer_2007}.
For example, given some function $F(\mathbf x)$ evaluated at $N$ points $\mathbf x_i$ where $i=1,\ldots, N$, the RBF approximation to $F$ is
\begin{linenomath*}
\begin{align}
    F(\mathbf x) \approx F^{RBF}(\mathbf x) = \sum_{i=1}^{N} w_i \Theta (\| \mathbf x - \mathbf x_i \|_2),
\end{align}
\end{linenomath*}
where $w_i$ are the interpolation coefficients, $\Theta(\cdot)$
is an interpolation kernel and $||\cdot||_2$ denotes the Euclidean distance.
In this study we use the quintic (polyharmonic) spline, defined as
\begin{linenomath*}
\begin{align}
    \Theta (r) = -r^5, \qquad \text{where } r = \| \mathbf x - \mathbf x_i \|_2.
\end{align}
\end{linenomath*}
We note that in order to ensure that the RBF interpolant when using the quintic spline is unique,
the interpolant is augmented with a complete polynomial basis of degree 2 for a space of dimension $\text{card}(\mathbf x)$ \cite{fasshauer_2007}.

Each row $i$ of $\mathbf C$ represents the parameter dependent weights for the $i^\text{th}$ POD basis,
hence in practice we fit an RBF through each of the $N$ rows of $\mathbf C$.
That is, we construct $c^{RBF}_i(\mathbf q)$ for $i=1, \dots, N$ using $c_i^{(1)}, \dots, c_i^{(N)}$.
Given the interpolated values for the POD coefficients the approximate temperature field is obtained by
multiplication with the spatial POD basis
\begin{linenomath*}
\begin{equation}
\widehat{\mathbf T}^{RBF}({\mathbf q}) = \mathbf U \left( c^{RBF}_1({\mathbf q}), c^{RBF}_2({\mathbf q}), \dots, c^{RBF}_N({\mathbf q}) \right)^\top.
\end{equation}
\end{linenomath*}

We quantify the approximation error of the reduced order model using a relative error measure ($E_{rel}$) and
an absolute error measure ($E_{\infty}$).
Given an arbitrary set of input parameters $\mathbf q^*_k$, $k=1, \dots, m$ 
let $\mathbb T = [\mathbf T( \mathbf q^*_1), \dots, \mathbf T(\mathbf q^*_m) ]$ 
be the collection of temperatures vectors obtained from the forward model $\mathcal F$
and let $\mathbb T_\text{ROM} = [\mathbf T^{RBF}( \mathbf q^*_1), \dots, \mathbf T^{RBF}(\mathbf q^*_m) ]$ 
be the set of approximate temperature vectors obtained from the RBF interpolation.
Given this, the error metrics are defined as
\begin{linenomath}
\begin{align}
E_{rel} =  \frac{\Vert \mathbb{T} - \mathbb{T}_\text{ROM}\Vert_F }{\Vert \mathbb{T}\Vert_F}, 
\label{eqn:E_rel}
\end{align}
\end{linenomath}
where the norm $\Vert \cdot \Vert_F$ is the Frobenius norm, defined as $\Vert \mathbf{X} \Vert_F = \sum_i \sum_j \vert \mathbf{X}_{ij} \vert ^2$
and
\begin{linenomath}
\begin{align}
E_{\infty} =  \sup \left| \mathbb{T} - \mathbb{T}_{ROM} \right|. 
\label{eqn:E_inf}
\end{align}
\end{linenomath}

\subsection{ROM Construction} \label{sec:construct} 

A crucial aspect of constructing the interpolated POD reduced order model is the method used to generate the input parameter
trajectories identified in Section~\ref{sec:POD-ROM} as $\mathbb Q$.
The issues to consider are two-fold: (i) How should one sample the parameter space $\boldsymbol q$?; and (ii) What value of $N$ (the number of forward models) should
be selected?
In this study we adopted a divide-and-conquer strategy driven by a desire to minimize the ROM approximation error 
of temperature and temperature dependent quantities of interest defined in Section~\ref{sec:qoi}.
Since the ROM we utilize is interpolatory in nature, we construct it to be accurate for cases when $\boldsymbol q \in \mathcal Q$.
We do not consider (during either the construction or evaluation phase) the extrapolation case, $\boldsymbol q \notin \mathcal Q$.

\begin{algorithm}
\caption{Error-driven ROM sampling strategy}\label{alg:rombuild}
\begin{algorithmic}
\Require  $\hat{\mathcal Q}$, where $\hat{\mathcal Q} \subset \mathcal Q$, $N_\text{inc}$, $N_\text{test}$, $\mathbb Z_\text{seed}$.
\State  ${\mathbb Q}_\text{all} = \emptyset$.
\State ${\mathbf X}_\text{all} = \emptyset$
\State $\mathcal H \gets \mathcal H(\mathbb Z_\text{seed})$ 	
		\Comment{Initiate a Halton sequence with seed $\mathbb Z_\text{seed}$.}
\State \texttt{converged} $\gets$ \texttt{False}
\While{\texttt{converged} $\ne$ \texttt{True}}
	\For{$i=1, \dots, N_\text{inc}$}
		\State $\hat{\mathbf q}^i \gets \mathcal H$ 
			\Comment{Draw vectors $\hat{\mathbf q}^i \in \hat{\mathcal Q}$ from the sampler $\mathcal H$.}
		\State $\mathbf q^i \gets \hat{\mathbf q}^i \cup ( \mathbf q_\text{base}  \setminus \hat{\mathbf q}^i  )$
			\Comment{Augment each $\hat{\mathbf q}^i$ with base values (see Table~\ref{table:notation_and_ranges}).}
	\EndFor
	
	\State ${\mathbb Q}^* \gets [  \mathbf q^1, \dots, \mathbf q^{N_\text{inc}} ]$ \Comment{Stack the parameter vectors.}

	\State $\mathbf X^* \gets \mathcal F(\mathbb Q^*)$ 
		\Comment{Generate data matrix by evaluating the forward model $\mathcal F$.}
	\State ${\mathbb Q}_\text{all} \gets {\mathbb Q}_\text{train} \cup \mathbb Q^*$.
	\State ${\mathbf X}_\text{all} \gets {\mathbf X}_\text{train} \cup \mathbf X^*$.
	
	\State $( 
			{\mathbb Q}_\text{train}, {\mathbf X}_\text{train}, 
			{\mathbb Q}_\text{test}, {\mathbf X}_\text{test}
			) \gets \text{split}({{\mathbb Q}_\text{all}, \mathbf X}_\text{all}, N_\text{test})$
		\Comment{Split the parameters and data into a test and training set.}
	
	\State $\mathcal R \gets \mathcal R^{\hat{\mathcal Q}}[\mathbb Q_\text{train}, {\mathbf X}_\text{train}]$  \Comment{Build the ROM.}

	\State \texttt{converged} $\gets$ EvaluateError$(\mathcal R, {\mathbb Q}_\text{test}, {\mathbf X}_\text{test})$

\EndWhile
\end{algorithmic}
\end{algorithm}

The sampling procedure we adopt is
\begin{enumerate}
\item Partition the entire parameter space $\mathcal Q$ into three sub-spaces, $\mathcal Q^\text{rheo, kin, bc}$,
where each sub-space is constructed from the Cartesian product of all $q_i \in \boldsymbol q^k$, $k = \{\text{rheo, kin, bc}\}$ and the intervals given in Table~\ref{table:notation_and_ranges} and Table~\ref{table:margin_ranges}.
\item For each $\mathcal Q^k$ sub-space, we apply Algorithm~\ref{alg:rombuild}, resulting in the construction of a ROM ($\mathcal R^{\mathcal Q^k}$) accurate over the sub-space.
We use values of $N_{inc} = 600$, $N_{test} = 200$.  
The function $\text{split}({{\mathbb Q}_\text{all}, \mathbf X}_\text{all}, N_\text{test})$ in Algorithm~\ref{alg:rombuild} takes 
the last $N_{test} = 200$ parameter and data vectors from ${\mathbb Q}_\text{all}$ and ${\mathbf X}_\text{all}$
and adds them to ${\mathbb Q}_\text{test}$ and ${\mathbf X}_\text{test}$,
while the remainder are added to ${\mathbb Q}_\text{train}$ and ${\mathbf X}_\text{train}$. 
The function EvaluateError$(\mathcal R, {\mathbb Q}_\text{test}, {\mathbf X}_\text{test})$ evaluates the ROM $\mathcal R$ on the test parameter vectors ${\mathbb Q}_\text{test}$ and directly compares to ${\mathbf X}_\text{test}$, the forward model output at the same parameter vectors. It checks four conditions: 
(i) whether the absolute error $E_{\inf}$ (Equation~\eqref{eqn:E_inf})) on the slab interface is accurate to $<20$ K,
(ii) whether the relative error $E_{rel}$ (Equation~\eqref{eqn:E_rel})) on the slab interface is accurate to $<10^{-3}$ K, and
(iii) whether the along-slab distance to where each isotherm (150{\textcelsius}, 350{\textcelsius}, 450{\textcelsius}, 600{\textcelsius}) intersects the slab interface varies by $<1$ km, and
(iv) whether the along-slab distance $D$ between isotherm intersections varies by $<1$ km.
If all four of these conditions are met, \texttt{converged}~$\gets$~\texttt{True}.
Otherwise \texttt{converged} remains \texttt{False} and the while loop continues. 

\item For each sub-space ROM, we perform Morris screening (see Section~\ref{sec:screening}) to identify the most important set of parameters within each sub-space.
We will denote these as $\boldsymbol q^k_\text{red}$.
\item Collect all important parameters identified from within each sub-space into a new parameter vector $\boldsymbol q_\text{red}$, where 
$\boldsymbol q_\text{red} = (\boldsymbol q^\text{rheo}_\text{red}, \boldsymbol q^\text{kin}_\text{red}, \boldsymbol q^\text{bc}_\text{red})$.
\item Create $\mathcal Q_\text{red}$, then apply Algorithm~\ref{alg:rombuild} over $\mathcal Q_\text{red}$, resulting in the final reduced order model $\mathcal R^{\mathcal Q_\text{red}}$.
We again initialize Algorithm~\ref{alg:rombuild} with values of $N_{inc} = 600$, $N_{test} = 200$. 
The functions $\text{split}({{\mathbb Q}_\text{all}, \mathbf X}_\text{all}, N_\text{test})$ 
and EvaluateError$(\mathcal R, {\mathbb Q}_\text{test}, {\mathbf X}_\text{test})$ behave as in step 2, with the same conditions for convergence.  
\end{enumerate}

\section{Sensitivity Analysis} \label{sec:SA}

Sensitivity analysis is the study of how variation in the output (e.g. quantities of interest -- see Section~\ref{sec:qoi}) of a model can be apportioned to model input parameters (see Equation~\eqref{eqn:param}) \cite{saltelli_book_2000}.
We use sensitivity analysis to determine (a) which parameters have more influence in determining the variability of the quantity of interest (see Section~\ref{sec:construct}), and (b) whether there are input parameters that affect the quantity of interest very little, to the extent that they can be held constant at a specific value.

\subsection{Parameter screening via the method of Morris} \label{sec:screening}

The aim of parameter screening is to identify which of a model's many input parameters can be held constant at a value with little effect on the quantity of interest.
The working assumption when using screening methods is that the number of important input parameters is small compared to the total number of input parameters.
We use the method of Morris, which is a one-at-a-time (OAT) design;
it evaluates the impact of changing the value of each input in turn \cite{morris_1991,saltelli_book_2000}.
It is also a global sensitivity experiment, covering the entire parameter range rather than just local variations around a nominal value.
The method is based on computing the \textit{elementary effect} of a parameter $i$ at a point $\mathbf{q} = (q_1,q_2, \ldots, q_k)$ in parameter space according to
\begin{linenomath*}
\begin{align}
    d_i(\mathbf{q}) = \frac{\text{QoI} (\mathbf{q} + \mathbf{e}_i \delta) - \text{QoI}(\mathbf{q})  }{\delta}.
\end{align}
\end{linenomath*}
where $\mathbf{q} + \mathbf{e}_i \delta = (q_1,\ldots,q_{i-1}, q_i + \delta, q_{i+1}, \ldots, q_k)$ and $\delta$
is small compared to the interval associated with the input parameter $i$.
By sampling the parameter space at many different values of $\mathbf{q}$,
we obtain a finite distribution of elementary effects associated with each input parameter.
From this finite distribution we compute the following sensitivity measures:
the standard deviation of the distribution ($\sigma$);
the mean of the distribution of absolute values of elementary effects $(\mu^*)$
which accounts for the potential effects of non-monotonic relationships between points in parameter space
and the QoI \cite{morris_1991, campolongo_2007}.
A high value of $\mu^*$ indicates an input parameter with a high overall influence on the QoI.
Meanwhile, a high value of $\sigma$ indicates that the influence of the input parameter on the QoI 
is strongly dependent on where the chosen value $\mathbf{q}$ lies in parameter space. 

As mentioned in Section~\ref{sec:construct} we use Morris screening to identify a reduced set of parameters for each of the sub-vectors
$\boldsymbol q^\text{rheo}, \boldsymbol q^\text{kin}, \boldsymbol q^\text{bc}$.
Since the ROMs are highly efficient to evaluate, the Morris screening is conducted (for each of the three parameter sets)
using $10^4$ parameter trajectories defined using a Latin Hypercube Sampling (LHS) design.
We first conduct Morris screening using the along-slab distances $D$ between isotherms as the QoIs.
For cases in which the screening results are unclear for a given parameter with index $k$,
we compute the maximum variation in $D$ as parameter $k$ varies.
If the variation in $D$ is less than 1 km, we consider it reasonable to hold that parameter constant at a given value.
In general, we err on the side of caution and retain parameters (i.e. we do not screen them out)
unless the screening results clearly indicate they can be held constant without notably affecting $D$.

\subsection{Parameter correlations} \label{sec:scatterplots}

To examine the relationships between input parameters and quantities of interest, we perform a simple regression analysis.
First we evaluate the ROM defined over $\mathcal Q_\text{red}$ at $O(10^5)$ samples in parameter space and compute the QoI: $D_{350}$, $D_{450}$, and $D_{600}$.
We then perform a curvilinear regression analysis where polynomial curves are fitted to the scattered points via the least-squares method.
For each polynomial regression, we complete the analysis of variance (ANOVA) and calculate the F-test at a 1\% level of significance to verify that the regression is appropriate.
Specifically, if the F-test statistic is above the critical value for given polynomial degree and degrees of freedom, we reject the null hypothesis that the variance about the regression curve is no different than the variance in the data.

\section{Results} \label{sec:results}

\subsection{Forward Model and Reduced Order Model Performance}
The task of running forward models for the parameter screening and for the final ROM
is part of what is referred to as the ``offline'' phase of the reduced order model.
One evaluation of the FOM takes about 44 minutes ($O(10^3)$ seconds) when run on one core of a machine 
with two AMD EPYC 7552 48-core processors (96 cores total) and 256 GB of system memory. 
For a typical required number of FOM evaluations of $N_\text{inc} = 1800$, 
the offline cost associated with each profile is $4.752 \times 10^6$ seconds.
The offline cost is expensive, and is partially offset by the fact that the compute resource used 
allows us to execute 92 parallel instances of the FOM simultaneously in order to generate the snapshot data.
Hence the walltime for the offline phase is approximately $4.752 \times 10^6$ / 92 = 51652 seconds, or $\sim$14.4 hours.
The time required to perform the SVD and build the RBF interpolant is also considered part of the offline phase.
For the numerical experiments we conducted, the typical size of the data matrix is $230000 \times 1600$.
The data matrix can be factorized on a personal laptop (MacBook Pro 2020 with a 2 GHz Quad-Core Intel Core i5 processor and 16 GB memory) 
using the SVD method from Numpy in $\sim$118 seconds.
We use the class \texttt{RBFInterpolator} from SciPy. For $N = 6$ the construction of the RBF
requires approximately 0.4 seconds.

The offline phase is expensive, however it is only required to be performed once, 
and thus its cost can be amortized over many ROM evaluations.
For example, evaluating the ROM $10^3$ times using a personal laptop takes $\sim 8$ seconds, or $\sim 0.01$ seconds per evaluation. 
Evaluating the ROM $10^4$ times takes $\sim 372$ seconds, or $\sim 0.03$ seconds per evaluation. 
This speedup factor of $10^4$--$10^5$ makes it possible to use global sensitivity analysis and uncertainty quantification techniques that require $O(10^5)$ or $O(10^6)$ evaluations of the ROM per analysis, which would be prohibitive if only using the FOM.

\subsection{Parameter Sensitivities}

We perform a sensitivity analysis for each of the nine profiles in this study, as described in Section~\ref{sec:screening}.
Within each subduction zone, the results of the screening analysis are consistent across the three profiles.
The screening results are summarized in Table~\ref{table:screening_results}.
For the kinematic parameter group $\boldsymbol q^\text{kin}$ (see Equation~\eqref{eqn:q_kin}),
we find that across all profiles, variations in the apparent coefficient of friction $\mu'$ and slab velocity $u_s$ do affect the estimated values of $D$.
Generally, the depth of decoupling $d_c$ and the degree of partial coupling $\lambda$ affect $D$ less than the other kinematic parameters, but they create variation in $D$ higher than the chosen 1 km threshold and therefore are retained in the screened sets.
The rheological parameters are screened out in all cases, since none of them cause sufficient variation in the estimated $D$.
To corroborate this result, we examine plots of the variation in isotherms and see that changes in rheological parameters significantly affect the thermal structure of the mantle wedge, but do not affect temperatures along the slab interface.
For the boundary condition parameters, the mantle inflow temperature $T_b$ and the slab age $a_s$ affect $D$ and are retained.
Their importance makes sense considering they determine the thermal structure of the incoming slab.
Meanwhile, the depth of the geotherm on the right boundary, $y_g$, can be screened out in all cases, indicating that the far-field boundary condition does not strongly affect temperatures along the slab interface for the steady-state models.

In summary, our screening analysis yielded the following reduced input parameter vectors:
\begin{linenomath}
\begin{align}
\boldsymbol q^\text{Cascadia}_\text{red} = \boldsymbol q^\text{Nankai}_\text{red} = \boldsymbol q^\text{Hikurangi}_\text{red} = \left(  u_s, d_c, \lambda, \mu', T_b, a_s   \right). 
\end{align}
\end{linenomath}

\begin{table}
\begin{center}
\caption{Parameter screening results for each margin, with parameters grouped as they were in the screening analysis. Checkmarks indicate a parameter was retained in the screened set, while crosses indicate a parameter was screened out. }
\label{table:screening_results}
\begin{tabular}{  c  l  l  c  c  c   }
~ & Param. &Description  & Cascadia & Nankai & Hikurangi  \\
\hline
  \multirow{4}{*}{\rotatebox[origin=c]{90}{$\boldsymbol q^\text{kin}$}} & $u_s$ & Slab velocity & \checkmark & \checkmark &  \checkmark \\
	 & $d_c$ & Depth of decoupling & \checkmark &  \checkmark &  \checkmark \\
	 & $\lambda$ & Degree of partial coupling &  \checkmark &  \checkmark &  \checkmark \\
	 & $\mu'$ & Apparent coefficient of friction &  \checkmark &  \checkmark &  \checkmark \\ \hline
  \multirow{5}{*}{\rotatebox[origin=c]{90}{$\boldsymbol q^\text{rheo}$}} & $A_\text{diff}$ & Diffusion creep pre-exp factor & \xmark & \xmark & \xmark \\
	 & $E_\text{diff}$ & Diffusion creep activation energy & \xmark & \xmark & \xmark \\
	 & $A_\text{disl}$ & Dislocation creep pre-exp factor & \xmark & \xmark & \xmark \\
	 & $E_\text{disl}$ & Dislocation creep activation energy & \xmark & \xmark & \xmark \\
	 & $n$ & Power law exponent & \xmark & \xmark & \xmark\\ \hline
  \multirow{3}{*}{\rotatebox[origin=c]{90}{$\boldsymbol q^\text{bc}$}} & $T_b$ & Mantle inflow temperature &  \checkmark &  \checkmark &  \checkmark \\
	 & $a_s$ & Slab age &  \checkmark &  \checkmark &  \checkmark \\
	 & $y_g$ & Depth of geotherm & \xmark & \xmark & \xmark \\ \hline
\end{tabular}
\end{center}
\end{table}

\subsection{Variability in Rupture Limits} \label{sec:variability_in_rupture_limits}

\begin{figure}
     \centering
     \includegraphics[width=\textwidth]{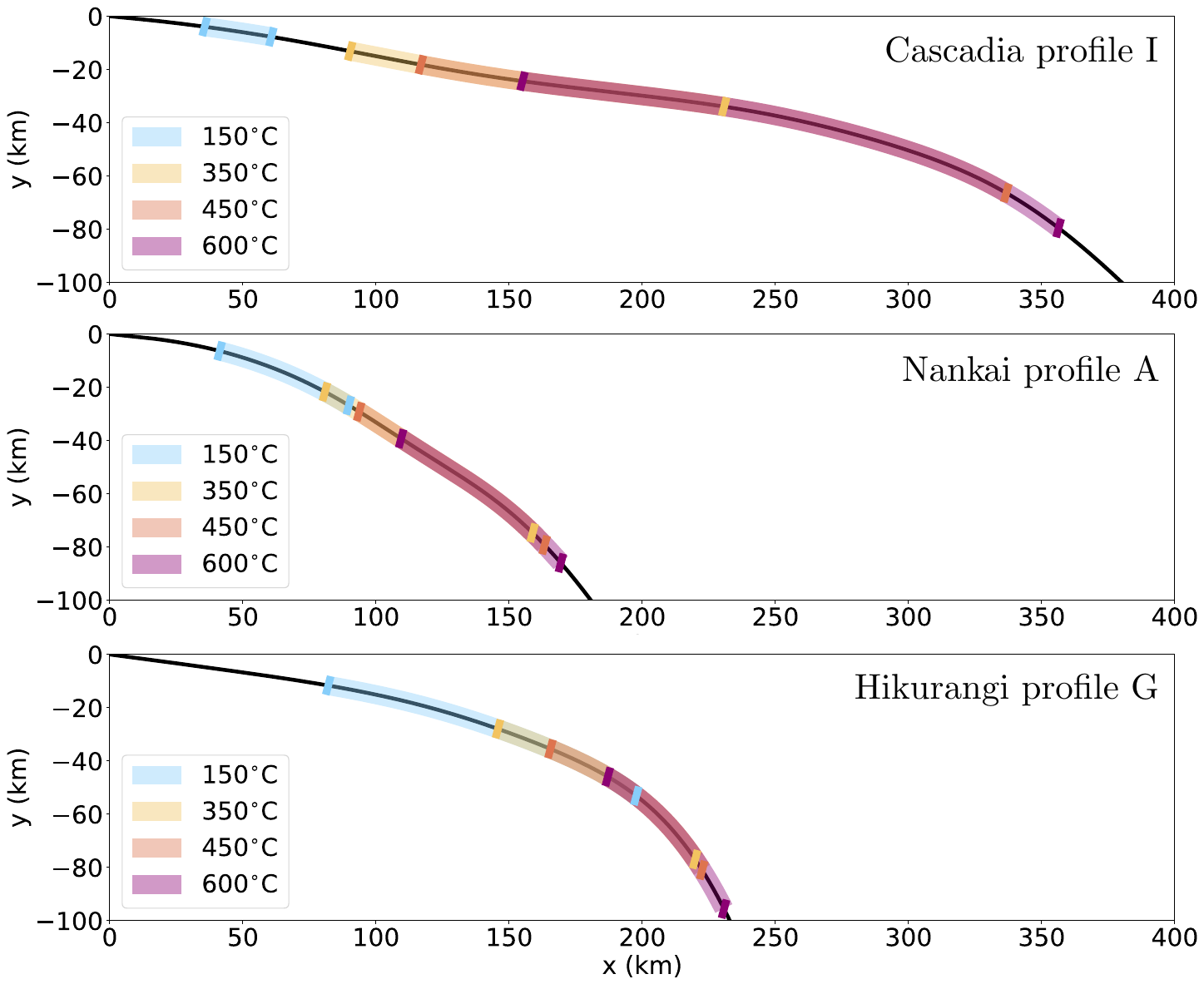}
     \caption{Variability in the locations where the 150{\textcelsius} (blue), 350{\textcelsius} (orange), 450{\textcelsius} (red) and 600{\textcelsius} (purple) isotherms intersect the slab interface, given variability in the screened sets of parameters. Cascadia profile I has the shallowest dip of the Cascadia profiles, while Nankai profile A has the steepest dip of the Nankai profiles. Hikurangi profile G is representative of the three Hikurangi profiles which have similar results for this analysis.}
     \label{fig:isotherm_intersection_uncertainty}
\end{figure}

In Figure~\ref{fig:isotherm_intersection_uncertainty} we illustrate the variability
in location where isotherms intersect the slab interface for one profile in each subduction zone.
This comparison illustrates some of the differences between the three margins.
For example, the Cascadia profile has on average a shallower dip than the other two subduction zones,
resulting in larger variability in where the isotherms intersect because more of the slab interface remains in the shallower depth range where these temperatures occur.
Within the Cascadia margin, there is a pattern of higher $D$ variability for the shallower dipping slab interfaces (see Table~\ref{table:D_var_by_profile}). 
Taking $D_{450}$ as an example, the variability is 124 km for Cascadia profile B, while the shallower dipping Cascadia profile G has a 177 km variability and the shallowest profile, Cascadia profile I, has a 204 km variability.
This indicates that profiles with shallow initial dip transitioning to steeper dip may be more sensitive to variability in input parameters than profiles that dip steeply closer to the trench.

Figure~\ref{fig:isotherm_intersection_uncertainty} also shows differences in the variability of the 150{\textcelsius} isotherm intersection, with low variability for Cascadia, higher for Nankai, and highest for Hikurangi. 
We interpret the low sensitivity of the Cascadia 150{\textcelsius} isotherm to the fact that the Cascadia slab is a hot end-member and therefore all of the isotherms are shallower; adding shear heating does not have as marked an effect of shallowing the isotherm, relative to the other, cooler slabs. 

In all three margins, we see some overlap in the variability ranges for isotherm intersections. 
Figure~\ref{fig:isotherm_intersection_uncertainty} shows that the variability ranges for the 350{\textcelsius}, 450{\textcelsius}, and 600{\textcelsius} 
isotherm intersections overlap for all three margins,
and the 150{\textcelsius} isotherm overlaps the others slightly in Nankai and significantly in Hikurangi. 
This overlap is of interest because it implies that for a given $D$ in the overlap, there exist points in parameter space that can result in that $D$ estimate 
whether we choose 350{\textcelsius}, 450{\textcelsius}, or 600{\textcelsius} as the downdip limit. 
Considering there is some uncertainty in which isotherm represents the downdip rupture limit, it is interesting to highlight the non-uniqueness of a given $D$ estimate caused by variability in the key parameters.

\begin{table}[h!]
\begin{center}
\caption{The variability in the screened input parameters corresponds to variability in $D$ presented in this table.}
\label{table:D_var_by_profile}
\begin{tabular}{ l  l  l  l  l  l  }
Margin | Profile & $D_{350}$  & $D_{450}$ &  $D_{600}$  \\
 \hline
 Cascadia | Profile B & $ [54, 152] $ & $[79, 203]$ &  $[111, 227]$ \\
 Cascadia | Profile G & $[61, 180]$ & $[91, 268]$ & $[134, 297]$  \\
 Cascadia | Profile I & $[54, 172]$ & $[80, 284]$ & $[119, 317]$   \\
 \hline
 Nankai | Profile C & $ [52, 113]$ & $ [69, 132]$ & $ [88, 148] $ \\
 Nankai | Profile A & $ [40, 104] $ & $ [54, 121] $ & $ [72, 136] $  \\
 Nankai | Profile D & $ [54, 126] $ & $ [71, 147] $ & $ [92, 163] $\\
 \hline
 Hikurangi | Profile F & $[27, 94] $ & $ [31, 112] $ & $ [45, 128] $ \\
 Hikurangi | Profile G & $[28, 93] $ & $ [32, 111] $ & $ [45, 130] $  \\
 Hikurangi | Profile H &  $[30, 91] $ & $ [34, 110] $ & $ [49, 129] $  \\
 \hline
\end{tabular}
\end{center}
\end{table}

\subsection{Regression Analysis}

Using the reduced-order models, here we examine the relationships
between input parameters $\boldsymbol q_\text{red}$ and the QoI $D$
using regression analysis (Section~\ref{sec:scatterplots}).
We perform a polynomial regression and find that a degree 4 polynomial fit
 of $D$ against $\mu'$ gives high $R^2$ values,
indicating this polynomial is a good estimator for the data. 
All other parameters are fitted with a degree 1 (linear) polynomial,
and the $R^2$ values are approximately zero.
This indicates that the range of $\mu'$ values generate such variability in $D$
that there is no apparent relationship between $D$ and any other parameter.

\begin{figure}
     \centering
     \includegraphics[width=\textwidth]{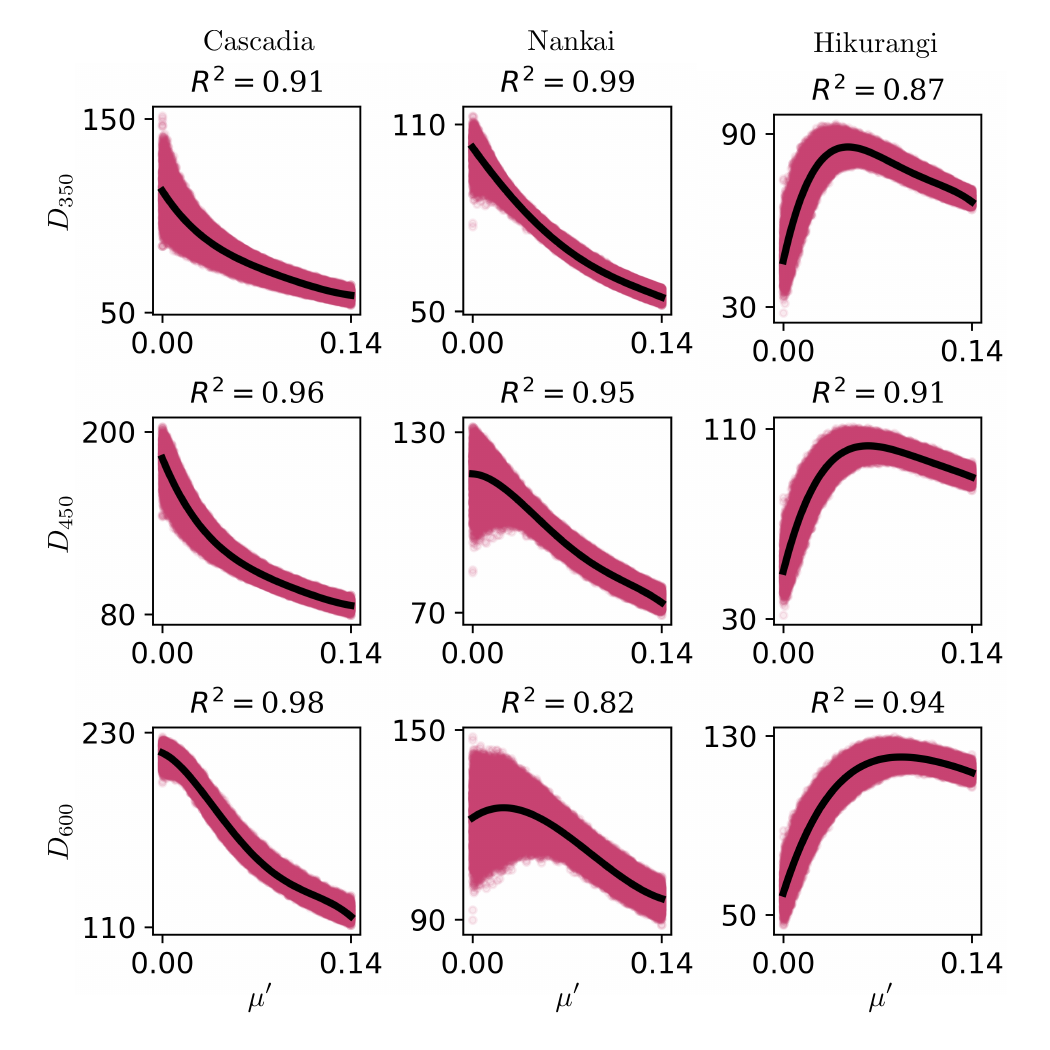}
     \caption{The relationship between $\mu'$ and $D$ for Cascadia profile B (left column), 
     Nankai profile C (centre column), and Hikurangi profile G (right column). 
     Each scatterplot shows $D$ values for $10^5$ samples in $\boldsymbol q_\text{red} $. 
     The thick line shows the fitted 4th order polynomial.}
     \label{fig:D_vs_mu_three_SZs}
\end{figure}

A selection of representative scatterplots and polynomial fits is shown in Figure~\ref{fig:D_vs_mu_three_SZs}, with $R^2$ values for all nine profiles reported in Table~\ref{table:param_R^2}. For the three Cascadia profiles, $\mu'$ is negatively related to $D$, with low $\mu'$ values corresponding to higher $D$ values.
For the three Nankai profiles, we find that $\mu'$ is more correlated with $D_{350}$ than $D_{450}$, and more correlated with $D_{450}$ than $D_{600}$.
The relationships between $\mu'$ and $D_{600}$ for Nankai profiles have comparatively low $R^2$ values, between 0.82 and 0.9.
This can be seen visually in the bottom center subplot of Figure~\ref{fig:D_vs_mu_three_SZs}, where there is more scatter around the fitted line than for $D_{350}$ and $D_{450}$.
We attribute this to a comparatively stronger influence from other parameters on the $D_{600}$ isotherm. 
Figure~\ref{fig:D_vs_mu_three_SZs} also shows that, for Nankai, the relationship between $\mu'$ and $D$ is generally negative but non-monotonic.
For $D_{450}$ and $D_{600}$ in particular, there is a slight increase in $D$ as $\mu'$ increases, followed by a marked decrease.
This non-monotonicity happens because the 150{\textcelsius}, 450{\textcelsius} and 600{\textcelsius} isotherms intersections get shallower as $\mu'$ increases, but they do not change at the same rate in different parts of the $\mu'$ parameter range.
For example, when $\mu'$ increases through the low end of its parameter interval, the 150{\textcelsius} isotherm gets shallower faster than the 600{\textcelsius} isotherm; this causes $D_{600}$ to briefly increase before decreasing again as $\mu'$ increases further.

The relationship between $\mu'$ and $D$ looks very different for Hikurangi: it is non-monotonic and generally positive. $D_{350}$ and $D_{450}$ increase as $\mu'$ increases initially, peak at intermediate $\mu'$ values, and decrease as $\mu'$ approaches the upper end of its range. The $D_{600}$ vs $\mu'$ relationship is closer to monotonic, with $D$ increasing as $\mu'$ increases but still slightly decreasing for high $\mu'$ values. The non-monotonicity comes from the 350{\textcelsius}, 450{\textcelsius} and 600{\textcelsius} isotherms getting deeper more slowly in the low range of $\mu'$ than in the high range, compared to the 150{\textcelsius} isotherm which gets deeper at a similar rate across the range of $\mu'$. This effect is similar for Nankai but is more pronounced for Hikurangi.

Considering that the Hikurangi profiles have slab dips between Cascadia and Nankai, we attribute the difference in sign of the relationship to the much older slab age for Hikurangi. The addition of shear heating along the interface of the older, colder Hikurangi slab moves the 150{\textcelsius} isotherm shallower. It also moves the 350{\textcelsius}, 450{\textcelsius}, and 600{\textcelsius} isotherms shallower,  but it affects 150{\textcelsius} more strongly and therefore widens $D$ as shear heating increases.

\subsection{Isolating the Effect of Shear Heating}

\subsubsection{Pointwise values for $\mu'$}
The variability in $D$ is primarily attributed to $\mu'$, to the extent that interpreting the relationship between $D$ and other parameters is difficult.
To explore the second-order relationships between $D$ and the other parameters, we hold $\mu'$ constant
at the values 0, 0.03 and 0.14 and reapply the regression analysis.
Overall, the $R^2$ values are low, in the 0.1--0.5 range, which is caused by multiple parameters generating similar amounts of variability (so each relationship contains scatter caused by the other parameters).
In Table~\ref{table:D_fix_mu} we report the intervals for $D$ for Cascadia profile B, Nankai profile C, and Hikurangi profile G,
along with the difference in $D$ values, denoted by $\Delta D$ and given by
\begin{linenomath}
\begin{equation}
\label{eqn:deltaD}
\Delta D = \max_{\boldsymbol p\in \mathcal Q} D(\boldsymbol p) - \min_{\boldsymbol p\in \mathcal Q} D(\boldsymbol p)
\end{equation}
\end{linenomath}

We find that even with $\mu'$ held constant, variation in the other parameters can still create variability in $D$ on the order of tens of kilometers. 
The value at which $\mu'$ is held constant affects the amount of variability in $D$ when the other parameters in $\boldsymbol q_\text{red}$ are varied. 
Taking Nankai profile C as an example, the variability on $D_{350}$ is 36 km for $\mu' = 0$, 12 km for $\mu' = 0.03$, and 6 km for $\mu' = 0.14$.

For Cascadia profile B, $T_b$ is most strongly correlated with $D$, followed by $u_s$ and $a_s$. For Nankai profile C, $a_s$, $u_s$ and $d_c$ are the strongest predictors of $D$, followed by $T_b$ which is less correlated with $D$. The coupling parameter $\lambda$ are weakly correlated with $D$ for all three $\mu'$ values. For Hikurangi profile G, $T_b$ and $a_s$ are strongly correlated with $D$, followed closely by $u_s$, for $\mu' = 0$ and 0.03; for $\mu' = 0.14$, $u_s$ is the strongest predictor of $D$ by far, a pattern which also holds for Nankai (and for $D_{450}$ and $D_{600}$ for Cascadia). See Supporting Information Figures~\ref{fig:linreg_cascadia_profile_B_mu_0.0}, \ref{fig:linreg_cascadia_profile_B_mu_0.03}, \ref{fig:linreg_cascadia_profile_B_mu_0.14}, \ref{fig:linreg_nankai_profile_C_mu_0.0}, \ref{fig:linreg_nankai_profile_C_mu_0.03}, \ref{fig:linreg_nankai_profile_C_mu_0.14}, \ref{fig:linreg_hikurangi_profile_G_mu_0.0}, \ref{fig:linreg_hikurangi_profile_G_mu_0.03}, \ref{fig:linreg_hikurangi_profile_G_mu_0.14} for scatterplots with $R^2$ values and slopes report for each linear regression of input parameters against $D$.

\begin{table}[h!]
\begin{center}
\caption{
Variability in $D$ with constant values of $\mu'$, whilst all other parameters in $\boldsymbol q_\text{red}$ are varied.
Values reported are: $[\min(D), \max(D)]$, $\Delta D$.
}
\label{table:D_fix_mu}
\begin{tabular}{ l  l  l  l  l }
Margin | Profile & QoI & $\mu' = 0$ &  $\mu' = 0.03$ & $\mu' = 0.14$ \\
 \hline
 Cascadia | Profile B & $D_{350}$ & $[84, 152]$, 68 & $[74, 102]$, 28 & $[54, 64]$, 10 \\
 
 & $D_{450}$ &  $[144, 204]$, 60 & $[117, 151]$, 34 & $[79, 92]$, 13 \\
 
 & $D_{600}$ &  $[199, 227]$, 28 & $[179, 207]$, 28 & $[111, 129]$, 18 \\
 \hline
 Nankai | Profile C & $D_{350}$ & $[77, 113]$, 36 & $[78, 90]$, 12 & $[52, 58]$, 6 \\
 
 & $D_{450}$ &   $[83, 133]$, 50 & $[95, 117]$, 22 & $[69, 79]$, 10 \\
 
 & $D_{600}$ &  $[90, 148]$, 58 & $[103, 141]$, 38 & $[88, 104]$, 16 \\
 \hline
 Hikurangi | Profile G & $D_{350}$ & $[28, 74]$, 46 & $[69, 94]$, 25 & $[64, 71]$, 7 \\
 
 & $D_{450}$ &  $[32, 81]$, 49 & $[77, 109]$, 32 & $[84, 96]$, 12 \\
 
 & $D_{600}$ &  $[44, 88]$, 44 & $[84, 118]$, 34 & $[107, 122]$, 15 \\
 \hline
\end{tabular}
\end{center}
\end{table}

\subsubsection{Sub-interval parameter space exploration} \label{sec:subdivision_of_param_space}

We now quantify how sub-intervals of parameter space contribute to the variability observed in $D$. 
This is done by sampling input parameters from within sub-intervals of their ranges and comparing the resulting variability in $D_{350}$, $D_{450}$, and $D_{600}$ to the variability when samples are drawn from the full parameter space.  
Given a quantity $X$, we denote its sub-interval as $\hat{\mathcal Q}_X \subset \mathcal Q_X$ and
then define a new parameter space given $\hat{\mathcal Q} = \hat{\mathcal Q}_X \times \Pi_{i \ne X} \mathcal Q_i$.
Thus $X$ varies in a subinterval but all other parameters vary in their full range.
From this, the variability in $D$ given the smaller parameter space $\hat{\mathcal Q}$ is given by
\begin{linenomath}
\begin{equation}
\label{eqn:deltaDred}
\Delta D^X = \max_{\boldsymbol p\in \hat{\mathcal Q}} D(\boldsymbol p) - \min_{\boldsymbol p\in \hat{\mathcal Q}} D(\boldsymbol p).
\end{equation}
\end{linenomath}

\begin{figure}
     \centering
     \includegraphics[width=\textwidth]{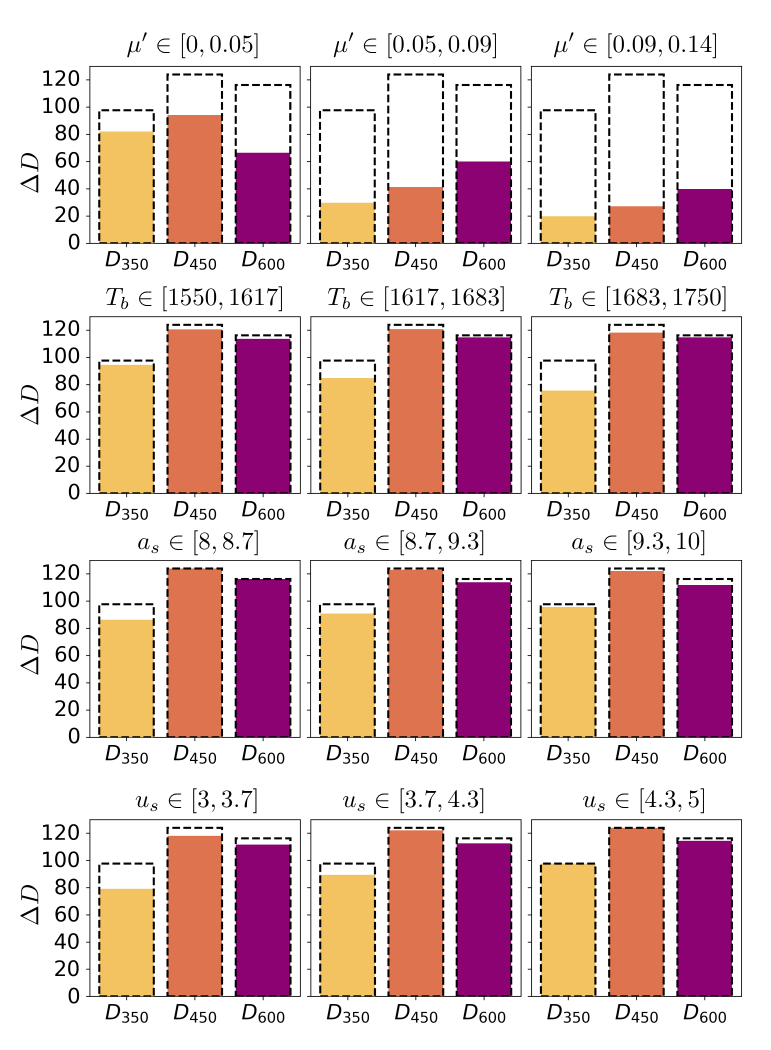}
     \caption{Bar plots of $\Delta D$ (km), the variability in $D$ as four parameters are varied in sub-intervals of the full parameter space. These results are for Cascadia profile B; black dotted outlines denote the variability in $D$ as the four parameters vary in the full parameter space. Yellow bars represent the maximum variability in $D_{350}$, orange corresponds to $D_{450}$, and purple corresponds to $D_{600}$.}
     \label{fig:subdivision_plots}
\end{figure}

In Figure~\ref{fig:subdivision_plots} we compare values for $\Delta D$ (dashed black lines), with $\Delta D^{\mu'}$, $\Delta D^{T_b}$, $\Delta D^{a_s}$, $\Delta D^{u_s}$ for Cascadia. 
The most striking pattern is that low $\mu'$ values account for most of the variability, with intermediate and high values accounting for significantly lower proportions of the variability (see uppermost row in Figure~\ref{fig:subdivision_plots}).
The same is observed for Nankai and Hikurangi (not shown).
The other parameters affect the magnitude of $D$ variability less than $\mu'$, 
but they still have sub-intervals in parameter space that contribute differently to $D$ variability.
For example, the high value range for $u_s$ accounts for almost all of the variability in $D$, while sampling $u_s$ in lower-valued ranges creates less variability.

Sampling degree of partial coupling $\lambda$ in sub-intervals does not appreciably affect the variability for Cascadia; however, for Nankai and Hikurangi low values of $\lambda$ account for most of the variability in $D$, while sampling in higher-valued subintervals slightly decreases the variability.
For Cascadia, the low range of $T_b$ accounts for most of the variability, while sampling $T_b$ in intermediate and high ranges seems to affect $D_{350}$ more than $D_{450}$ and $D_{600}$.
For Nankai, sampling non-$\mu'$ parameters in sub-intervals affects $D_{600}$ proportionally much more than $D_{350}$ and $D_{450}$ (except $\lambda$ which has only a slight effect).
This indicates that the higher scatter and lower $R^2$ values for the relationship between $D_{600}$ and $\mu'$ in Figure~\ref{fig:D_vs_mu_three_SZs} (column 2, Nankai) are due to the stronger effect of variation in non-$\mu'$ parameters on $D_{600}$.
For Hikurangi, we find that sampling non-$\mu'$ parameters within sub-intervals has an even less pronounced effect on $D$ variability than for Cascadia and Nankai. Overall, from this analysis we learn that some sub-intervals of parameter space contribute significantly more to $D$ variability than other, equally-sized sub-intervals.

\subsection{Variability in Isotherm Intersection Depths}

\begin{figure}
     \centering
     \includegraphics[width=\textwidth]{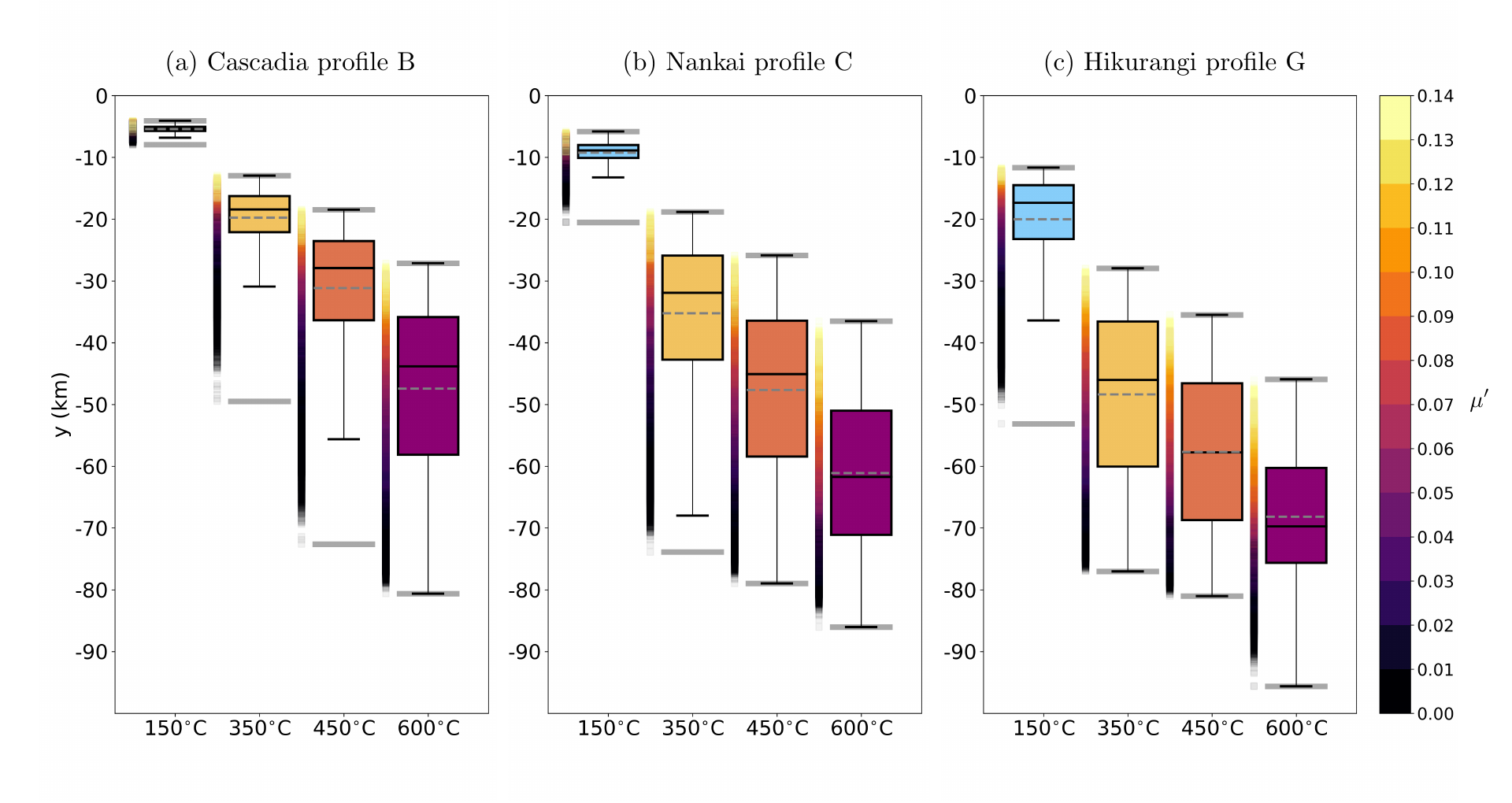}
     \caption{Boxplots for the distribution of depths where the 150{\textcelsius}, 350{\textcelsius}, 450{\textcelsius}, and 600{\textcelsius} isotherms intersect the slab interface. The solid boxes indicate the interquartile range (IQR), the black whiskers indicate 1.5 times the IQR, and gray bars indicate the minimum and maximum depths. Within the IQR box, the solid black line indicates the median value and the dotted gray line indicates the mean. The points next to each boxplot are the $\mu'$ values against the depth of intersection for each point in parameter space. Subplot (a) is for Cascadia profile B, subplot (b) is for Nankai profile C, and subplot (c) is for Hikurangi profile G.}
     \label{fig:selected_boxplots}
\end{figure}

We examine the distribution of isotherm intersection depths, $I$, via boxplots as shown in Figure~\ref{fig:selected_boxplots}.
Within each margin, the distributions of $I$ are similar across the three profiles.
Comparing $I$ distributions across margins, we see that Hikurangi has the deepest interquartile ranges (IQRs), with Nankai in the middle and Cascadia the shallowest.
The deep IQRs for Hikurangi may be attributed to Hikurangi's much older incoming lithosphere, resulting in a colder slab overall and deeper isotherms.
The Nankai IQRs are deeper than Cascadia, with two opposing factors at play: the Nankai profiles have steeper dip than the Cascadia profiles, making the slab interface hotter at shallower depths.
However, the incoming plate age for Nankai is older than Cascadia which makes the overall slab thermal structure, and therefore the slab interface, colder.
The effect of plate age is stronger than the effect of dip in this case, so Nankai isotherms are deeper than Cascadia, though not as deep as they would be if the Nankai profiles had shallower dips.

In both the variability ranges in Figure~\ref{fig:isotherm_intersection_uncertainty} and the $I$ boxplots in Figure~\ref{fig:selected_boxplots}, we see some overlap in variability ranges for isotherm intersections. 
Figure~\ref{fig:isotherm_intersection_uncertainty} shows overlap for the 350{\textcelsius}, 450{\textcelsius}, and 600{\textcelsius} isotherm intersections for all three margins, 
though Figure~\ref{fig:selected_boxplots} shows that Cascadia has no overlap in the IQRs. 
Nankai has 350{\textcelsius}, 450{\textcelsius}, and 600{\textcelsius} IQRs that overlap by about a third of each range, 
while Hikurangi has significant overlap of about two thirds of each IQR. 
As discussed in Section~\ref{sec:variability_in_rupture_limits}, this overlap implies that for a given $D$ in the overlap, there exist points in parameter space that can result in the same $D$ estimate for different choices of downdip limit isotherm. 

As previously noted (see Section~\ref{sec:subdivision_of_param_space}), the sensitivity of $D$ to variations in $\mu'$ is dependent on where in its range $\mu'$ is varying.
This characteristic is also shown in Figure~\ref{fig:selected_boxplots}, where the $\mu'$ values for each isotherm intersection depth is plotted next to each isotherm's boxplot.
From these plots it is clear that the deeper range of the boxplots can be attributed to low $\mu'$ values, in approximately the 0--0.03 range. Especially for Cascadia and Nankai, the change in depth as $\mu'$ increases is heavily skewed, with most of the change in depth happening in a relatively small interval of the $\mu'$ range. For Hikurangi, the distribution of $\mu'$ is less skewed, with the change in depth as $\mu'$ increases through the low sub-interval still present but less pronounced.

In cases where the deeper 1.5 IQR whisker is not aligned with the maximum depth bound of the distribution, we infer the existence of relatively few points in the distribution which lie near the maximum depth bound.
Typically these points would be considered outliers when interpreting boxplots, but all of the data points come from valid thermal fields and therefore should be retained in the distribution.
Instead we interpret these points as being a few points in parameter space where $\mu'$ is a very low value ($\mu' \approx 0$), driving the isotherm much deeper.
For Hikurangi in particular, the isotherms associated with $\mu' \approx 0$ are unrealistically deep.
This may be due to the use of the half-space cooling model for the incoming lithosphere thermal boundary condition, which may cause the slab to be unrealistically cold for plate ages near or greater than 100 Myr \cite{stein_stein_1992}.

\subsection{Variability From the Steady-State Assumption} \label{sec:time_dep_nankai}

The steady-state assumption is reasonable for ``sufficiently old'' subduction zones such as Cascadia and Hikurangi, 
which are estimated to have been active for 185 Myr and 70 Myr, respectively \cite{schellart_2020}.
Numerical experiments of kinematically-driven subduction models have demonstrated that 
the mantle wedge cools most rapidly over the first 100 Myr \cite{hall2012thermal}, hence quantifying
that ``sufficiently old'' implies ages of $\sim 100$ Myr.
Nankai, however, is estimated at about 18 Myr old (best estimate, with an upper limit of 20 Myr) \cite{schellart_2020},
which is significantly younger than the age where a steady-state model is appropriate.

To assess the variability associated with the steady-state assumption we evolve the time-dependent forward model
forward in time to $t = 18$ Myr. 
We take the temperature field at 18 Myr as the snapshot vector and use this to construct the ROM.
The ROM is not time-dependent, rather it defines the temperature field at a fixed point in time (18 Myr).
We perform Morris screening with ROMs built on temperature fields taken at $t = 18$ Myr.
The Morris screening for the time-dependent model yields the same result as for the steady-state model, 
with the addition of $y_g$ being identified as an important parameter.
Hence we find that $\boldsymbol q^{\text{Nankai},18}_\text{red} = \left(   u_s, d_c, \lambda, \mu', T_b, a_s, y_g  \right)$, where the super-script $\text{,18}$ implies a time-dependent model evolved up to $t = 18$ Myr.
The fact that variation in $y_g$ affects the temperature field at $t = 18$ Myr reflects the fact that the thermal structure has not reached steady state and is still affected by the initial condition.

\begin{figure}
     \centering
     \includegraphics[width=\textwidth]{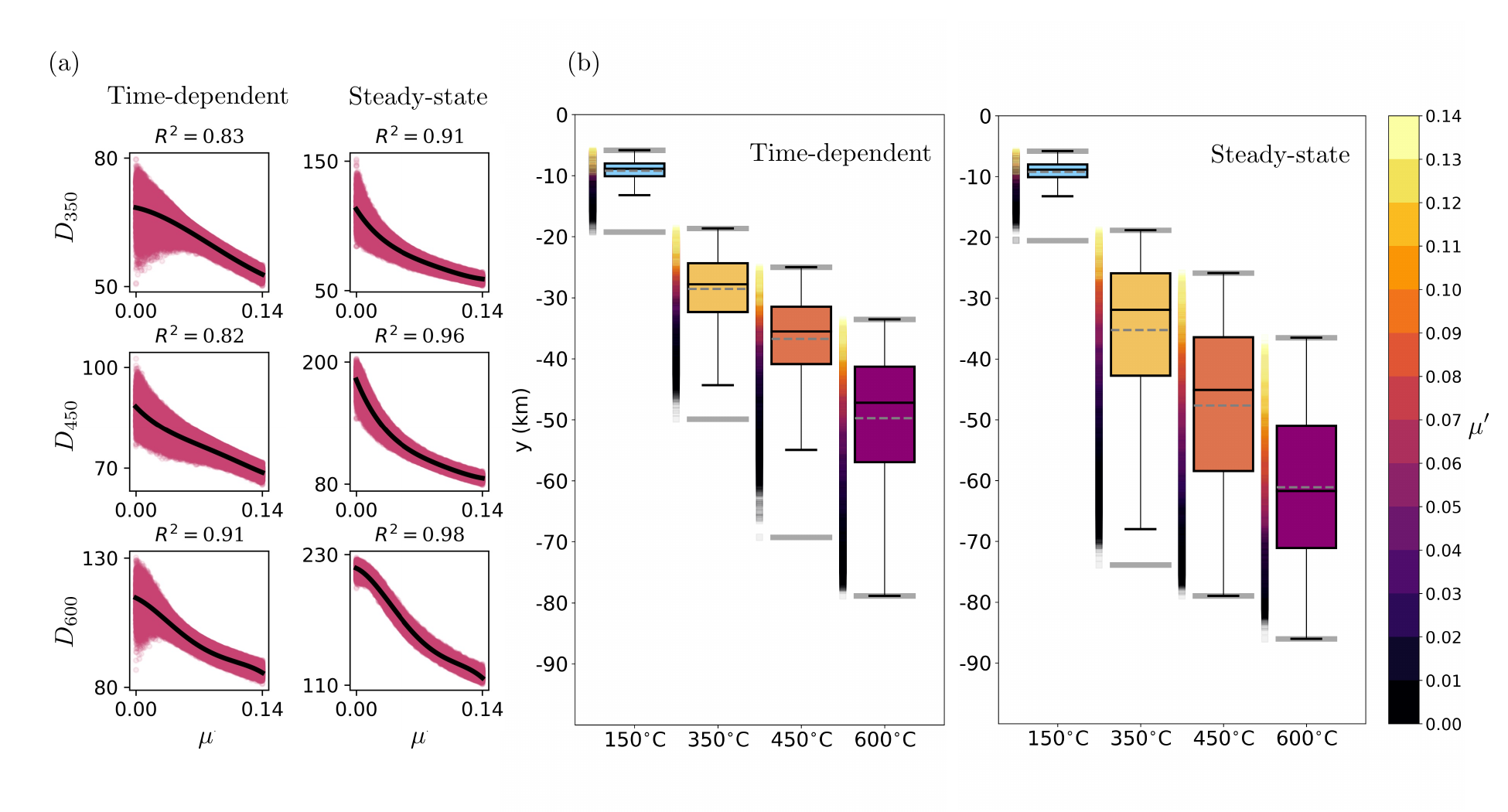}
     \caption{Comparisons between steady-state and time-dependent results for Nankai profile C. 
     (a) Relationships between $D$ and $\mu'$. Plot style is identical to that used in Figure~\ref{fig:selected_boxplots}.
     (b) Boxplots for the distribution of depths $I$.}
     \label{fig:timedep_figures}
\end{figure}

When all parameters intervals in $\boldsymbol q^{\text{Nankai},18}_\text{red}$ are explored, $D_{350}$ varies between 46--87 km, $D_{450}$ varies between 63--105 km, and $D_{600}$ is between 70--121 km. This variability is similar to that of the steady-state model (see Table~\ref{table:D_var_by_profile}) but the isotherms are generally shallower in the time-dependent model output due to the overall hotter thermal structure, as shown in Figure~\ref{fig:timedep_figures}(b). The estimated $D$ from the time-dependent case is still most sensitive to variability in $\mu'$, with relationships similar to those observed for steady-state case (see Figure~\ref{fig:timedep_figures}a).

We evaluate ROMs built on time-dependent and steady-state model output at the same $10^5$ sampled points in parameter space. For the purposes of making a direct comparison to the steady-state model, we hold $y_g$ constant at $-35$ km when evaluating the time-dependent model ROM. We find that the steady-state estimates of $D$ are generally higher than the time-dependent estimates, by as much as 23 km for $D_{350}$, 22 km for $D_{450}$, and 16 km for $D_{600}$. The difference between estimates is highest for low values of $\mu'$ and tends to decrease as $\mu'$ increases. In light of this, we argue that $D$ estimates for young subduction zones such as Nankai that are made using the steady-state assumption should be considered to have an uncertainty on the order of 20 km.

\section{Discussion} \label{sec:discussion}

Having quantified the variability in estimated rupture extent that results from input parameter variability and certain modeling choices, we turn to an examination of the implications for estimated moment magnitude of megathrust events. 
This is followed by a comparison between model-estimated rupture limit isotherm depths and the observed maximum depth of thrust faulting events.
Finally, we consider variability in model input parameters when comparing model-estimated slab interface P-T conditions to P-T conditions inferred from a global compilation of exhumed rocks. 
Given our analysis in Section~\ref{sec:time_dep_nankai}, all of the Nankai results presented in this discussion (Section~\ref{sec:discussion}) are for the time-dependent model. 

\subsection{Quantifying Variability in Estimated Moment Magnitude}

We take the empirical scaling relationship for large and great earthquakes from \citeA{ye_2016}, who observe that moment scales with $A^{3/2}$, where $A$ is the effective rupture area.
We assume that $A=D^2$ as in \citeA{england_2018}, which is conservative, and
compute $M_w = \frac{2}{3} \left( \log_{10}(M_0) - 9.1\right)$ to obtain the estimated moment magnitude.
Given a choice of the 350{\textcelsius} isotherm as the downdip rupture limit, variability in key parameters results in $M_w$ estimates that vary from 7.6--8.6 for Cascadia, 7.3--8.2 for Nankai, and 6.6--8.0 for Hikurangi.
Similarly, $M_w$ estimates for Cascadia vary between 8.0--9.0 for a 450{\textcelsius} isotherm rupture limit and 8.3--9.1 for a 600{\textcelsius} isotherm rupture limit.
Nankai and Hikurangi have $M_w$ estimates with similar variability, though smaller magnitudes (see Table~\ref{table:Mw_variability}).
This analysis highlights the significant impact that the choice of downdip rupture limit isotherm has on $M_w$ estimates.

\begin{table}[h!]
\begin{center}
\caption{Margin specific $M_w$ intervals given variability in $\boldsymbol q_\text{red}$, for each choice of the downdip limit isotherm $D$.}
\label{table:Mw_variability}
      \begin{tabular}{lSSS}
        \toprule
        \multirow{2}{*}{Margin} &
          \multicolumn{3}{c}{$M_w$} \\
          & {$D_{350}$} & {$D_{450}$} & {$D_{600}$}  \\
          \midrule
          \midrule
	{Cascadia} & {$ [7.6, 8.6] $}  & {$[8.0 , 9.0] $} & {$[8.2 , 9.1]$} \\
 	{Nankai} & {$[7.8 , 8.2]$} & {$[7.9, 8.4]$} & {$[7.9 , 8.4]$} \\
 	{Hikurangi} & {$[7.0 , 8.1]$} & {$[7.1, 8.2]$} & {$[7.4 , 8.4]$} \\
        \bottomrule
      \end{tabular}
\end{center}
\end{table}

\subsection{Comparison to Max Depth of Thrust Faulting and SSE Observations}

\begin{figure}
     \centering
     \includegraphics[width=\textwidth]{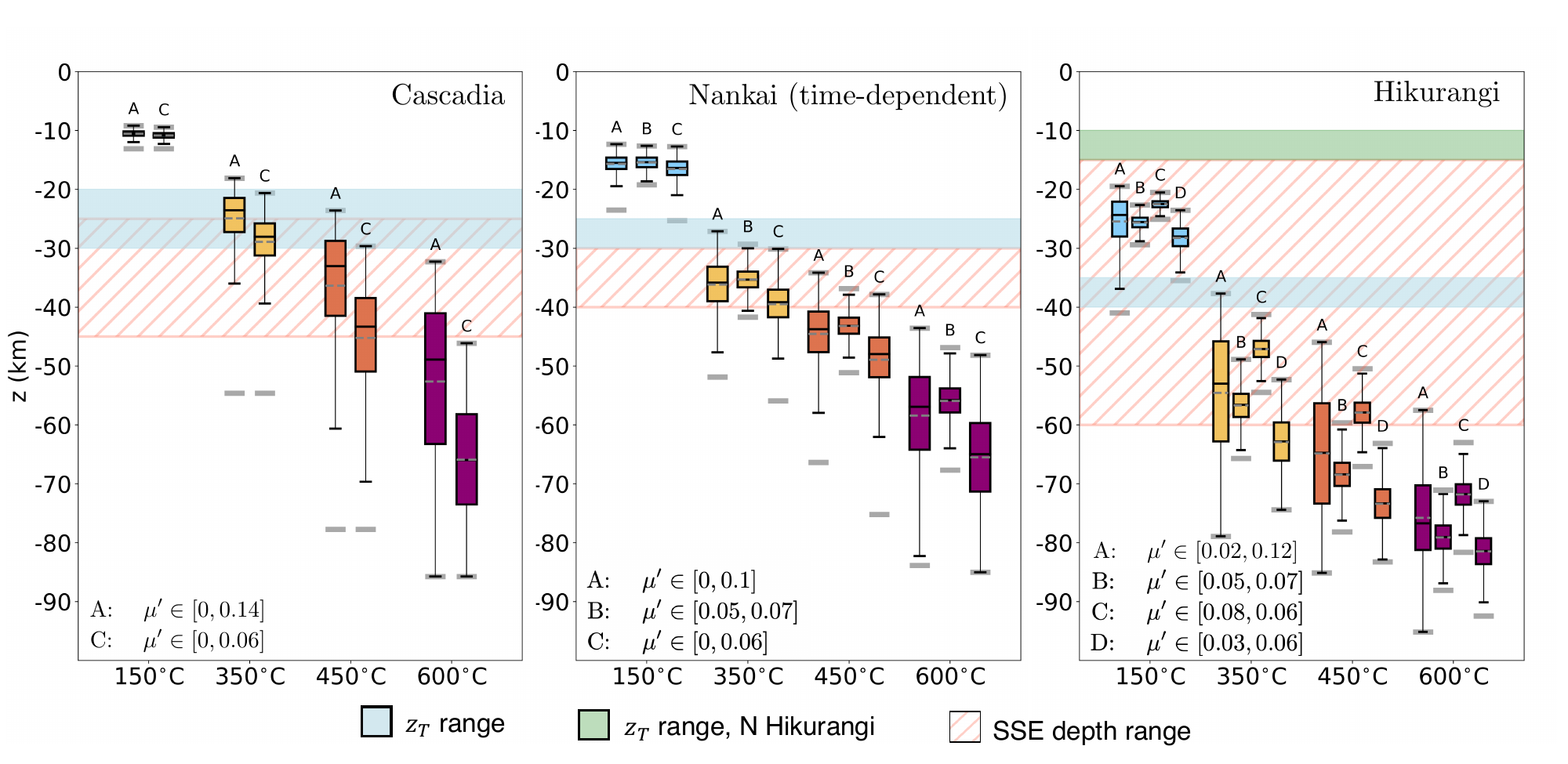}
     \caption{Boxplots of depths at which isotherms intersect the slab interface, given estimated ranges for $\mu'$, for $10^5$ samples in parameter space. A is the widest interval from \citeA{england_2018}, B is the preferred interval from \citeA{england_2018}, C is the interval from \citeA{gao_wang_2014}, and D is the interval inferred by \citeA{england_2018} from \citeA{mccaffrey_2008}.}
     \label{fig:zt_boxplots_application}
\end{figure}

We examine a selection of estimated ranges for $\mu'$ from the literature and compare model-estimated isotherm intersection depths to observations in each margin.
Estimated ranges for $\mu'$ are drawn from \citeA{england_2018}, \citeA{gao_wang_2014}, and \citeA{mccaffrey_2008}.
We find that these estimated ranges result in large variability in estimated isotherm intersection depths, as shown in Figure~\ref{fig:zt_boxplots_application}.
We compare isotherm intersection depths to the maximum depth of thrust faulting $z_T$, from \citeA{england_2018}.
We also compare to depths of observed slow-slip events (SSEs), focusing on SSEs occurring downdip of the locked zone.
SSEs are understood to occur in regions where frictional behavior transitions from seismic (velocity-weakening) to aseismic (velocity-strengthening) and where effective stress is low due to high pore fluid pressure \cite{liu_rice_2007, audet_et_al_2009, kodaira_et_al_2004}.
Specifically, we are comparing $z_T$ to the 350{\textcelsius} isotherm depths and comparing SSE depths to the 450{\textcelsius} and 600{\textcelsius} isotherm depths.

\citeA{england_2018} concluded that, for Cascadia, $\mu'$ could not be estimated with confidence from heat flux data, and reported an interval of [0.0, 0.14].
It is not possible to compare the depths of isotherm intersections with seismicity, due to a lack of focal mechanisms for Cascadia \cite{england_2018}.
If we make the assumption that the maximum depth of thrust faulting follows the line of 20\% locking \cite{schmalzle_2014}, the estimated $z_T$ is 20--30 km.
Given this is an assumption and not a direct observation, we use caution in drawing conclusions;
however, we note that the IQR of the 350{\textcelsius} isotherm lies within this depth range,
and the upper end of the 450{\textcelsius} isotherm IQR for the \citeA{england_2018} interval falls at the downdip end of this range (see Figure~\ref{fig:zt_boxplots_application}, left).
Short term SSEs are observed along the entire Cascadia margin at 25--45 km depths \cite{brudzinski_allen_2007}.
This depth range encompasses the depths of the model-estimated 450{\textcelsius} isotherm IQR, with the 600{\textcelsius} IQR for \citeA{england_2018} also partially overlapping this region.
We think this partial overlap is interesting because it indicates that when variability is taken into account, not only the 450{\textcelsius} isotherms but some 600{\textcelsius} isotherms agree with observations of the downdip transition zone depths.
Overall these results show good agreement for Cascadia between observed seismicity depths and model-estimated isotherm intersection depths.

In the case of Nankai, we compare isotherm depths to $z_T$ of 25--30 km depth which is determined from the downdip rupture limits of the 1944--1946 $M_w$ 8.1 earthquakes, and geodetically determined regional locking depths \cite{england_2018}.
The bounds of the 350{\textcelsius} isotherm depth distribution overlap the $z_T$ depth range, but the IQR is deeper than $z_T$.
Nankai hosts short-term SSEs downdip of the locked zone, occurring between 30--40 km depth, with the most slip concentrated at 30--35 km depths \cite{kano_kato_2020}.
This overlaps the 350{\textcelsius} and 450{\textcelsius} isotherm IQRs for the time-dependent model  (Figure~\ref{fig:zt_boxplots_application}, center).
The model-estimated 600{\textcelsius} isotherm distributions are significantly deeper than the SSE depths.

The Hikurangi margin displays significant variability in seismic styles and geophysical characteristics across the margin.
For Hikurangi, we take $z_T$ to be the estimated downdip limit of interseismic coupling \cite{wallace_2009}.
Estimates are highly variable between the southern portion, which has deep interseismic coupling, and the central and northern portion where interseismic coupling is shallower \cite{wallace_2009}.
In the northern and central margin (north of 40°S), $z_T$ is estimated at 10--15 km, while at the southern end of the North Island $z_T$ is estimated to be 35--40 km \cite{wallace_2009}.
In the southern end of the margin, the downdip limit of interseismic coupling occurs where temperatures have been estimated at 300--400{\textcelsius}, indicating there may be a thermal control on the limit in this region \cite{mccaffrey_2008, fagereng_2009}.
However, if the downdip limit of interseismic coupling were also thermally controlled in the central and northern margin, it would have to occur between 35--60 km depths, instead of 10--15 km.
 Therefore temperature is not thought to be the primary control of interseismic coupling in the central and northern regions \cite{mccaffrey_2008, fagereng_2009}.
\citeA{fagereng_2009} showed that variation in seismicity across the margin can be explained by differences in the fluid pressure state along-strike.
However \citeA{wallace_2009} argue that the along-margin variability is likely due to the interplay of different factors including fluid pressure, upper and lower plate structure, subducting sediments, the regional tectonic stress regime, and thermal effects.

Our results are consistent with previous work \cite{mccaffrey_2008}; there is some agreement between the shallowest 350{\textcelsius} isotherm depths and the southern margin's deeper $z_T$ range, but all isotherm depths are much deeper than the northern margin's shallower $z_T$ range. We focus our SSE comparison on the deep (25--60 km), long-lived SSEs observed adjacent to the deeply locked patch in the southern margin \cite{wallace_beavan_2010, wallace_et_al_2012, wallace_2020}. The model-estimated 350{\textcelsius} isotherm IQRs lie almost entirely within the SSE depth range, while the three shallower 450{\textcelsius} IQRs overlap or lie within the range. Only the shallowest 600{\textcelsius} isotherms overlap the SSE depth range. We also note the pronounced differences in IQRs for the 350{\textcelsius}, 450{\textcelsius}, and 600{\textcelsius} isotherms that result from the different published estimated ranges for $\mu'$, particularly for the Hikurangi margin. In addition, some IQRs for different isotherms overlap each other. We refer to parameter range EW for Hikurangi as an example (see Figure~\ref{fig:zt_boxplots_application}, right). The overlap implies that we could choose the 350{\textcelsius} isotherm as the downdip rupture limit, and there is a point in parameter space that yields, say, a 55 km depth estimate; equally, we could choose the 450{\textcelsius} isotherm as the limit, and a different point in parameter space would also yield a 55 km depth estimate. We argue that this non-uniqueness means that a straightforward choice of one isotherm as the rupture limit should be treated with caution.

\subsection{Comparison to a global compilation of exhumed rock P-T conditions}

Exhumed rocks from past subduction zones can be analyzed to infer the pressure-temperature (P-T) conditions they were subject to during subduction.
\citeA{pennistonDorland_2015} compare P-T conditions inferred from a global compilation of exhumed rocks to P-T conditions estimated by thermal models.
Their work reveals a discrepancy of over 100{\textcelsius}--200{\textcelsius} at lower pressures, and up to 300{\textcelsius} at certain depths, with exhumed rocks recording hotter temperatures than were estimated by models in \citeA{gerya_et_al_2002} and \citeA{syracuse_2010}.
\citeA{pennistonDorland_2015} argue that thermal models do not include several important sources of heat.
The sources they propose include shear heating, convection of rock in the shallow mantle and subduction channel,
and hydration reactions due to fluid release from the subducting slab.
They highlight shear heating as the source which could contribute the most heat.

Efforts to reconcile the discrepancy have included adding shear heating to models, as well as modeling slab interface temperatures during subduction initiation \cite{van_keken_et_al_2018, holt_condit_2021, van_keken_wilson_2023, peacock_review_2020}.
\citeA{van_keken_et_al_2018} demonstrate that adding shear heating increases slab interface temperatures, but does not sufficiently heat the deeper subducting plate to match exhumed rocks.
They argue that for shear heating to explain the discrepancy, rocks would have to be exhumed from on or near the slab interface, rather than deeper in the slab where temperatures are cooler.
Instead they propose that the rock record was preferentially exhumed from anomalously warm subduction conditions, such as during subduction initiation or during subduction of young lithosphere.
\citeA{holt_condit_2021} modeled dynamically evolving (time-dependent) slab interface temperatures and showed that modeled slab top P-T conditions during subduction initiation span much of the P-T space of exhumed rocks.
However, \citeA{pennistonDorland_2015} show that the rock record does not correlate with age or convergence rate, and they argue that it is unlikely that every exhumed rock records anomalous conditions and that none of the exhumed rocks record ``average" subduction conditions.
\citeA{whitney_2020} note that some subduction zones contain both high-temperature and very-low-temperature rocks,
which indicates that those exhumed rocks were subject to thermal conditions that varied considerably in space or time within the same subduction zone.

\begin{figure}
     \centering
     \includegraphics[width=\textwidth]{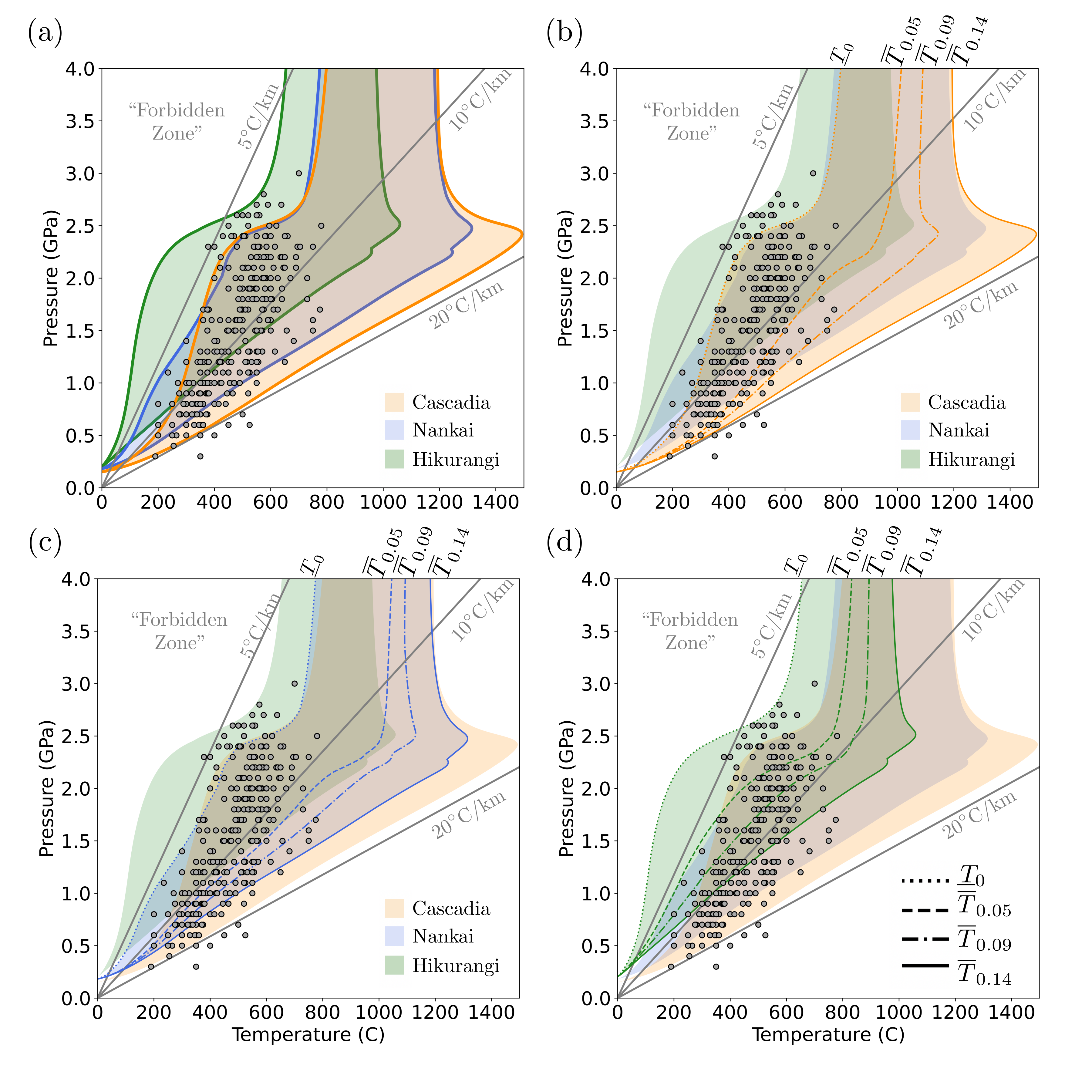}
     \caption{P-T curves for slab interface temperatures for each margin; each line is one of $10^4$ samples in parameter space. 
     Gray scattered data points are from the global catalog of P-T data for exhumed rocks from \citeA{whitney_2020}. 
     The label $\overline T_{\bar\mu'}$ denotes the maximum of an envelope of temperatures for an interval $[0,\bar\mu']$, 
     defined as $\overline T_{\bar\mu'} = \max\limits_{\mu' \in [0, \bar\mu']}  T_{\Gamma_I} \left( \boldsymbol q_\text{red} \right)$, 
     where $T_{\Gamma_I}\left( \boldsymbol q_\text{red} \right)$ refers to the $10^4$ samples of temperature on the slab interface 
     taken in the $\boldsymbol q_\text{red}$ parameter space.  
     The label $\underline{T}_0$ denotes the minimum of a temperature envelope for $\mu' \in [0,0.14]$ and 
     is defined as $\underline T_{0} = \min\limits_{\mu' \in [0, 0.14]}  T_{\Gamma_I} \left( \boldsymbol q_\text{red} \right)$. 
     Depth $z$ was converted to pressure using a constant density of 3000 $\text{kg/m}^3$.
     }
     \label{fig:PT_envelopes}
\end{figure}

We examine whether considering variability in model input parameters results in P-T estimations that span observed P-T conditions from the rock record. 
Figure~\ref{fig:PT_envelopes} shows P-T paths for $10^4$ samples in parameter space for a profile of each margin,
as well as the global catalog of P-T data for exhumed rocks from \citeA{whitney_2020} which is updated from \citeA{pennistonDorland_2015}.
We show that considering a family of models that includes variability in input parameters results in a wide range of estimated P-T conditions, which encompass most of the observed P-T conditions in the global exhumed rock catalog.
As in the rest of this study, there is a strong relationship with shear heating: the higher the value of $\mu'$, the hotter the temperature at a given depth, as visualized in Figure~\ref{fig:PT_vs_mu}.

Due to Hikurangi's old incoming lithosphere, its slab interface temperatures lie in a colder region of P-T space.
Some modeled P-T curves for Hikurangi lie in the ``forbidden zone", a low temperature - high pressure region in P-T space generally considered unrealistic in Earth conditions, and not represented in the exhumed rock record \cite{hacker_2006}.
However, we note that the coldest Hikurangi estimates are for model evaluations with low or zero shear heating.
Adding a reasonable amount of shear heating brings the Hikurangi estimates back out of the ``forbidden zone", as shown in Figure~\ref{fig:PT_envelopes}.
Cascadia, the warm end-member, lies in the hotter region of P-T space, and adding reasonable shear heating (within published estimated ranges) results in P-T conditions that explain the hotter exhumed rocks.
P-T curves for Nankai overlap Cascadia and Hikurangi but generally lie between the two.
Taken together, the three margins span most of the variability observed in the rock record.

There remain many aspects to the comparison between exhumed rocks and modeled P-T conditions that are not precisely known, 
including whether exhumed rocks were located on the subduction interface, 
whether they were altered during exhumation, 
and whether they were preferentially exhumed during subduction initiation. 
We do not comprehensively examine these possibilities, nor do we argue for a specific explanation for the discrepancy between exhumed rocks and modeled P-T conditions. 
Rather we consider the exhumed rock record to be an imperfect but useful set of information representing the diversity of global subduction conditions. 
We argue that variability in model input parameters should be considered when making comparisons between the global catalog of exhumed rocks and model estimates.

\subsection{Limitations}

The thermal models used in this study have several limitations.
The effects of fluids, brought into the system by the downgoing slab, is not captured by the model; nor are incoming sediments, which can insulate and lubricate the plate interface.
Thermal and compositional buoyancy is not included in the model, meaning that we cannot capture processes sometimes observed in fully dynamic models,
such as subduction-channel flow and diapirism \cite{peacock_review_2020}.
Radiogenic heat production is also not included as a heat source.
The slab convergence rate and the interface geometry are held constant.
The thermal models in this study are 2D and thus unable to capture processes which may be occurring in the along-margin direction. 
Subduction zones have natural geometric variability in all three dimensions, and may exhibit along-margin mantle flow effects
which cause 3D variation in slab surface temperatures. 
Three-dimensional kinematic-dynamic models using generic geometries indicate that the thermal structure is affected by oblique subduction and along-margin
variation in slab geometry, which generate along-margin mantle flow, \cite<e.g.>{kneller_van_keken_2008, bengtson_van_keken_2012}. 
While these limitations are important to consider, the 2D kinematic-dynamic models allow us to investigate the primary controls on the overall thermal structure.

The framework we present here is built for the purpose of quantifying variability, not for constraining input parameters. Better constraints on $\mu'$, for example, would likely require more surface heat flow measurements, combined with inversions such as those in, \cite<e.g.>{england_2018, gao_wang_2014, morishige_kuwatani}.

\section{Conclusions} \label{sec:conclusions}

We have presented a framework that leverages fast, accurate ROMs to quantify variability in slab interface temperatures and potential rupture extent.
We find that the estimated potential extent of rupture $D$ is sensitive to variability in the apparent coefficient of friction $\mu'$.
Variability in the low range of $\mu'$ values contributes most of the variability in $D$ estimates.
Other key parameters such as $T_B$, $a_s$, and $u_s$ also contribute to variability in $D$ by as much as 10--60 km (see Table~\ref{table:D_fix_mu}).
Overall, variability in all key parameters results in $D_{350}$ estimates between 54--180 km for Cascadia, 40--126 for Nankai, and 27--94 km for Hikurangi.
Taking 450{\textcelsius} or 600{\textcelsius} as the downdip rupture limit isotherm significantly increases the estimated extent of rupture, especially for shallow-dipping slabs such as Cascadia (see Table~\ref{table:D_var_by_profile}).
This variability translates to significant differences in estimated moment magnitude for megathrust events.
A comparison between model-estimated rupture limit depths and observed max depths of thrust faulting $z_T$ reveals some overlap between the 350{\textcelsius} isotherm depths and $z_T$ ranges;
the 450{\textcelsius} and 600{\textcelsius} isotherm depths are generally deeper than the $z_T$ ranges.
This analysis also highlights the noticeable differences in estimated isotherm depths that result from different previously-published estimated ranges for $\mu'$.
Finally, considering a family of models that includes variability in input parameters results in a wide range of estimated P-T conditions, encompassing most of the observed P-T conditions in the global exhumed rock catalog.

\appendix

\section{Comparison of thermal model with analytical expressions} \label{sec:appendix_verif_analytical}

\citeA{england_may_2021} present analytical expressions for temperatures on curved plate interfaces, 
both with and without shear heating on the interface. 
We refer the reader to \citeA{england_may_2021} for full details of the analytical expressions. 
We compare the slab interface temperatures estimated using our numerical model to those approximated by the analytical expressions. 
To do the comparison, we create meshes using the polynomial expression for the slab interface profile in Equation 1 of \citeA{england_may_2021}, 
for both $a = 0.002$ (shallower dip) and $a = 0.004$ (steeper dip). 
We match the model input parameters as closely as possible to the numerical calculations in \citeA{england_may_2021}, 
as reported in Table~\ref{table:analytical_expression_numerical_params}. 

\begin{figure}
     \centering
     \includegraphics[width=\textwidth]{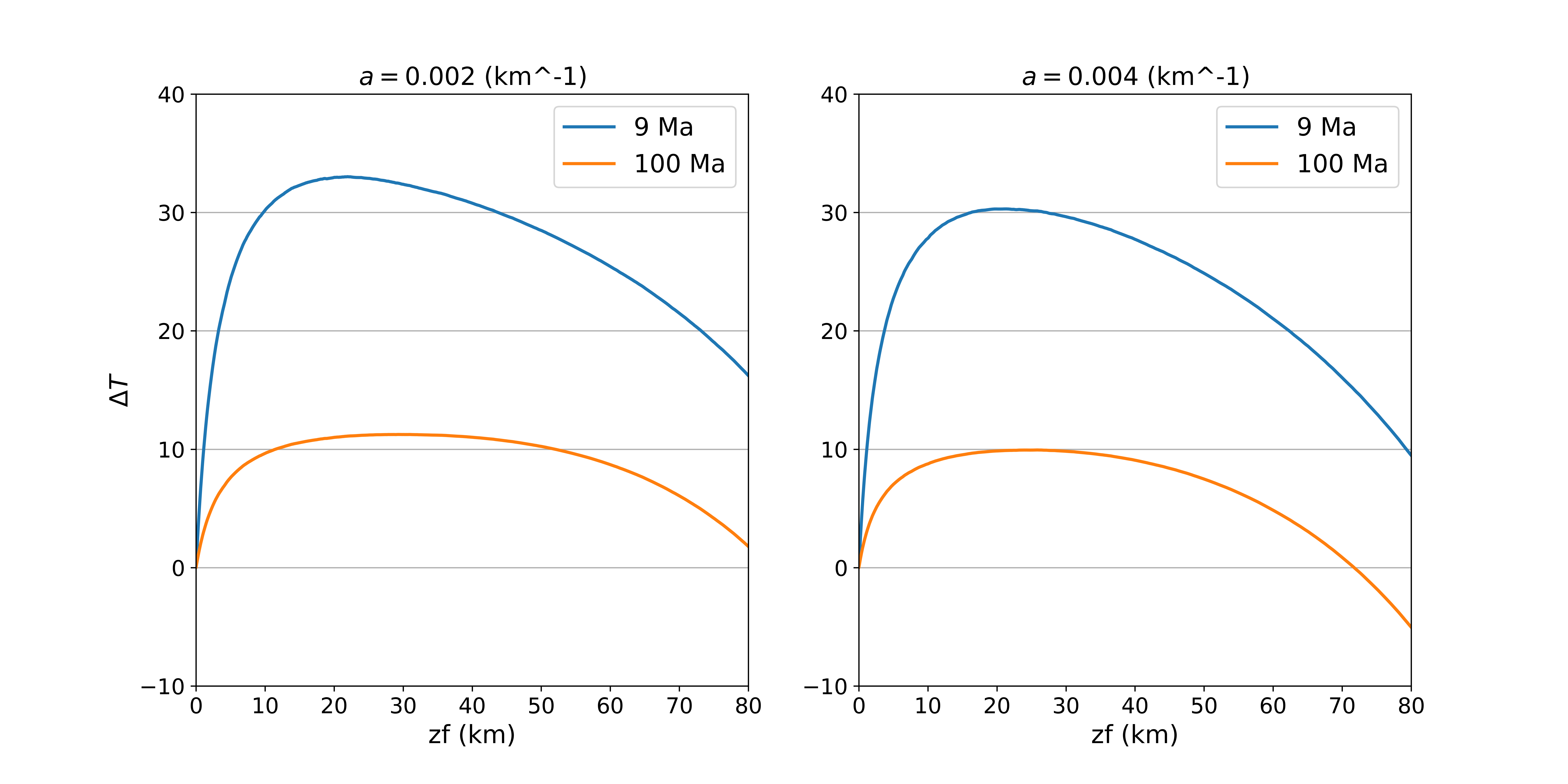}
     \caption{The difference in slab interface temperatures between analytical expressions and numerical calculations, for the case with no shear heating.}
     \label{fig:analytical_caseA}
\end{figure}

For the case with no shear heating (Case A in Table~\ref{table:analytical_expression_numerical_params}), 
we compare to the analytical expression calculated using Equation~11 from \citeA{england_may_2021}. 
Above 80 km depth, we find agreement to within 35{\textcelsius} for a slab age of 9 Myr, 
and to within 12{\textcelsius} for a slab age of 100 Myr. 
The difference between analytically and numerically calculated temperatures is shown in Figure~\ref{fig:analytical_caseA}. 

\begin{figure}
     \centering
     \includegraphics[width=\textwidth]{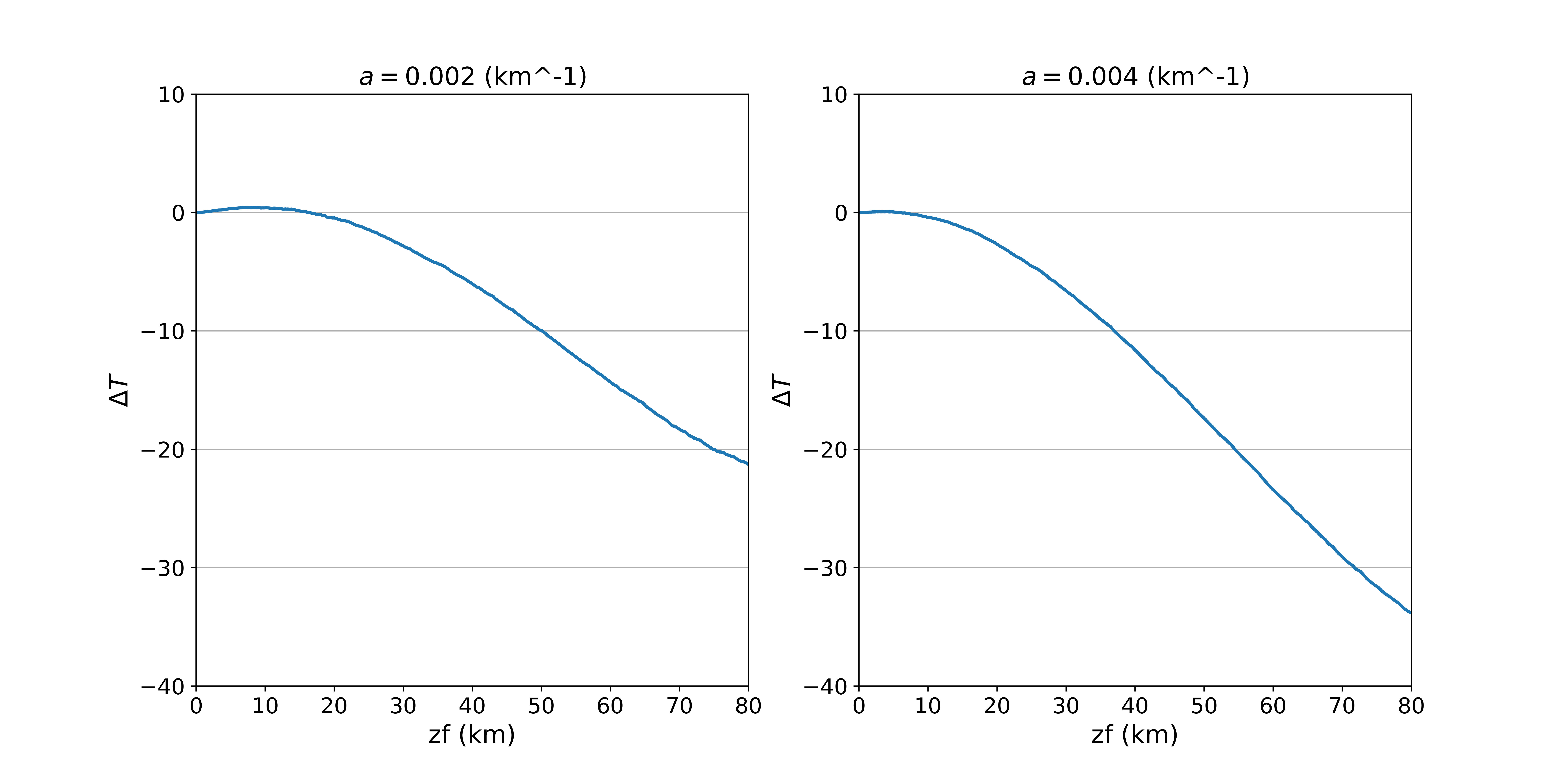}
     \caption{The difference in slab interface temperatures between analytical expressions and numerical calculations, for the case with only shear heating present.}
     \label{fig:analytical_caseB}
\end{figure}

We also examine the case with shear heating (Case B in Table~\ref{table:analytical_expression_numerical_params}), in the absence of all other heat sources. 
The analytical expression is calculated using Equation~17 from \citeA{england_may_2021}.
In our numerical calculations we set $T_s = 273$K and $T_b = 274$K, and using a slab age of $10^{10}$ Myr, 
to effectively "turn off" the overlying plate linear geotherm and the heat flow from the slab. 
Above 80 km depth, we find agreement to within 22{\textcelsius} for the profile with $a = 0.002$ 
and within 35{\textcelsius} for the profile with $a = 0.004$.
This is shown in Figure~\ref{fig:analytical_caseB}. 

\begin{figure}
     \centering
     \includegraphics[width=\textwidth]{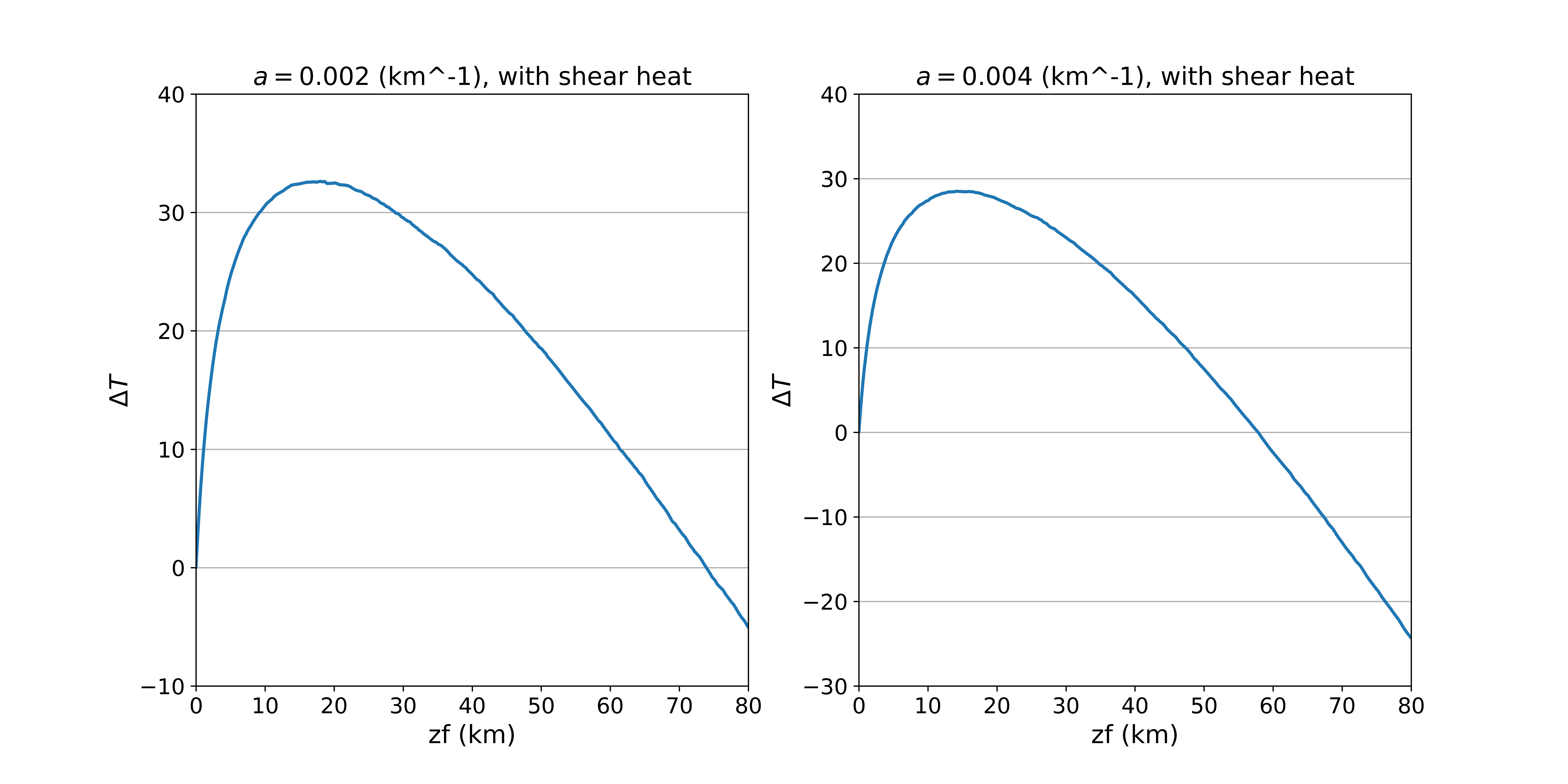}
     \caption{The difference in slab interface temperatures between analytical expressions and numerical calculations, for the case where the no-shear heating analytical approximation (from Equation 11 of \citeA{england_may_2021}) is summed with the approximation with only shear heating present (from Equation 17 of \citeA{england_may_2021}). }
     \label{fig:deltaT_sum}
\end{figure}

\begin{figure}
     \centering
     \includegraphics[width=\textwidth]{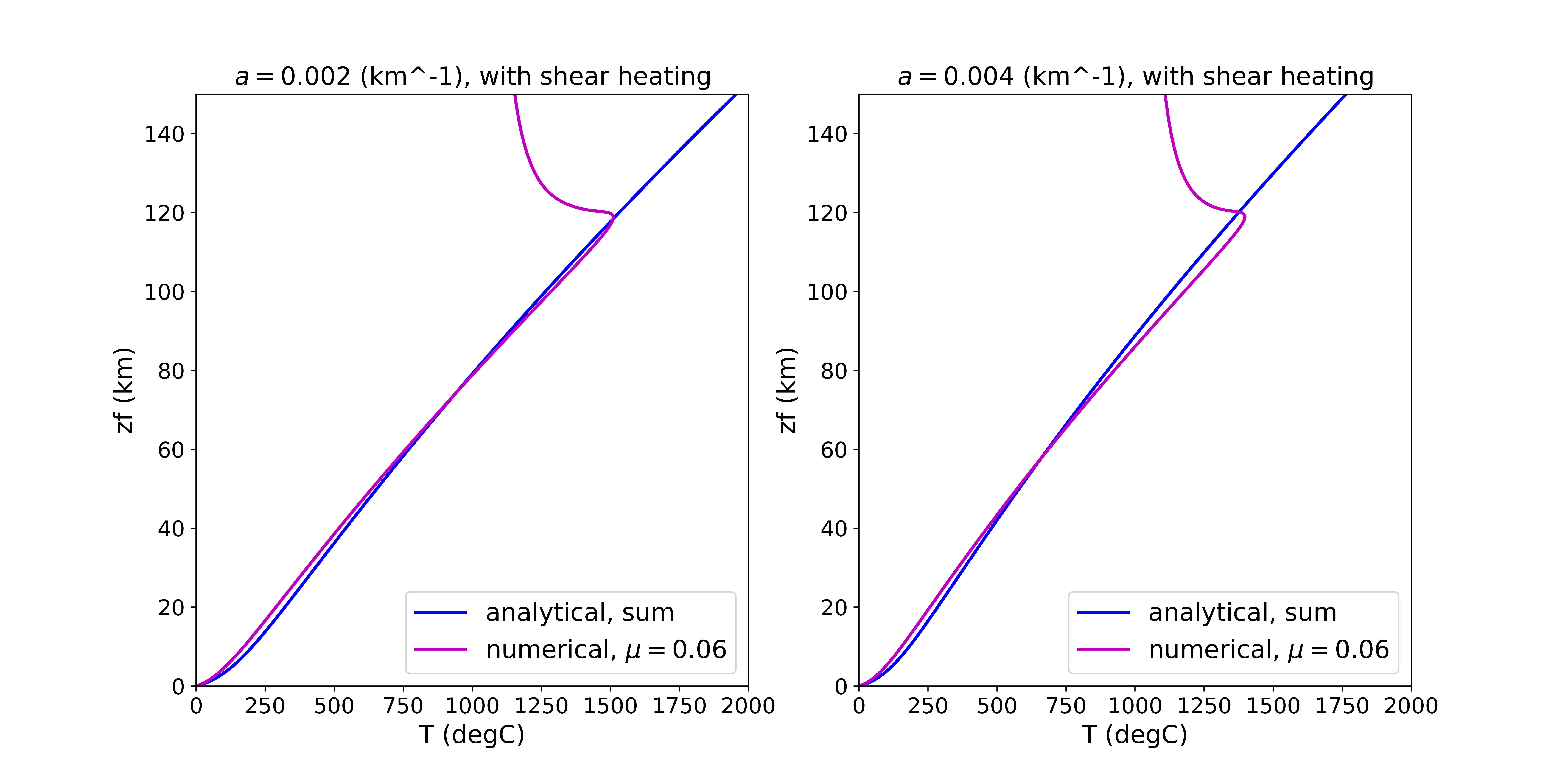}
     \caption{Slab interface temperatures for a numerical calculation with $\mu' = 0.06$, compared with the case where the no-shear heating analytical approximation (from Equation 11 of \citeA{england_may_2021}) is summed with the approximation with only shear heating present (from Equation 17 of \citeA{england_may_2021}). }
     \label{fig:T_comparison_sum}
\end{figure}

The temperature calculated by Equation~11 from \citeA{england_may_2021} can be summed with 
the temperature calculated by their Equation~17 to approximate temperature on the interface in the case with shear heating and the presence of other heat sources. 
We run a numerical model with a slab age of 9 Myr and $\mu' = 0.06$, and compare to the summed analytical expression temperatures. 
We find agreement to within 35{\textcelsius} for the profile with $a = 0.002$, and within 30{\textcelsius} for $a = 0.004$. 
The temperatures are shown in Figure~\ref{fig:T_comparison_sum}, and the difference between analytically and numerically calculated temperatures
is shown in Figure~\ref{fig:deltaT_sum}.  

\begin{figure}
     \centering
     \includegraphics[width=\textwidth]{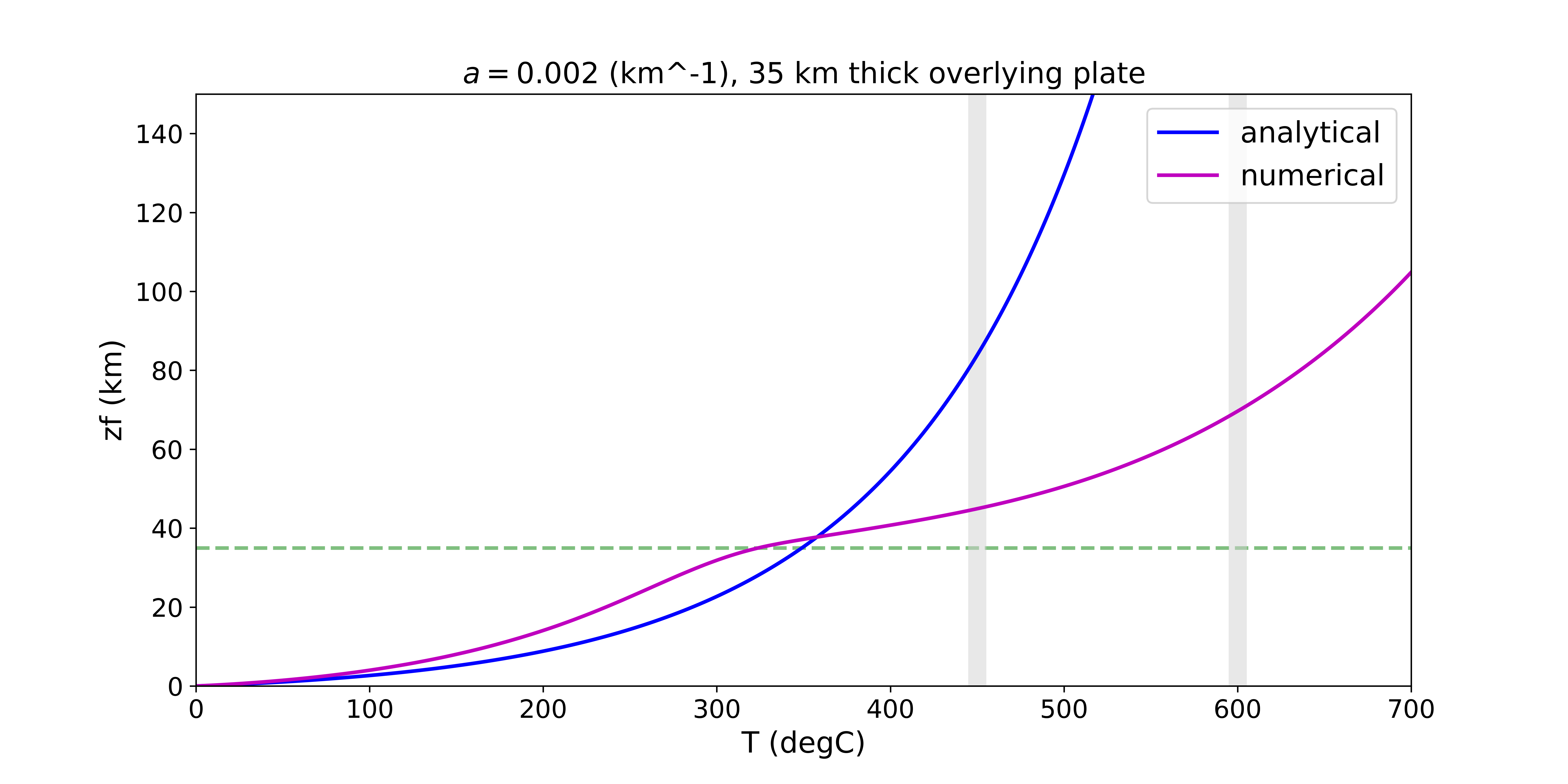}
     \caption{Slab interface temperatures for the example with $a = 0.002$, a 35 km thick overlying plate, and Cascadia-like parameters. The vertical grey bars show the 450{\textcelsius} and 600{\textcelsius} isotherms, while the horizontal green dotted line shows the depth of the mantle wedge corner, equivalent to the depth of decoupling in this example.}
     \label{fig:analytical_caseC}
\end{figure}

In this study, we examine quantities of interest that lie close to or below the mantle wedge corner, at deeper depths than where the analytical expressions are valid. 
We can see from Figure~\ref{fig:isotherm_intersection_uncertainty} and Figure~\ref{fig:selected_boxplots} that several of our QoIs, specifically the 450{\textcelsius} and 600{\textcelsius} isotherms and sometimes the 350{\textcelsius} isotherm,  lie below the mantle wedge corner (at 35 km depth). 
To demonstrate the need for the numerical models, we present a case with $a = 0.002$ and parameters matching our Cascadia calculations; 
namely, a 35 km thick overlying plate, a plate convergence velocity of 5 cm/yr, and a slab age of 8 Myr (Case C in Table~\ref{table:analytical_expression_numerical_params}). 
In Figure~\ref{fig:analytical_caseC}, we show temperature on the slab interface given by both the analytical expressions and the numerical calculations. 
The analytical expressions agree well to about 35 km depth, near the mantle wedge corner, and then diverge significantly. 
The gray vertical bars show the 450{\textcelsius} and 600{\textcelsius} isotherms, which lie in the region not well captured by the analytical expressions. 

\section{Comparison of thermal model with community benchmark} \label{sec:appendix_comm_benchmark}

\begin{figure}
     \centering
     \includegraphics[width=\textwidth]{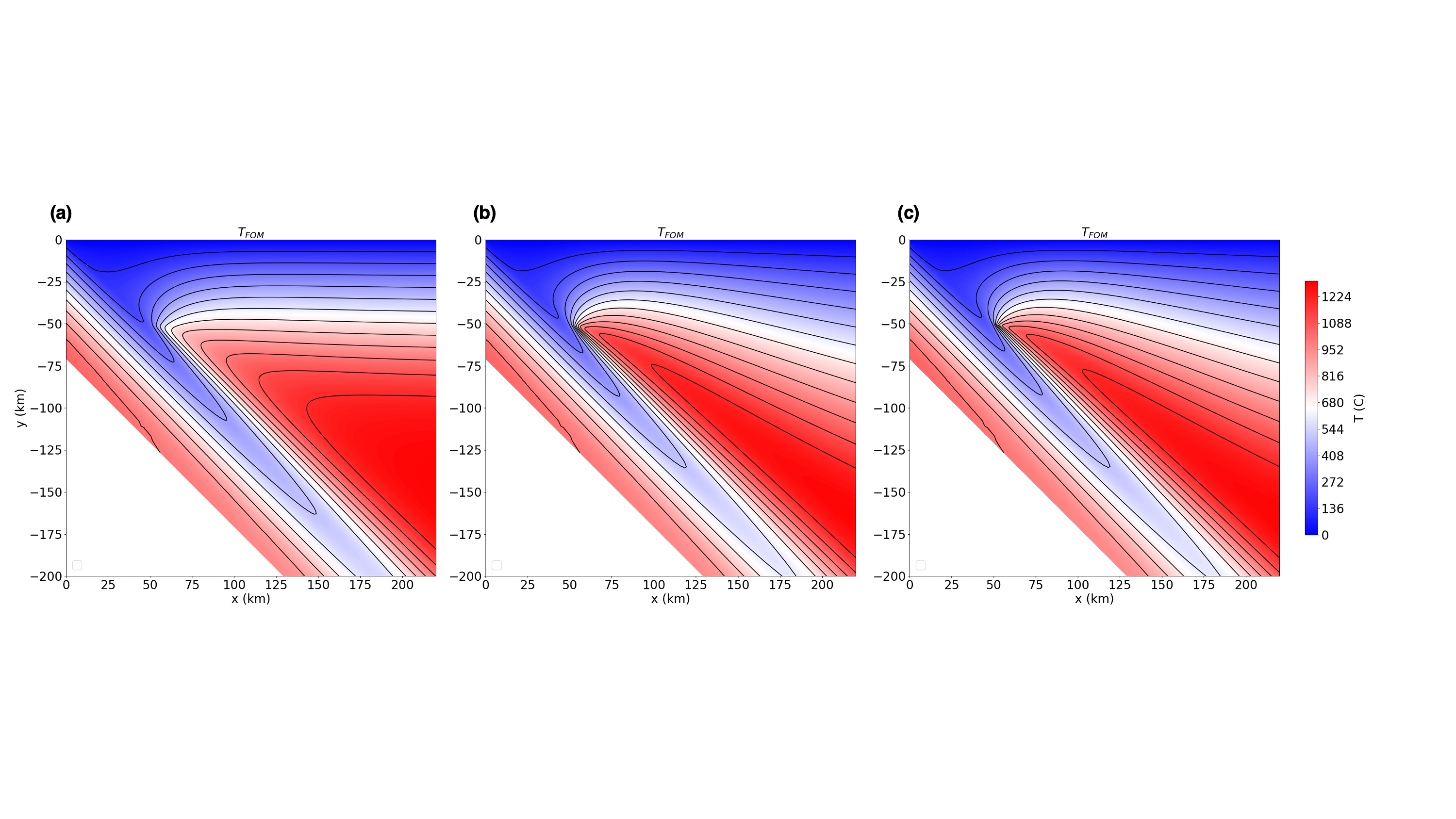}
     \caption{Inset temperature fields for cases matching the community benchmark \cite{community_benchmark_van_keken_2008}, with (a) showing case 1a with an isoviscous wedge, (b) showing case 2a, diffusion creep, and (c) showing case 2b, dislocation creep. }
     \label{fig:benchmark_T_fields}
\end{figure}

We compare results from our forward model to the community benchmark presented by \citeA{community_benchmark_van_keken_2008}. To make the comparison, we create a geometry with a straight slab dipping at 45 degrees; see Supporting Information Section~\ref{sec:benchmark_info} for details. 
We find good agreement between our forward models and the community benchmark. 
The benchmark uses three metrics for comparison between all contributed codes: 
(1), the temperature at (60, 60 km), in the mantle wedge near the corner point, denoted as $T_{11,11}$, 
(2), the L2 norm of the slab-wedge interface temperature between 0 and 210 km depth, denoted as $||T_\text{slab}||$ and defined in Equation~\ref{eqn:T_slab}, 
and (3), the L2 norm of the temperature in the corner of the wedge, between 54 and 120 km depth, denoted as $||T_\text{wedge}||$ and defined in Equation~\ref{eqn:T_slab}.  
We compute these three metrics for each of the three relevant cases: case 1c, dynamical flow in an isoviscous wedge, case 2a, dynamical flow with diffusion creep, and case 2b, dynamical flow with dislocation creep. 
See Table~\ref{tab:benchmarking} for the values. Figure~\ref{fig:benchmark_T_fields} shows insets of the temperature fields for each case, permitting a visual comparison to the results in \cite{community_benchmark_van_keken_2008}. 

\begin{table}[ht!]
\begin{center}
\caption{Computed metrics for three cases to compare to \cite{community_benchmark_van_keken_2008}. All values reported in degrees C, and community benchmark ranges denoted as CB.}
\begin{tabular}{ |c|c|c|c|c|c|c| } 
 \hline
 Metric & Case 1c, CB & Case 1c & Case 2a, CB & Case 2a & Case 2b, CB & Case 2b \\
 \hline
 $T_{11,11}$ & $388$--$397$ & 390.4 & 570--581 & 584.2 & 550--583 & 585.7 \\ 
 $||T_\text{slab}||$ & 503--511 & 488.0 & 607--614 & 592.8 & 593--608 & 591.3 \\
 $||T_\text{wedge}||$ & 850--854 & 847.7 & 1002--1007 & 1000.0 & 994--1000 & 996.6 \\ \hline
\end{tabular}
\label{tab:benchmarking}
\end{center}
\end{table}

\section{Software capability and extensions}

The framework used in this study is available as open-source software. It was designed to be flexible in several respects. Firstly, the code that generates the slab interface profiles can take other geometries as input, not just Slab2 data as used in this study. Secondly, this study focuses on Cascadia, Nankai and Hikurangi, but the framework is designed to be extendable to other margins; both in how profiles are made and in how input parameters are handled, to facilitate experiments in margins with very different parameter ranges. Finally, thermal model output, and therefore ROM evaluations, are T fields over the whole model domain, not just the slab interface. In this work, we focus on slab interface temperature in our uncertainty quantification, but the framework could be used to study other aspects of the thermal structure.

\section{Supporting Information}

\subsection{Transition From Partial to Full Coupling Along Slab Interface}

\begin{figure}
     \centering
     \includegraphics[width=\textwidth]{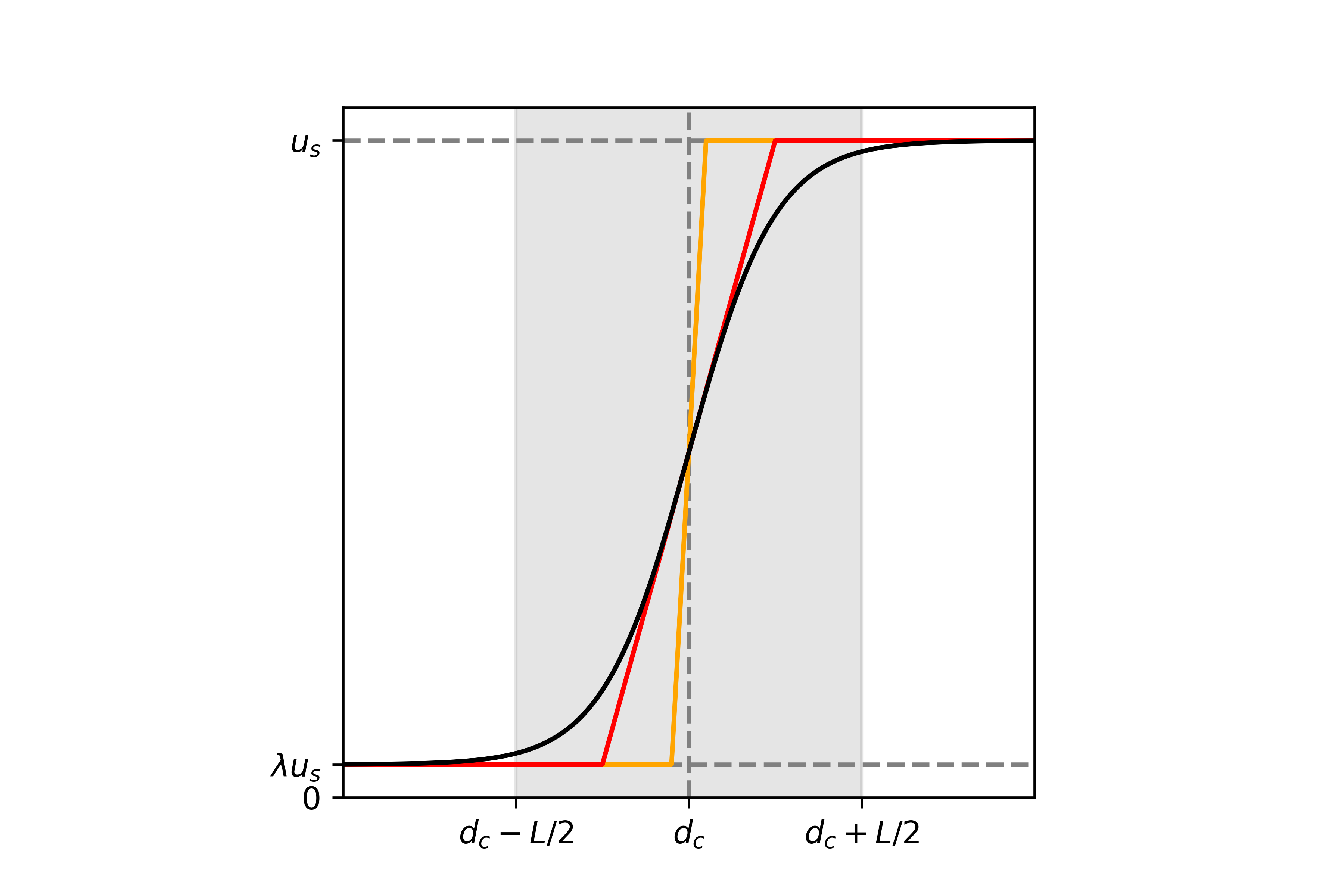}
     \caption{The transition from partial to full coupling along the slab interface. The tanh function used in this study (black line) allows for a smooth increase in velocity over an $L=10$ km interval centered at the $d_c$. For comparison purposes we plot ramp functions where the velocity increase happens over 5 km (red line) or 1 km (orange line). The $L=10$ km transition distance used in this study is longer than the ramps imposed at the mantle wedge corner in, e.g. \cite{van_keken_kiefer_peacock_2002, community_benchmark_van_keken_2008}, but produces a smooth increase in velocity.}
     \label{fig:plot_tanh_transition}
\end{figure}

Close to the depth of decoupling, the transition from partially coupled to fully coupled is implemented using a $\tanh$ function, 
\begin{linenomath*}
\begin{align}
    \mathbf{u}_{trans} = \frac{1}{2} \mathbf{u} \left( (\lambda+1) + (1-\lambda) \tanh \left( \frac{D-c}{L/4} \right) \right),  \label{eqn:tanh}
\end{align}
\end{linenomath*}
where $D$ is the along-slab distance, $c$ is the along-slab distance to the depth of decoupling location, and $L$ is a parameter describing the length scale of the transition (see Table~\ref{table:notation_and_ranges}). 
In this study we use a transition distance of $L = 10$ km. 
Figure~\ref{fig:plot_tanh_transition} shows the transition imposed using Equation~\ref{eqn:tanh}. 

\subsection{Slab Profiles and Mesh Generation} \label{sec:profiles_mesh}

We create slab profiles by taking 2D slices of 3D surfaces generated from Slab 2.0 data, which represent the slab interfaces for the Cascadia, Nankai and Hikurangi subduction zones. The 2018 Slab 2.0 model provides 3D geometries for all currently subducting slabs across the globe \cite{slab2}. 
A set of start and end points for each profile is selected, in geographic coordinates (latitude and longitude). 
We create a 3D surface from the Slab 2.0 data using \verb|scipy|'s \verb|RBFInterpolator()| class \cite{scipy_2020}. 
Profiles of the interpolated surface are taken between the start and end points, and the profiles are converted such that the horizontal axis is the along-slab distance, offering a true view of the profile from side-on. 
The depth coordinates are re-scaled such that the trench point is the zero coordinate for depth, rather than using the raw depth values from Slab 2.0 data. 
Depth values in our model domains should be interpreted as depth below the trench. 

\begin{figure}
     \centering
     \includegraphics[width=\textwidth]{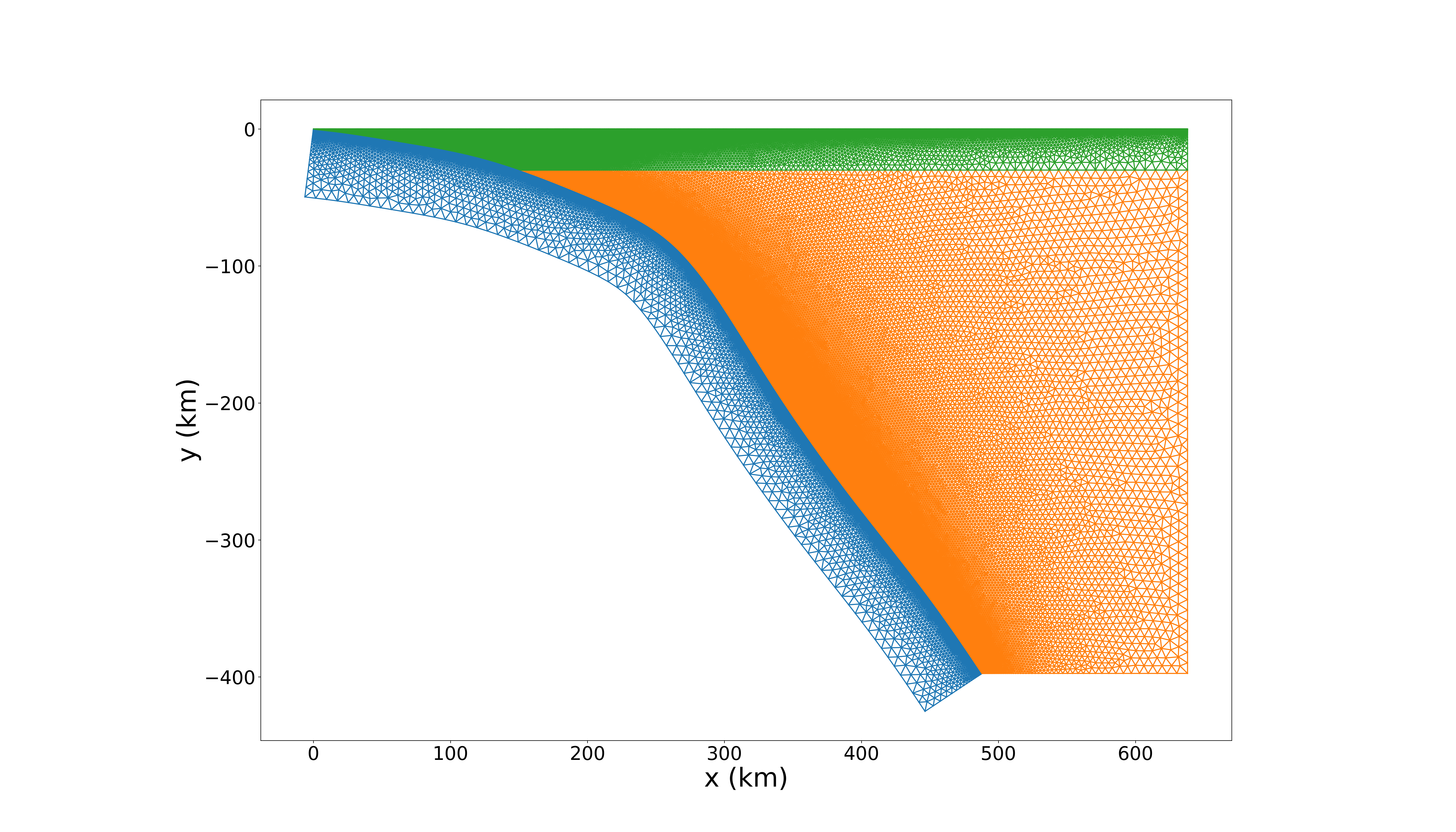}
     \caption{An example of a 3D unstructured triangular mesh with local refinement. This mesh corresponds to Cascadia profile B.}
     \label{fig:mesh_colored}
\end{figure}

We use GMSH version \verb|4.10.1| to create an unstructured triangular mesh with subdomains for the slab, overlying plate, and mantle wedge regions. 
GMSH is an open-source meshing tool that generates two- and three-dimensional finite element meshes using a build-in CAD engine and post-processor \cite{gmsh}. 
The mesh creation process is standardized using Python to facilitate changing the slab thickness, the overlying plate thickness, and the mesh size as desired. 
The mesh is locally refined to have a mesh size of 0.5 km near the slab interface, growing to 8 km far from the slab. 
An example of a locally-refined mesh is shown in Figure~\ref{fig:mesh_colored}. 

\begin{figure}
     \centering
     \includegraphics[width=\textwidth]{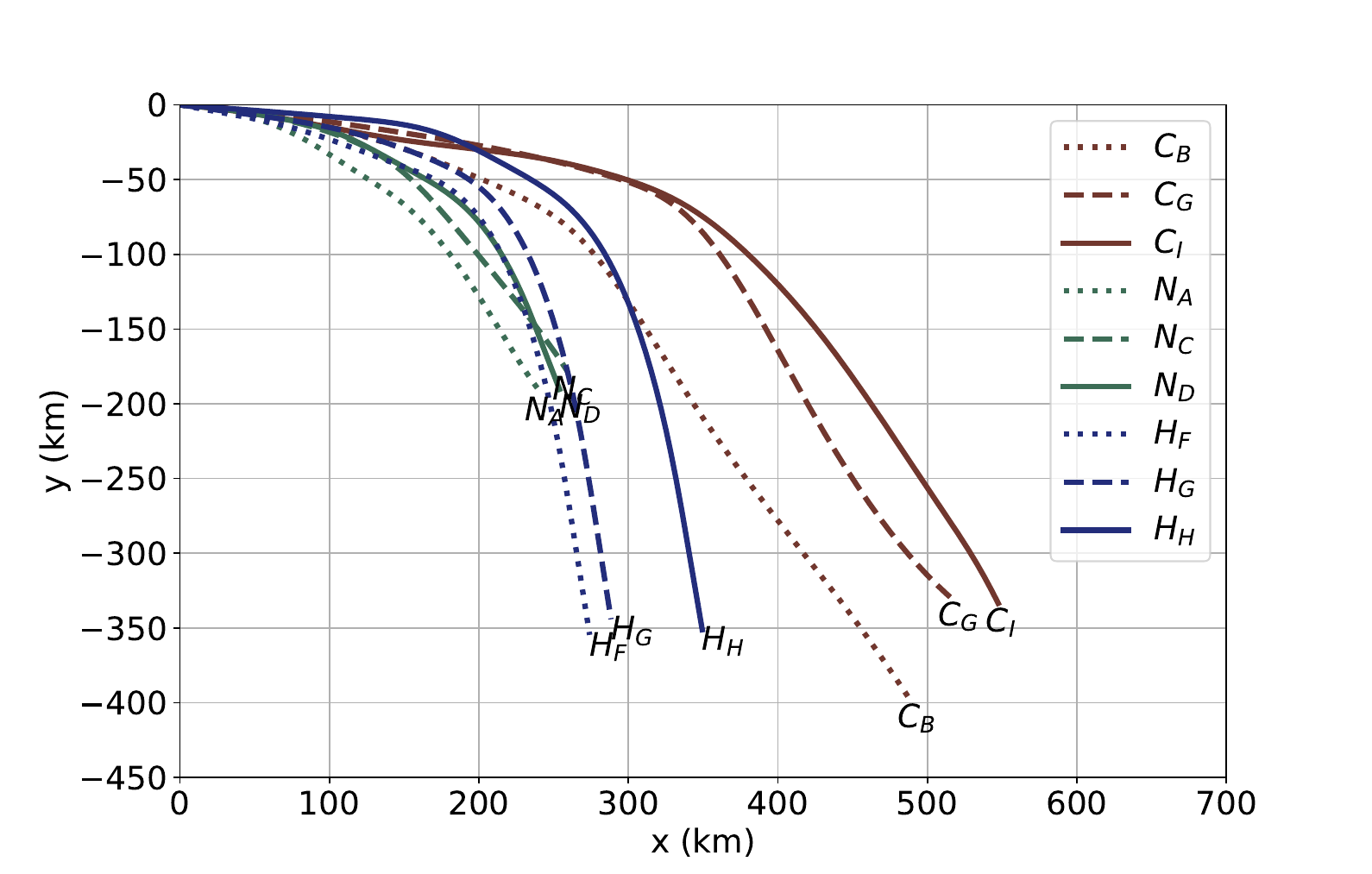}
     \caption{Superimposed profiles of the Cascadia, Nankai and Hikurangi subduction zones.}
     \label{fig:superimposed_profiles}
\end{figure}

\subsection{Comparing to community benchmark} \label{sec:benchmark_info}

As discussed in \ref{sec:appendix_comm_benchmark}, we compare temperature estimated by our forward model to the community benchmark presented by \citeA{community_benchmark_van_keken_2008}. To make the comparison, we create a geometry with a straight slab dipping at 45 degrees. Since our method requires that the slab inflow boundary is normal to the slab interface, our geometry does not have a vertical slab inflow boundary as does \cite{community_benchmark_van_keken_2008}. The impact of this difference is that the half-space cooling model used for the slab inflow temperature boundary condition is computed using shallower depths than it would for a vertical boundary. To correct for this, in the benchmark case only, we include a factor of two in the half space cooling model equation so that temperatures along the slab inflow are artificially set as if we were using a vertical boundary. The factor of two comes from the fact that with a 45 degree angled boundary, each point on the inflow boundary is at half the depth of the corresponding point on the vertical line originating at the trench. 

We define here for reference the three metrics used in \citeA{community_benchmark_van_keken_2008} for comparison between all contributed codes. 
When computing these measurements, they use temperature solutions discretized on an equidistant 6 km grid, $T_{ij}$. 
The first measure is the temperature at (60, 60 km), in the mantle wedge near the corner point. 
The second measure is the L2 norm of the slab-wedge interface temperature between 0 and 210 km depth: 
\begin{align} 
    || T_{\text{slab}} || = \sqrt{\frac{\sum_{i=1}^{36} T_{ii}^2}{36}}.   \label{eqn:T_slab}
\end{align}
The third measure is the L2 norm of the temperature in the corner of the wedge, between 54 and 120 km depth:
\begin{align}
    || T_{\text{wedge}} || = \sqrt{\frac{\sum_{i=10}^{21} \sum_{j=10}^{i} T_{ij}^2}{78}}. \label{eqn:T_wedge}
\end{align}

\subsection{Implementations of shear heating} \label{sec:shear_heating}

In our study, we present results calculated using shear heating imposed as a line source along the subduction interface. 
In the code, this source is implemented by adding a line integral along $\Gamma_I$ to the weak formulation. 
The code framework provided alongside this study \cite{github_repo} also includes the option to implement shear heating volumetrically using a properly scaled Gaussian centered on the slab interface. 
In this case, the parameterization for shear heating is given by a 1D Gaussian function centered on $\Gamma_I$ and applied in the directions $\pm \boldsymbol n_I$, i.e. normal to the subduction interface.
The Gaussian at each point along the slab is equal to $\mu' [[\boldsymbol u]]_{\Gamma_I} P_{litho}$,
where $\mu'$ is the apparent coefficient of friction and  $P_{litho} = \rho g y$ is the lithostatic pressure.
Hence the source term in Equation~\eqref{eqn:energy} is given by
\begin{linenomath*}
\begin{equation}
    H_{\text{sh}} = \mu' [[ \boldsymbol u]]_{\Gamma_I} P_{litho} \frac{1}{s\sqrt{2 \pi}} \exp \left( - \frac{w^2}{2 s^2} \right),
    \quad \text{where} \quad \int_{-\infty}^{\infty}  \frac{1}{s\sqrt{2 \pi}} \exp \left( - \frac{w^2}{2 s^2} \right) dw = 1.
     \label{eqn:H_sh_gaussian}
\end{equation}
\end{linenomath*}
In Equation~\eqref{eqn:H_sh_gaussian}, $w$ is the distance normal to the slab interface and $s$ defines the width of the Gaussian.

The Gaussian shear heating source is equivalent to the line source, provided the width of the Gaussian is appropriately resolved by the mesh. 
The Gaussian width is parameterized using the parameter $s$, and we refer to the mesh size as $h$. 
We perform some resolution tests using code that implements both the line source and Gaussian source for shear heating in a 2D box domain. 
We find that the ratio $s/h$ must be greater than approximately 0.5 for the Gaussian to be well resolved by the mesh. 
If the Gaussian is too narrow relative to the mesh size, the source may contribute too much shear heating along the interface. 
The code for the resolution tests is provided in the Zenodo repository that accompanies this paper. 
They can be used to test that the line source and Gaussian implementations give equivalent results, and that the Gaussian is well resolved by the mesh.

\subsection{Comparison of thermal model with analytical expressions} \label{sec:analytic_soln}

In \ref{sec:appendix_verif_analytical}, we detail the comparison between our thermal model and the analytical expressions presented in \citeA{england_may_2021}. 
This section contains the figures and table referred to in \ref{sec:appendix_verif_analytical}.

\begin{table}[h!]
\begin{center}
\caption{Parameters used in the numerical calculations made for comparison to the analytical expressions. Case A refers to the case with no shear heating, and Case B refers to the case with only shear heating present, in the absence of all other heat sources. Case C was made using a 35 km thick overlying plate to demonstrate that quantities of interest related to the 450{\textcelsius} and 600{\textcelsius} isotherms are in a region where temperature is not well approximated by the analytical expressions.}
\label{table:analytical_expression_numerical_params}
\begin{tabular}{ l  l  l  l  l  l } 
 \hline
 Parameter & Definition & Units & Case A & Case B & Case C \\
 $a$ & Coeff. from \citeA{england_may_2021}, Eq 1 & - & 0.002, 0.004 & 0.002, 0.004 & 0.002 \\
 
 - & Overlying plate thickness & km & 120 & 120 & 35 \\
 
 $u_s$ & Plate convergence velocity & cm/yr & 10 & 10 & 5 \\
 
 $d_c$ & Max depth of decoupling & km & 120 & 120 & 35 \\
 
 $\lambda$ & Degree of partial coupling & -  & 1.0 & 1.0 & 1.0 \\

$\mu'$ & Apparent coefficient of friction & -  & 0 & 0.06 & 0.06 \\

$T_b$ & Mantle inflow temperature & K & 1623 & 1623 & 1623 \\

$a_s$ & Age of oceanic lithosphere at trench & Myr & 9, 100 & $10^{10}$ & 8\\

$y_g$ & Depth of linear geotherm & km  & 120 & 120 & 35 \\ 

$L$ & Partial to full coupling transition length & km & 10 & 10 & 10 \\

$\rho_\text{mantle}$ & Density of the mantle & kg m$^{-3}$ & 3300 & \\

$\rho_\text{crust}$ & Density of the crust & kg m$^{-3}$ & 3300 & \\

$\rho_\text{slab}$ & Density of the slab & kg m$^{-3}$ & 3300 &\\

$g$ & Gravitational acceleration & m s$^{-2}$ & 9.8  & \\

$k_\text{mantle}$ & Thermal conductivity of mantle & W m$^{-1}$ K$^{-1}$ & 3.0 &  3.0 & 3.0 \\

$k_\text{crust}$ & Thermal conductivity of crust & W m$^{-1}$ K$^{-1}$ & 3.0 &  3.0 & 3.0  \\

$k_\text{slab}$ & Thermal conductivity of slab & W m$^{-1}$ K$^{-1}$ & 3.0 &  3.0 & 3.0  \\

$c_p$ & Specific heat capacity & J kg$^{-1}$ K$^{-1}$ & 1250 & \\
 
\end{tabular}
\end{center}
\end{table}

\subsection{Sampling in Parameter Space} \label{sec:sampling_param_space}

We run the forward model given parameter trajectories sampled in parameter space using a Halton sequence. The Halton sequence is a deterministic but low-discrepancy and quasi-random sequence \cite{halton_1960}. We use the Halton sequence generator implemented in \verb|scipy|'s \verb|Quasi-Monte Carlo| submodule \cite{scipy_2020}. This implementation uses scrambling to ensure good performance in higher dimensions. The advantage of using Halton for our forward model parameter samples is that the sequence can be continued if an insufficient number of forward models was initially run. 

When performing the global sensitivity analysis, as described in Section~\ref{sec:SA}, we use Latin Hypercube sampling. This method generates a near-random sample of parameter values where each sample is the only one in each axis-aligned hyperplane. It does not require more samples for higher dimensions. This method suits our purposes when drawing samples for sensitivity analysis because it ensures the parameter space is evenly and completely sampled. We again use the Latin Hypercube class implemented in \verb|scipy|'s \verb|Quasi-Monte Carlo| submodule \cite{scipy_2020}.

\subsection{Viscosity flow law parameters} \label{sec:visco_params}

Experimental data have been used to arrive at an equation for the relationship between strain $\Dot \epsilon$ and applied differential stress $\sigma_\text{d}$ which is general to both diffusion and dislocation creep \cite{hirth_kohlstedt_2003}. 
This flow law has the form:

\begin{align} \label{eqn:hk_flow_law_strain}
\Dot{\epsilon} = A (\sigma_\text{d})^n d^{-p} (fH2O)^{r} \exp(\alpha \phi) \exp \left( - \frac{E_a + V_a P}{RT}  \right)
\end{align}

In this flow law, $P$, $T$, and $R$ are the pressure, temperature and gas constant, respectively; $d$ is the grain size and $fH2O$ is water fugacity. 
The other parameters are experimentally determined: 
$A$ is the pre-exponential factor, 
$n$ is the stress exponent (1 for diffusion creep, $>1$ for dislocation creep), 
$p$ is the grain size exponent, 
$r$ is the water fugacity exponent, 
$\phi$ is the melt fraction,
$\alpha$ is a constant, 
 $E_a$ is the activation energy (J/mol), 
 and $V_a$ is the activation volume (J/Pa). 

We require an expression for the effective viscosity of the mantle, $\eta_{eff}$, in terms of the second invariant of the deviatoric stress $\sigma_\text{II}$ or the second invariant of the strain rate $\Dot{\epsilon}_\text{II}$. 
We can reformulate the expression in equation \ref{eqn:hk_flow_law_strain} using the following relation for isotropic, incompressible materials: $\sigma_\text{II} = 2 \eta_\text{eff} \Dot{\epsilon}_\text{II}, \quad \text{or} \quad \eta_\text{eff} = \frac{\sigma_\text{II}}{2 \Dot{\epsilon}_\text{II}}$.

We assume zero melt fraction, $\alpha = 0$, and thus $\exp(\alpha \phi) = 1$. 
The resulting expression for effective viscosity is:

\begin{align}  \label{eqn:hk_flow_law}
 \eta_\text{eff} = F_2 (A)^{-1/n} (d)^{p/n} (fH2O)^{-r/n}    \exp \left( \frac{E_a + V_a P}{nRT}  \right) (\Dot{\epsilon}_\text{II})^{(1 - n)/n}
\end{align}

The factor $F_2$ is necessary to account for the dependence of the relationship between  $\sigma_\text{d}$ and $\sigma_\text{II}$, and between $\Dot{\epsilon}$ and $\Dot{\epsilon}_\text{II}$, on the type of experiment performed. 
We will use parameter ranges from \citeA{hirth_kohlstedt_2003} and to our understanding, they use data from both tri-axial and uni-axial experiments. 
For our purposes we will use the tri-axial definition of $F_2$, which is $F_2 = 1 / \left( 2^{(n-1)/n} 3^{(n+1)/2n}  \right)$ \cite{gerya_book}. 
The difference between factors is small for dislocation creep (0.3008 for a triaxial experiment and 0.3048 for simple shear), and the diffusion creep factors are similar enough for our purposes (0.333 for triaxial and 0.5 for simple shear). 

For the prefactors $A$ reported in \cite{hirth_kohlstedt_2003}, the other parameters in the flow law must have specific units that are not necessarily SI base units. They use stress in MPa, fH2O in MPa, and grain size in $\mu$m. When using the flow law in equation \ref{eqn:hk_flow_law}, we take care to match the necessary units. 

For diffusion creep, $n=1$ and $p=3$ (see \cite{hirth_kohlstedt_2003}, Table 1). 

\begin{align}  \label{eqn:eta_diff_v1}
 \eta_\text{diff} = F_2 (\Hat{A}_\text{diff})^{-1} (d)^{p} (fH2O)^{-r}    \exp \left( \frac{E_\text{diff} + V_\text{diff} P}{RT}  \right)
\end{align}

For dislocation creep, $n > 1$ and $p = 0$ (see \cite{hirth_kohlstedt_2003}, Table 1).

\begin{align}  \label{eqn:eta_disl_v1}
 \eta_\text{disl} = F_2 (\hat{A}_\text{disl})^{-1/n} (fH2O)^{-r/n}    \exp \left( \frac{E_\text{disl} + V_\text{disl} P}{nRT}  \right) (\Dot{\epsilon}_\text{II})^{(1 - n)/n}
\end{align}

\begin{table}
\begin{center}
\caption{Values from \citeA{hirth_kohlstedt_2003}}
\label{table:values_from_lit}
\begin{tabular}{  l l l l }
Parameter & Unit & Range \\
\hline
Diffusion creep & & & \\
$\Hat{A}_\text{diff}$ & - & [ 1.0 x $10^6$ , 1.5 x $10^9$]   \\
$r$ & - & [0.7, 1.0] \\
$E_\text{diff}$ & J/mol & [260 x $10^3$, 450 x $10^3$] \\
$d$ & $\mu$m & [10,1000]  \\
$fH2O$ & MPa & [500, 1500] \\
\hline
Dislocation creep & &  \\
$\Hat{A}_\text{disl}$ & - & [ 90 , 1.1 x $10^5$]   \\
$r$ & - & [0.8, 1.6]  \\
$E_\text{disl}$ & J/mol & [440 x $10^3$, 560 x $10^3$]  \\
$n$ & - & [3.2, 3.8]  \\
$fH2O$ & MPa & [500, 1500] \\
\end{tabular}
\end{center}
\end{table}

For the purposes of our parametric study, we combine the ranges of certain parameters into a single prefactor that encompasses the full range of each parameter. 
For diffusion creep, we combine $\Hat{A}_\text{diff}$, $r$, $d$, and $fH2O$ into a single prefactor $A_\text{diff}$:

\begin{align} 
 A_\text{diff} = (\Hat{A}_\text{diff})^{-1} (d)^{p} (fH2O)^{-r}
\end{align}

For dislocation creep, we combine $A_\text{disl}$, $fH2O$, and $r$ into the prefactor $\Hat{A}_\text{disl}$:

\begin{align} 
 A_\text{disl} = (\Hat{A}_\text{disl})^{-1} (fH2O)^{-r} 
 \end{align}

To compute the ranges of $A_\text{diff}$ and $A_\text{disl}$, we take the Cartesian product of the intervals of each parameter in their respective expressions and take the minimum and maximum $A_\text{diff}$ and $A_\text{disl}$ values computed for each ordered pair of parameter values.

The final expressions for effective viscosity used in our implementation are:

\begin{align}
 \eta_\text{diff} &= A_\text{diff}  F_2  \exp \left( \frac{E_\text{diff} + V_\text{diff} P}{RT}  \right), \label{eqn:eta_diff}\\
 \eta_\text{disl} &= (A_\text{disl})^{1/n}  F_2  \exp \left( \frac{E_\text{disl} + V_\text{disl} P}{nRT}  \right) (\Dot{\epsilon}_\text{II})^{(1 - n)/n}.\label{eqn:eta_disl} 
\end{align}

The resulting ranges of $A_\text{diff}$ and $A_\text{disl}$ are reported in Table~\ref{table:interval_details}. With these ranges and reasonable choices for $T, P$, and $\Dot{\epsilon}_\text{II}$, it is possible to evaluate Equation~\ref{eqn:eta_diff} and Equation~\ref{eqn:eta_disl} and obtain effective viscosities orders of magnitude smaller and larger than the expected viscosities of Earth's mantle. As such, when implementing the composite diffusion-dislocation creep flow law, we limit the viscosity above and below to $\eta_\text{max} = 10^{26}$ and $\eta_\text{min} = 10^{18}$. 

\begin{align}
\frac{1}{\eta_\text{comp}(\boldsymbol u, T)} &= \frac{1}{\eta_{\text{diff}}(T) } + \frac{1}{\eta_{\text{disl}}(\boldsymbol u, T) } + \frac{1}{\eta_{\text{max}}}, \\
\eta &= \max \left\{ \eta_\text{comp}, \eta_\text{min} \right\}.
\end{align}

\subsection{Additional Results/Content}

\renewcommand*{\arraystretch}{2.0}

\begin{longtable}{ p{.1\textwidth}  p{.1\textwidth}  p{.17\textwidth} p{.63\textwidth} } 
\caption{Input parameter intervals that are used for all three margins. $^*$Scaled for stress in Mpa, water fugacity in Mpa, and grain size in $\mu$m (See Supporting Information Section~\ref{sec:visco_params}).}
\label{table:interval_details}
\endfirsthead
\endhead
Param. & Units & Value & Citation and Rationale \\ \hline
$d_c$ & km & [70, 80] & \citeA{wada_wang_2009} present a study of 17 subduction zones, showing that for three of the profiles, surface heat flux data are consistent with maximum depths of decoupling in the $\sim$~60~--~90 km range, and for the remaining 14 profiles a depth of 75 km is not inconsistent with the surface heat flux data. They hypothesize that the maximum depth of decoupling is between 70 and 80 km in most subduction zones. We vary $d_c$ in the 70--80 km range in this study, but note that this range is not tightly constrained and may be broader. It can be assumed that the plate interface extends at least to the maximum depth of thrust faulting earthquakes ($\sim$~30~--~70 km, \cite<e.g>{slab2, heuret_2011}), and presumably extends deeper, to depths where low-frequency earthquakes and episodic tremor and slip (ETS) occur (with ETS typically distributed over a further $\sim$~10 km depth range, \cite<e.g>{wech_2021}). \\

$\lambda$ & - & [0, 0.1] & Range encompasses values in \cite{kneller_et_al_2007}; they test local sensitivity to variations between [0.01, 0.09]. \cite{abers_et_al_2006} use a range of [0, 0.1], saying this produces a strong vertical division between cold and hot regions as seen in Q tomography. \\

$\mu'$ & - & [0, 0.14] & 
\citeA{gao_wang_2014} predicted a relatively weak subduction interface, with preferred values of $\mu' = 0.03$ for six out of ten subduction zones studied, with large uncertainties of up to $\pm$ 0.03. Their range for Cascadia was [0, 0.06], while \cite{england_2018} reports a Cascadia uncertainty interval of [0, 0.14]. 
For Nankai, \citeA{gao_wang_2014} have an uncertainty range of [0, 0.06], while \citeA{england_2018} reports [0.01, 0.1]. 
For Hikurangi, \citeA{fagereng_2009} use a different formulation for shear heating but the equivalent $\mu'$ varies between 0.03 and 0.36, while \citeA{england_2018} considers a range of [0.02, 0.1] for Hikurangi.
We use a range of [0, 0.1] since it encompasses most of these uncertainty ranges and is representative of $\mu'$ values used previously in the literature. 
The uncertainty in estimates of $\mu'$ comes from scatter in surface heat flow data and from uncertainty in the amount of radiogenic heat production in the crust.
\\

$A_\text{diff}$ & Pa s$^*$  & [4.4 $ \times 10^{-10}$, 1.3 $ \times 10^{1}$] & Includes stated uncertainty ranges from \cite{hirth_kohlstedt_2003} for both dry and wet diffusion creep. \\

$E_\text{diff}$ & kJ/mol & [$300$, $450$] & Includes stated uncertainty ranges from \cite{hirth_kohlstedt_2003} for both dry and wet diffusion creep.\\

$A_\text{disl}$ & Pa s$^*$  & [7.5 $\times 10^{-11}$, 7.7 $ \times 10^{-5}$] & Includes stated uncertainty ranges from \cite{hirth_kohlstedt_2003} for both dry and wet dislocatin creep.\\

$E_\text{disl}$ & kJ/mol & [$480$, $560$] & Includes stated uncertainty ranges from \cite{hirth_kohlstedt_2003} for both dry and wet dislocation creep. \\

$n$ & ~ & [0, 3.5] & Upper limit of 3.5 is the value for dry olivine from \cite{karato_wu_1993}. We choose a lower limit of 0 to account for any amount of contribution from the strain rate term.\\

$T_b$ & K & [1550, 1750] &  Parsons and Sclater report temperatures of 1606 $\pm$ 274 K for the N Pacific, and 1638 $\pm$ 276 K for the N Atlantic. \citeA{stein_stein_1992} report a best-fit value of 1723 K. We select a slightly broader range to encompass values used in the literature, e.g. 1723 K in \cite{wada_wang_2009}, 1523 K in \cite{hyndman_wang_1993}; 1723 K in \cite{van_keken_kiefer_peacock_2002}; 1573 K in \cite{community_benchmark_van_keken_2008}; 1723 K in \cite{currie_et_al_2004}. \\

$y_g$ & km & [10, 60] & Typical continental crustal thickness is 30--40 km \cite{laske_et_al_2013}, we consider $\pm 20$ km on this range to account for broader variability in the upper-plate geotherm. \\ \hline
\hline
\end{longtable}


\begin{longtable}{ p{.08\textwidth}  p{.08\textwidth}  p{.12\textwidth} p{.12\textwidth} p{.1\textwidth} p{.5\textwidth} } 
\caption{Margin-specific input parameter intervals and the base values used when that parameter is held constant.}
\label{table:margin_ranges_longtable}
\endfirsthead
\endhead
& Units & Margin & Range & Value & Citation / Rationale \\ \hline
$u_s$ & cm/yr & Cascadia & [3.0, 5.0] & 5.0 & See \citeA{demets_et_al_2010}; range encompasses values used in \cite{van_keken_kiefer_peacock_2002, currie_hyndman_2006, syracuse_2010}.\\

& & Nankai & [3.0, 5.0] & 5.0 & \cite{syracuse_2010} use a value of 4.3 cm/yr, while \cite{wada_wang_2009} use 4.6 cm/yr. The plate convergence rate (not trench normal) reported in \cite{demets_et_al_2010} is 5.8 cm/yr, and by projecting this in the along-profile direction we estimate a convergence rate of 4.4 cm/yr.  
\\

& & Hikurangi & [3.0, 4.0] & 3.0 & Range of values estimated from \cite{wallace_et_al_2004}. \\

$a_s$ & Myr & Cascadia & [8, 10] & 8 & See \cite{muller_et_al_1997, seton_et_al_2020}. Range encompasses values used in \cite{currie_hyndman_2006, van_keken_kiefer_peacock_2002, wada_wang_2009}. We select 8 as the base value since it is most commonly used in the literature. \\

& & Nankai & [15, 26] & 20 & See \cite{okino_et_al_1994, muller_et_al_1997, seton_et_al_2020}. We select an approximately average value of 20, as in e.g. \cite{syracuse_2010}. \\

& & Hikurangi & [90, 110] & 100 & See \cite{muller_et_al_1997, seton_et_al_2020}. \citeA{wada_wang_2009} use an age of 100 Myr, we add $\pm10$ Myr to obtain our range and use 100 Myr as the base value.  \\

\hline
\end{longtable}

\renewcommand*{\arraystretch}{2.0}

\begin{longtable}{ p{.08\textwidth}  p{.14\textwidth}  p{.10\textwidth} p{.72\textwidth} } 
\caption{Reference input parameter values. For parameters in $\mathbf q_\text{base} = ( u_s, d_c, \lambda, \mu', A_\text{diff},  E_\text{diff}, A_\text{disl}, E_\text{disl}, n, T_b, a_s, y_g)$, the values in this table are used whenever that parameter is held constant. For parameters not in $\mathbf q_\text{base}$, the values reported here are used in all experiments in this study.}
\label{table:base_params_longtable}
\endfirsthead
\endhead
Param. & Units & Value & Citation / Rationale \\ \hline

$d_c$ & km & 80 & Deepest value in range from \cite{wada_wang_2009}. \\

$\lambda$ & - & 0.05  & \cite{syracuse_2010} use the value 0.05, the average of the values in \cite{kneller_et_al_2007}. \\

$\mu'$ & - & 0.03 & Value used by \citeA{wada_wang_2009}. \citeA{gao_wang_2014} have a preferred value for Nankai of 0.03, while their range for Cascadia is [0, 0.06] with a preferred value of 0.02. For Hikurangi, \citeA{fagereng_2009} use a different formulation for shear heating but the equivalent $\mu'$ varies between 0.03 and 0.36; \citeA{england_2018} considers a range of [0.02, 0.1] for Hikurangi. We select 0.03 since it is a common estimate for the three margins and provides moderate shear heating. 
\\

$A_\text{diff}$ & GPa s & $1.32043 $ & \cite{community_benchmark_van_keken_2008} \\

$E_\text{diff}$ & J/mol & $335  \ \times  \ 10^3$ & \cite{karato_wu_1993, community_benchmark_van_keken_2008} \\

$A_\text{disl}$ & Pa s$^{(1/\text{n})}$ & 28968.6 & Value from \cite{karato_wu_1993}; also used in, e.g. \cite{van_keken_kiefer_peacock_2002, community_benchmark_van_keken_2008, syracuse_2010}. \\

$E_\text{disl}$ & J/mol & $540  \ \times \  10^3$ & Value from \cite{karato_wu_1993}; also used in, e.g. \cite{van_keken_kiefer_peacock_2002, community_benchmark_van_keken_2008, syracuse_2010}.  \\

$n$ & - & 3.5 & Value used for dislocation creep for dry olivine in \cite{karato_wu_1993}. \\

$T_b$ & K & 1573 & \cite{community_benchmark_van_keken_2008} \\

$y_g$ & km & 35  & \cite{turcotte_schubert_2014}, typical value for continental crust thickness. \\

$T_s$ & K & 273 & \cite{community_benchmark_van_keken_2008}  \\

$L$ & km & 10  & Chosen value for this study. \\

$\rho_\text{mantle}$ & kg m$^{-3}$ & 3300 & \cite{turcotte_schubert_2014}, typical mantle density; used in, e.g. \cite{van_keken_kiefer_peacock_2002, community_benchmark_van_keken_2008, morishige_kuwatani}\\

$\rho_\text{crust}$ & kg m$^{-3}$ & 2700 & \cite{turcotte_schubert_2014}, typical crustal density; used in, e.g. \cite{morishige_kuwatani} \\

$\rho_\text{slab}$ & kg m$^{-3}$ & 3300  & Value chosen to match $\rho_\text{mantle}$. \\

$g$ & m s$^{-2}$ & 9.8  & Common value for gravitational acceleration, e.g. \cite{turcotte_schubert_2014}. \\

$k_\text{mantle}$ & W m$^{-1}$ K$^{-1}$ & 3.1  & \cite{van_keken_kiefer_peacock_2002, gao_wang_2014}\\

$k_\text{crust}$ & W m$^{-1}$ K$^{-1}$ & 2.5  & \cite{van_keken_kiefer_peacock_2002, gao_wang_2014} \\

$k_\text{slab}$ & W m$^{-1}$ K$^{-1}$ & 3.1 & Value chosen to match $k_\text{mantle}$. \\

$c_p$ & J kg$^{-1}$ K$^{-1}$ & 1250 & \cite{van_keken_kiefer_peacock_2002, community_benchmark_van_keken_2008, syracuse_2010} \\

$s$ & km & 0.2 & Chosen value for this study, so that almost all of the shear heating is generated in a narrow region $<1$ km wide. \\
\hline
\end{longtable}


\begin{table}[h!]
\begin{center}
\caption{$R^2$ values for a 4th order polynomial fit to the relationship between $\mu'$ and along-slab distances $D_{350}$, $D_{450}$, and $D_{600}$, for each profile. $R^2$ values are 0 for the relationship between all other input parameters and $D_{350}$, $D_{450}$, and $D_{600}$, for sets of parameters that include $\mu'$. }
\label{table:param_R^2}
\begin{tabular}{ l  l  l  l  } 
 \hline
 $R^2$ for $\mu'$ vs: & $D_{350}$ & $D_{450}$ & $D_{600}$ \\
 Cascadia profile B & 0.91 & 0.96 & 0.98 \\
 Cascadia profile G &  0.91 & 0.95 & 0.98 \\
 Cascadia profile I & 0.9 & 0.94 & 0.98 \\
 Nankai profile C & 0.99 & 0.95 & 0.82 \\
 Nankai profile A & 0.99 & 0.97 &  0.9 \\
 Nankai profile D & 0.99 & 0.97 & 0.89 \\
 Hikurangi profile F & 0.91 & 0.91 & 0.94 \\
 Hikurangi profile G & 0.87 & 0.91 & 0.94 \\
 Hikurangi profile H &  0.76 &  0.86 & 0.92 \\
\end{tabular}
\end{center}
\end{table}

\begin{figure}
     \centering
     \includegraphics[width=\textwidth]{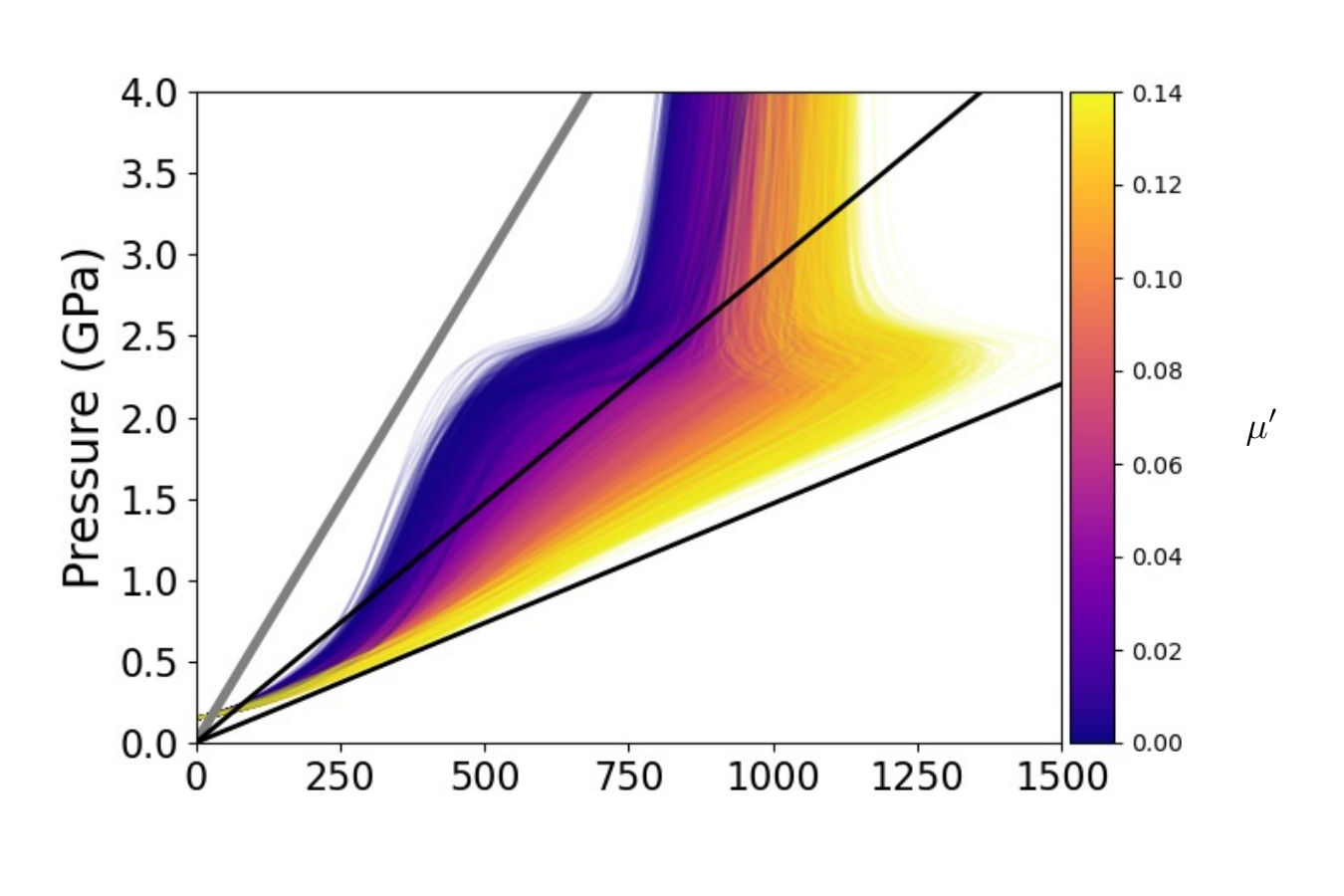}
     \caption{P-T curves for slab interface temperatures for Cascadia profile B; each line is one of $10^4$ samples in parameter space and is colored according to the $\mu'$ value for that sample. This shows the relationship between increasing $\mu'$, and therefore increasing shear heating, and higher temperature conditions for a given pressure.}
     \label{fig:PT_vs_mu}
\end{figure}

\begin{figure}
     \centering
     \includegraphics[width=\textwidth]{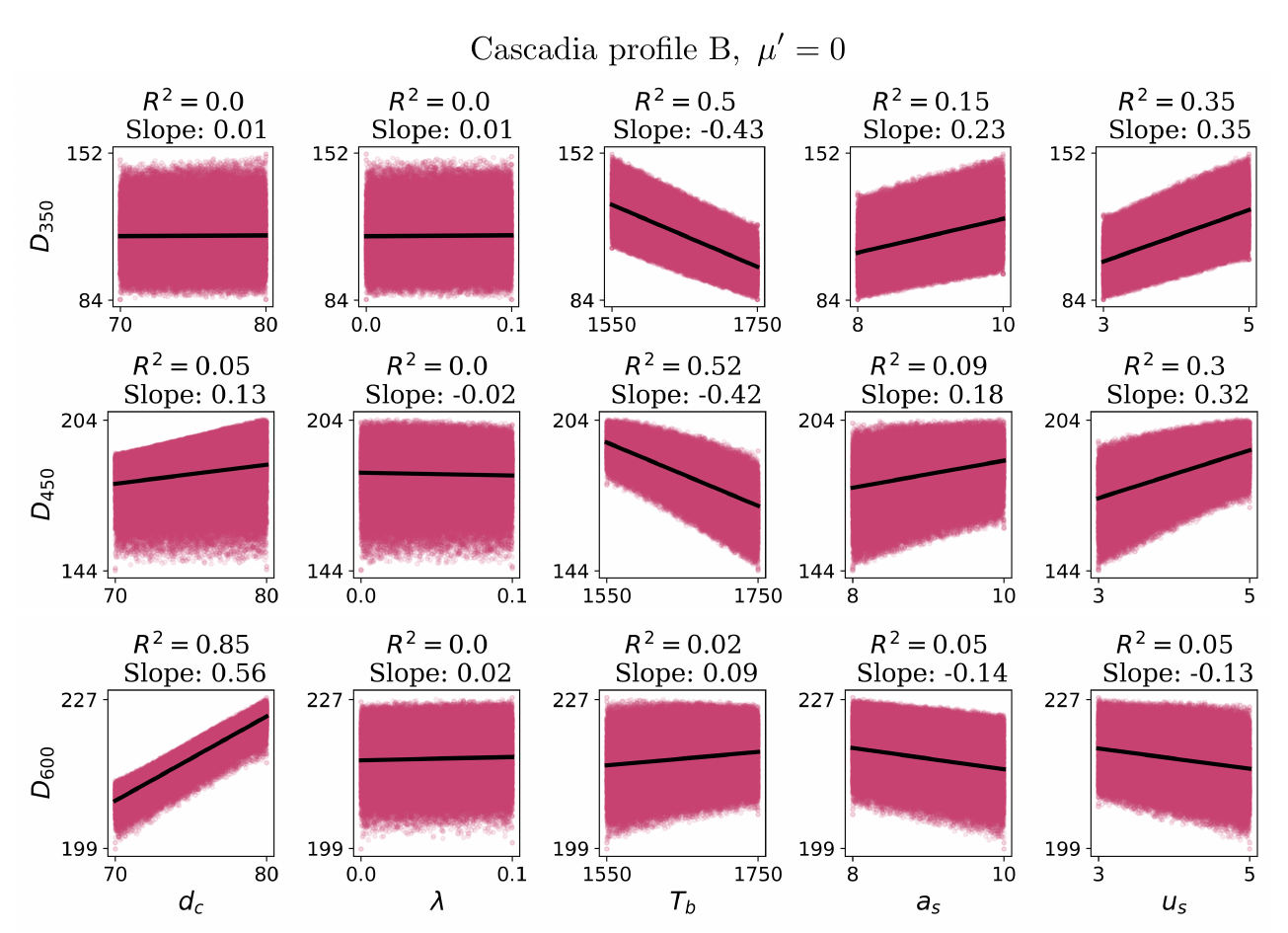}
     \caption{Linear regression for $D$ against input parameters, for Cascadia profile B with $\mu'=0$. The $R^2$ value and slope of each linear relationship are reported. }
     \label{fig:linreg_cascadia_profile_B_mu_0.0}
\end{figure}

\begin{figure}
     \centering
     \includegraphics[width=\textwidth]{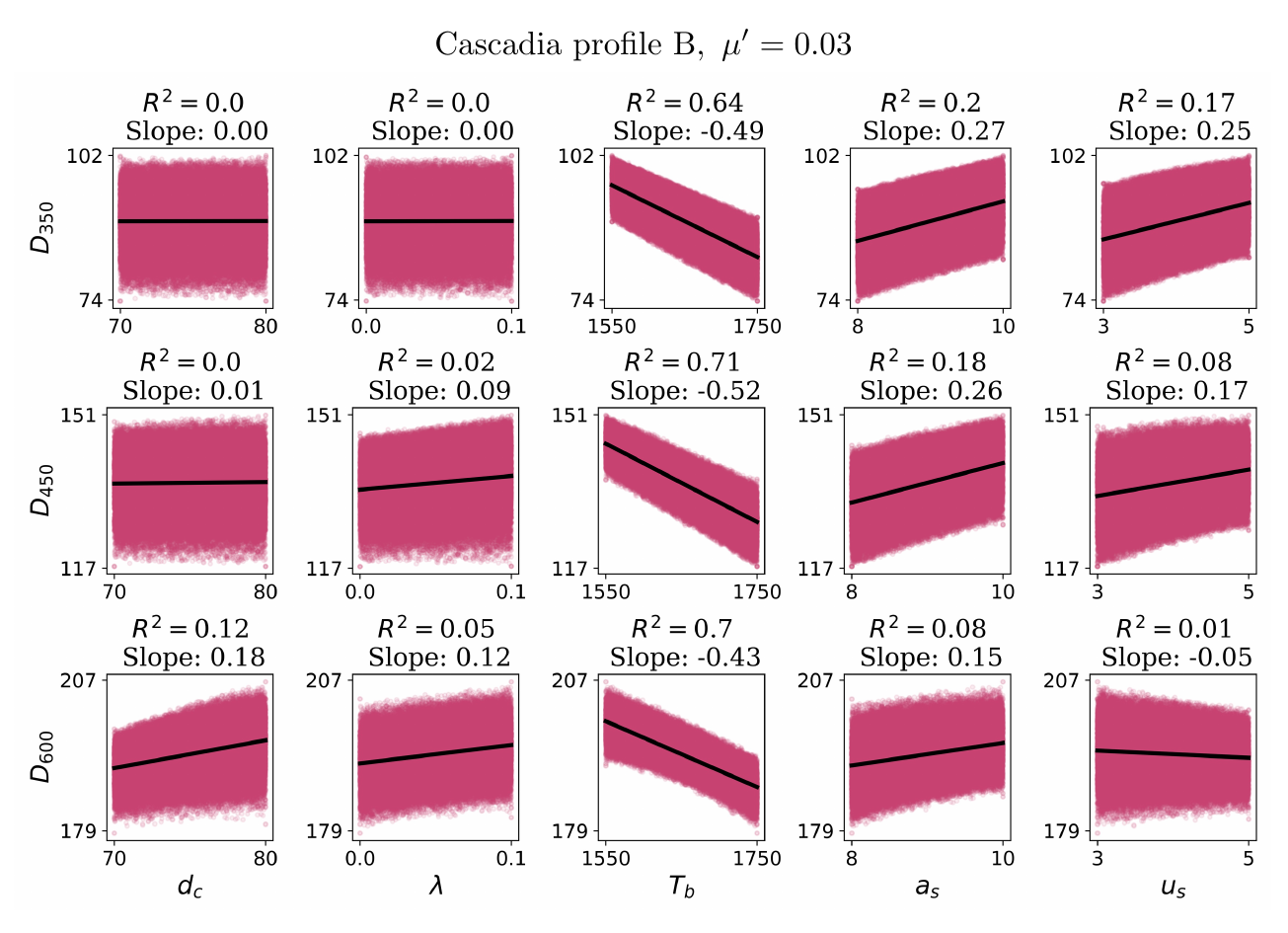}
     \caption{As in Figure~\ref{fig:linreg_cascadia_profile_B_mu_0.0}, but with $\mu' = 0.03$.}
     \label{fig:linreg_cascadia_profile_B_mu_0.03}
\end{figure}

\begin{figure}
     \centering
     \includegraphics[width=\textwidth]{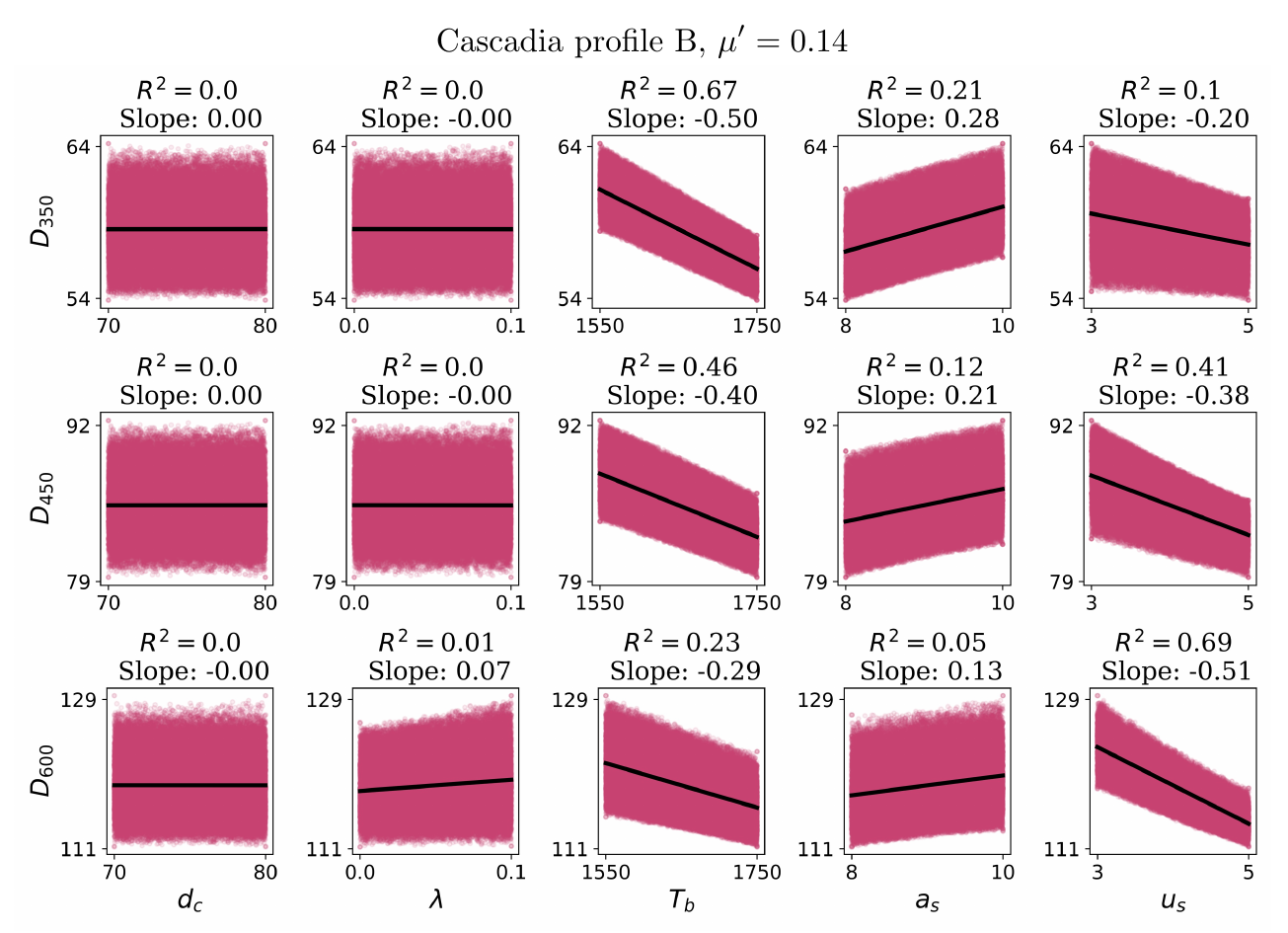}
     \caption{As in Figure~\ref{fig:linreg_cascadia_profile_B_mu_0.0}, but with $\mu' = 0.14$.}
     \label{fig:linreg_cascadia_profile_B_mu_0.14}
\end{figure}

\begin{figure}
     \centering
     \includegraphics[width=\textwidth]{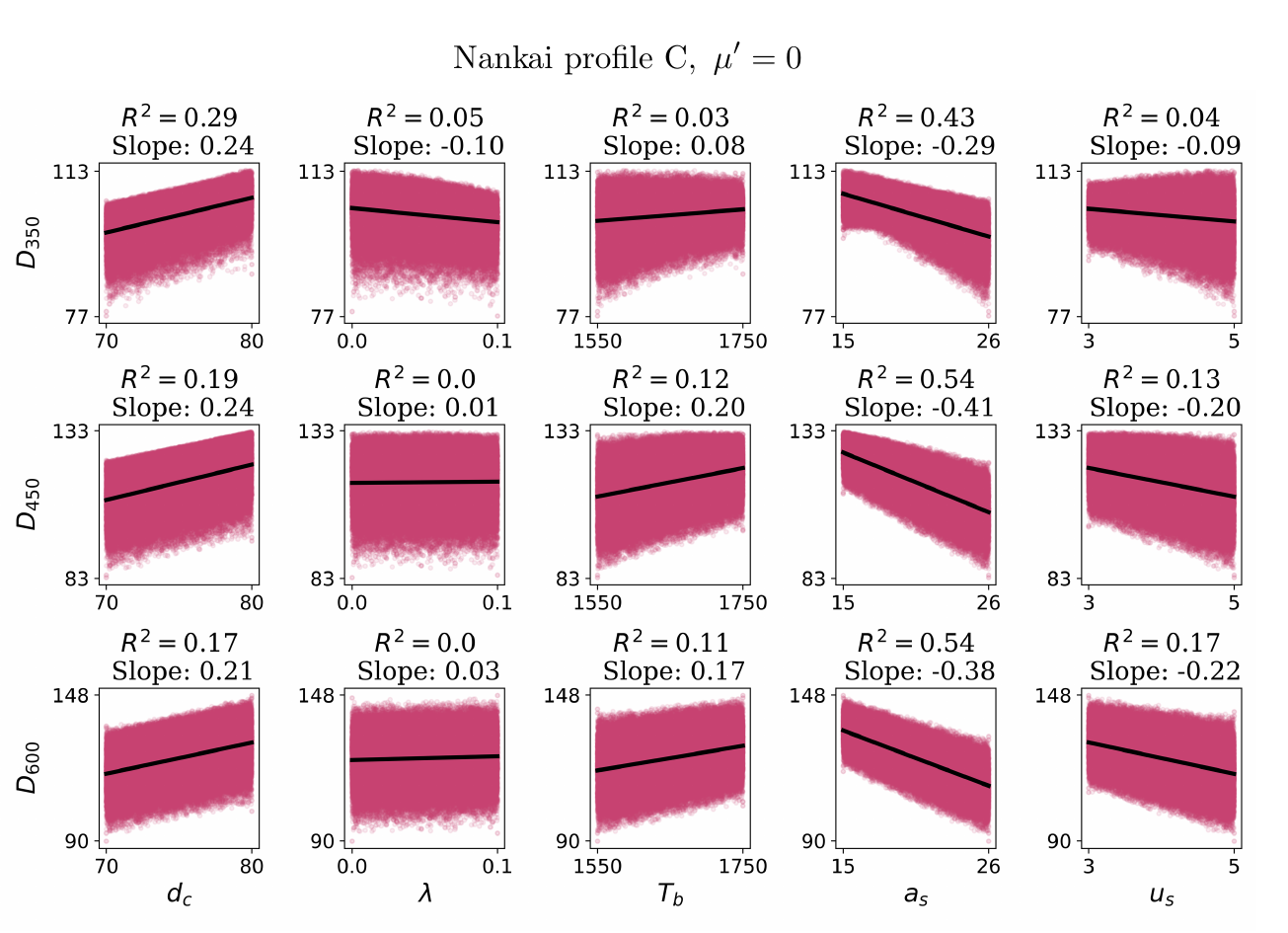}
     \caption{As in Figure~\ref{fig:linreg_cascadia_profile_B_mu_0.0}, but for Nankai profile C with $\mu' = 0$.}
     \label{fig:linreg_nankai_profile_C_mu_0.0}
\end{figure}

\begin{figure}
     \centering
     \includegraphics[width=\textwidth]{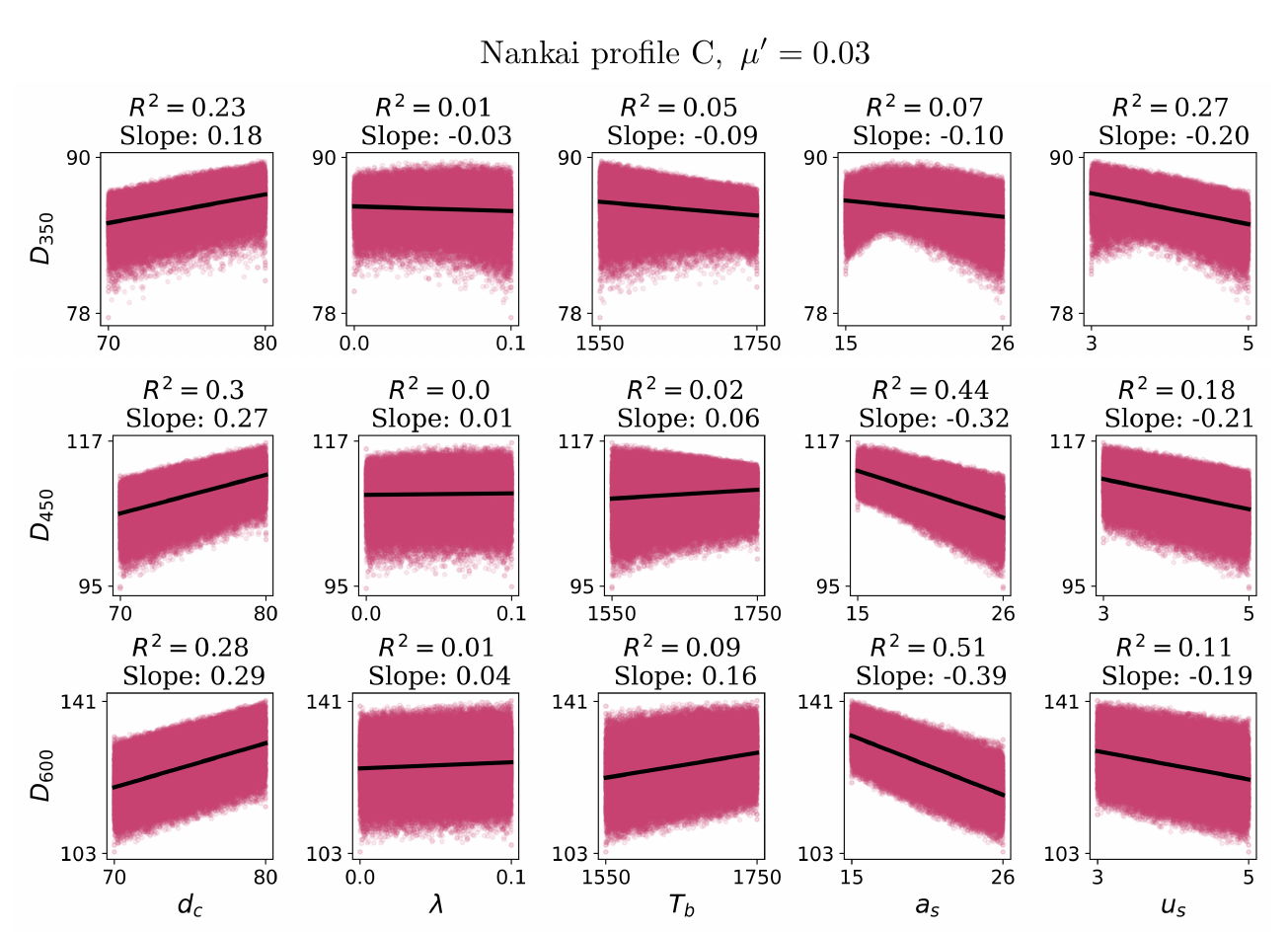}
     \caption{As in Figure~\ref{fig:linreg_cascadia_profile_B_mu_0.0}, but for Nankai profile C with $\mu' = 0.03$.}
     \label{fig:linreg_nankai_profile_C_mu_0.03}
\end{figure}

\begin{figure}
     \centering
     \includegraphics[width=\textwidth]{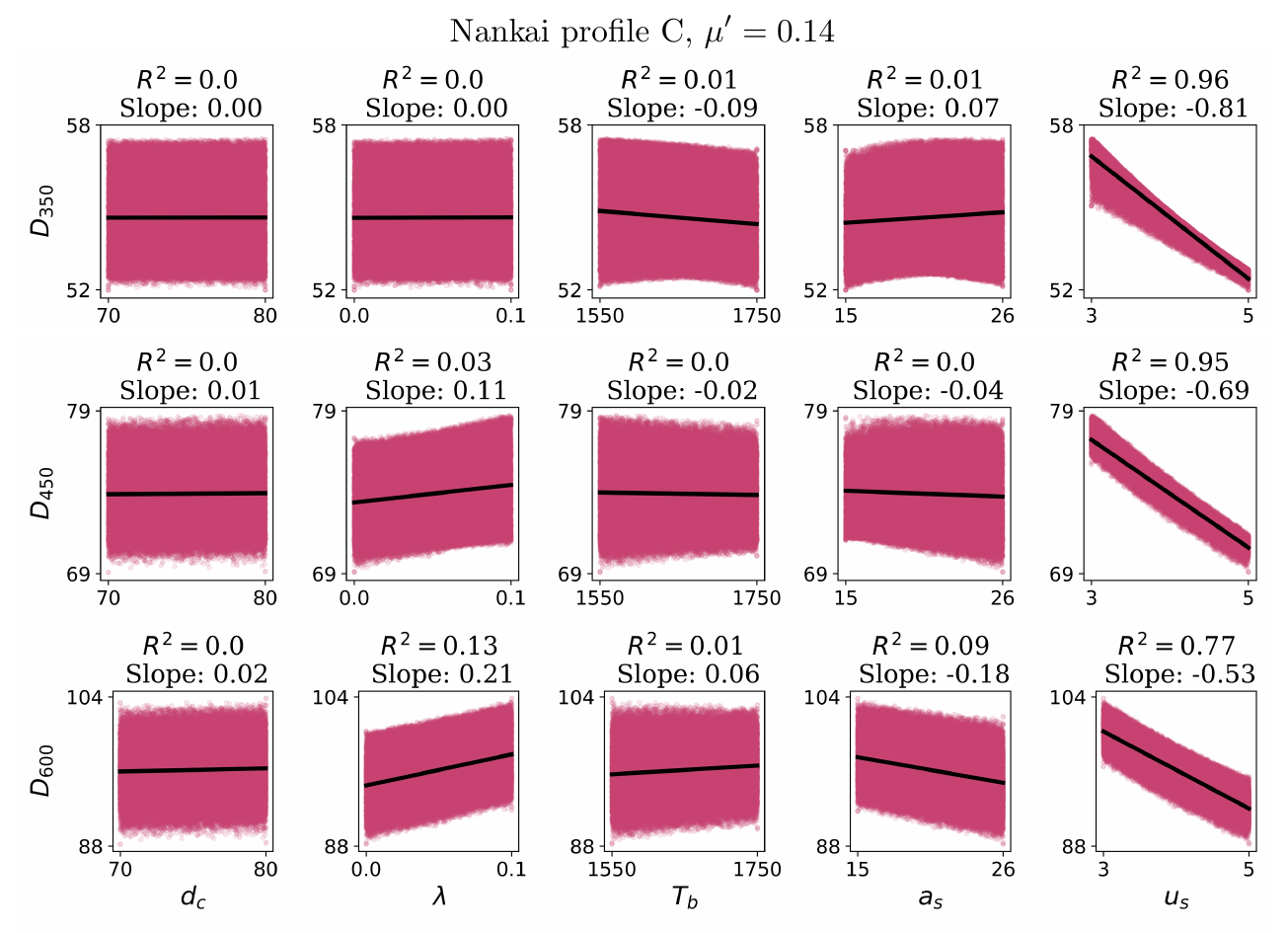}
     \caption{As in Figure~\ref{fig:linreg_cascadia_profile_B_mu_0.0}, but for Nankai profile C with $\mu' = 0.14$.}
     \label{fig:linreg_nankai_profile_C_mu_0.14}
\end{figure}

\begin{figure}
     \centering
     \includegraphics[width=\textwidth]{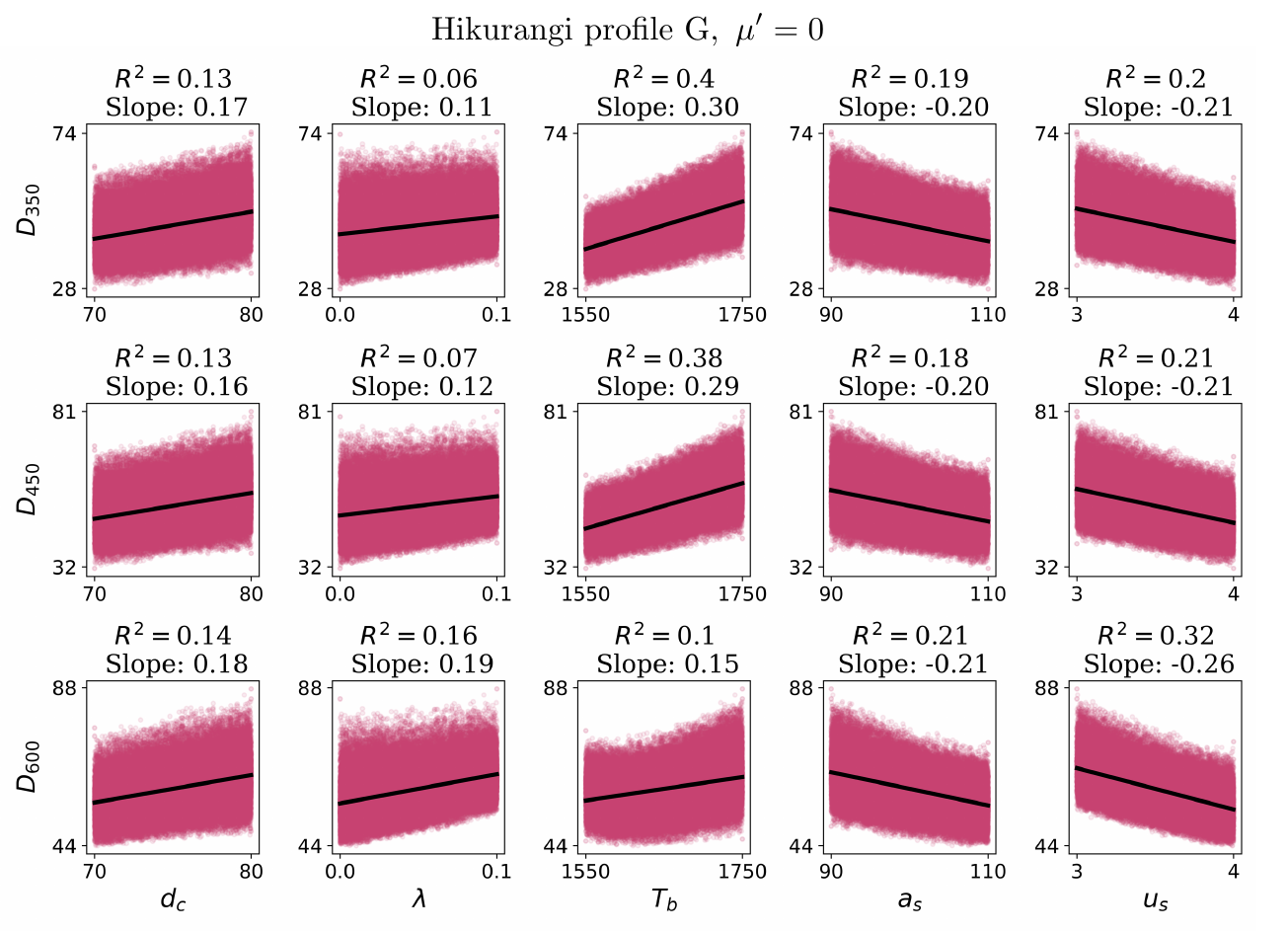}
     \caption{As in Figure~\ref{fig:linreg_hikurangi_profile_G_mu_0.0}, but for Hikurangi profile G with $\mu' = 0$.}
     \label{fig:linreg_hikurangi_profile_G_mu_0.0}
\end{figure}

\begin{figure}
     \centering
     \includegraphics[width=\textwidth]{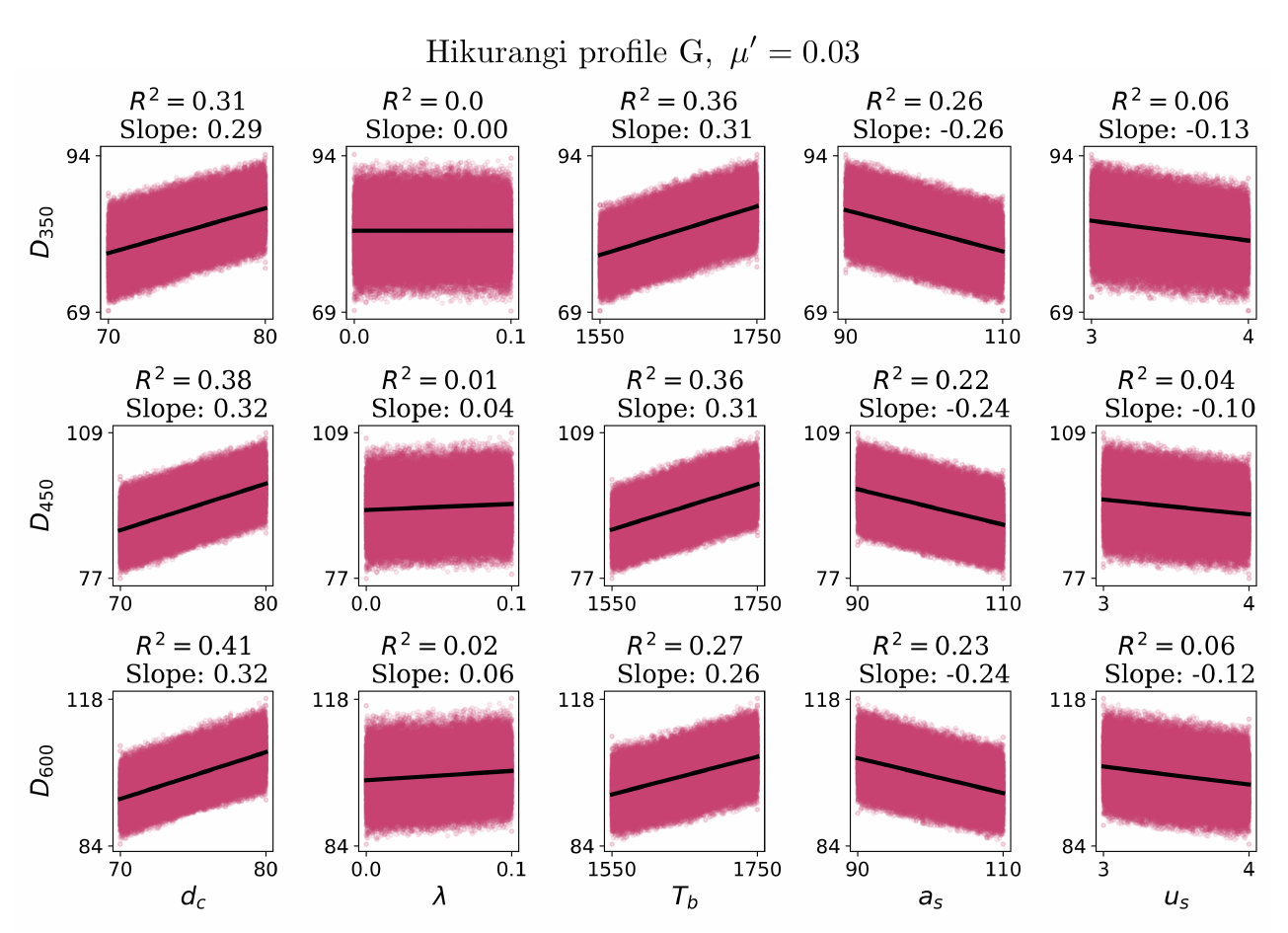}
     \caption{As in Figure~\ref{fig:linreg_hikurangi_profile_G_mu_0.0}, but for Hikurangi profile G with $\mu' = 0.03$.}
     \label{fig:linreg_hikurangi_profile_G_mu_0.03}
\end{figure}

\begin{figure}
     \centering
     \includegraphics[width=\textwidth]{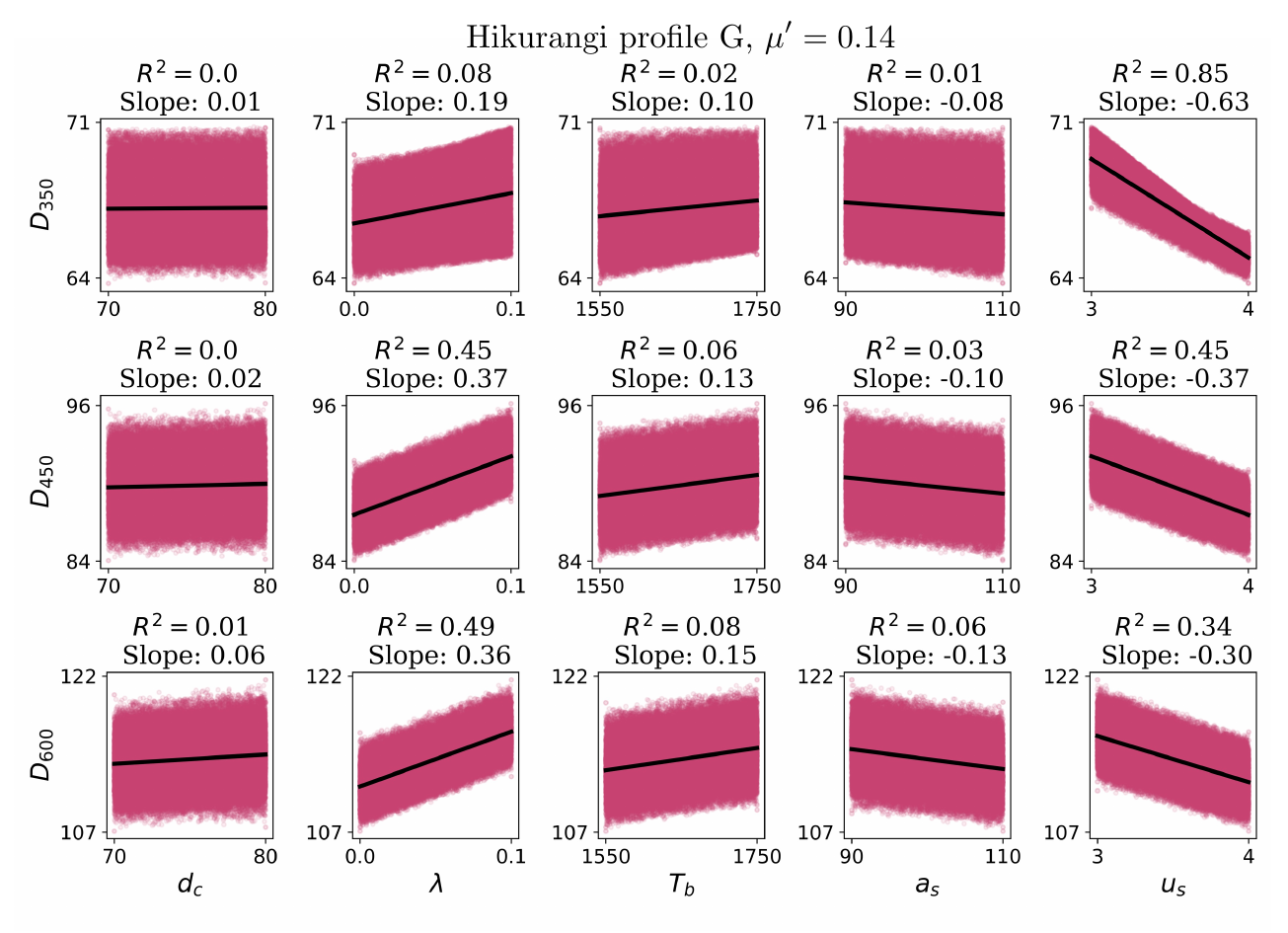}
     \caption{As in Figure~\ref{fig:linreg_hikurangi_profile_G_mu_0.0}, but for Hikurangi profile G with $\mu' = 0.14$.}
     \label{fig:linreg_hikurangi_profile_G_mu_0.14}
\end{figure}

\section{Open Research}

The code required to run the thermal models in this study is available open-source in a Github repository \cite{github_repo} under the GPL-3.0 license.  
The Zenodo repository that accompanies this paper contains: input files for running the thermal models; code demonstrating the shear heating implementations; data files and jupyter notebooks for the comparison with analytical solutions in \ref{sec:appendix_verif_analytical}; and the data files for each ROM presented in this study, along with Jupyter notebooks demonstrating how to evaluate the ROMs and plot the results \cite{zenodo_repo}. 

\acknowledgments
Both authors acknowledge financial support from the National Science Foundation grant No. EAR-2121568.
T.~W. Becker and Dunyu Liu are thanked for providing the computational resources required for the offline stage of the ROM construction.
We thank Philip England for valuable feedback on an earlier version of this manuscript.

\bibliography{references}

\end{document}